\documentclass[twocolumn,pre,10pt,showpacs,superscriptaddress,amsmath,aps]{revtex4-1}
\usepackage{graphicx,hyperref}
\usepackage[caption=false]{subfig}
\usepackage{floatrow}
\usepackage{epsfig}
\usepackage{color}
\usepackage[normalem]{ulem}
\definecolor{green}{rgb}{0.0, 0.44, 0.0}
\definecolor{red}{rgb}{1.0, 0.13, 0.32}
\definecolor{blue}{rgb}{0.06, 0.2, 0.65}
\definecolor{magenta}{rgb}{1.0, 0.0, 1.00}
\definecolor{purple}{rgb}{0.7, 0.0, 0.7}
\definecolor{cyan}{rgb}{0.0, 1.0, 1.0}

\newcommand{\res}[1]{{\color{black} #1}}
\newcommand{\be}{\begin{equation}}
\newcommand{\ee}{\end{equation}}
\newcommand{\bea}{\begin{eqnarray}}
\newcommand{\eea}{\end{eqnarray}}
\def\le{\left}
\def\ri{\right}
\def\PF{\phi}
\def\qc{q_{\rm c}}

\begin{document}

\title{Configurational entropy measurements in extremely supercooled liquids that break the glass ceiling}

\author{L. Berthier}  
\affiliation{Laboratoire Charles Coulomb, CNRS-UMR 5221, 
Universit\'e de Montpellier, Montpellier, France}

\author{P. Charbonneau} 
\affiliation{Department of Chemistry, Duke University, Durham,
North Carolina 27708, USA}

\affiliation{Department of Physics, Duke University, Durham,
North Carolina 27708, USA}

\author{D. Coslovich} 
\affiliation{Laboratoire Charles Coulomb, CNRS-UMR 5221, 
Universit\'e de Montpellier, Montpellier, France}

\author{A. Ninarello} 
\affiliation{Laboratoire Charles Coulomb, CNRS-UMR 5221, 
Universit\'e de Montpellier, Montpellier, France}

\author{M. Ozawa} 
\affiliation{Laboratoire Charles Coulomb, CNRS-UMR 5221, 
Universit\'e de Montpellier, Montpellier, France}

\author{S. Yaida} 
\affiliation{Department of Chemistry, Duke University, Durham,
North Carolina 27708, USA}


\begin{abstract}
Liquids relax extremely slowly upon approaching the glass state. One explanation is that an entropy crisis, due to the rarefaction of available states, makes it increasingly arduous to reach equilibrium in that regime. Validating this scenario is challenging, because experiments offer limited resolution, while numerical studies lag more than eight orders of magnitude behind experimentally-relevant timescales. In this work we not only close the colossal gap between experiments and simulations but manage to create {\it in-silico} configurations that have no experimental analog yet. Deploying a range of computational tools, we obtain four estimates of their configurational entropy. These measurements consistently confirm that the steep entropy decrease observed in experiments is also found in simulations, even beyond the experimental glass transition. Our numerical results thus extend the new observational window into the physics of glasses and reinforce the relevance of an entropy crisis for understanding their formation. 
\end{abstract}

\maketitle

\emph{Introduction--}
In his landmark 1948 paper, Kauzmann gathered
experimental data for several glass-forming liquids
and found that they all showed a steep 
decrease of their equilibrium configurational entropy upon 
lowering temperature towards their glass transition~\cite{Kauzmann48}.
Theoretically, the nature of a thermodynamic glass transition associated with a vanishing configurational entropy is well-understood at the mean-field level~\cite{KT87,KW87,CKPUZ14}, suggesting that glass formation is accompanied by a rarefaction of available disordered states~\cite{BB11}. Its pertinence beyond the mean-field framework, however, remains controversial~\cite{KTW89,stillinger_supercooled_1988,GC03,BB11}. 
In particular, it is still not known whether such entropy reduction is the
core explanation for glass formation. Experimental 
measurements are carried out over too limited a temperature range, 
within boundaries that have remained essentially unchanged since Kauzmann's work 
and thus form a solid glass ceiling.
In addition, experimental determinations of the configurational entropy
are marred by approximations that influence their 
physical interpretation~\cite{Dyre,Wyart,Wolynes}.
Computer simulations can potentially provide 
more precise estimates~\cite{sciortino1999inherent,BC14}, 
but have so far been restricted to a temperature range that 
is not experimentally relevant.

Can the debate over the role of configurational entropy
ever be settled? At first sight, closure appears 
unlikely for two main reasons. (i) Measuring the configurational 
entropy below the experimental 
glass transition seems logically impossible, because experiments 
are constrained by their own duration, which fixes an 
upper limit to the accessible thermalization timescale, $\tau$.
Specifically, $\tau/\tau_0 \sim 10^{13}$ for molecules~\cite{rossler} 
(where the relaxation time at the onset temperature is 
$\tau_0 \approx 10^{-10}{\rm s}$) and $\tau/\tau_0 \sim 10^{5}$ 
for colloids~\cite{BEPPSBC08} (where $\tau_0 \approx 10^{-1}{\rm s}$). 
The situation for computer simulations is even worse. 
Current approaches access at most $\tau/\tau_0 \sim 10^{5}$, which
is eight orders of magnitude behind molecular liquid experiments, and
numerical progress has been slow. The two to three decades gained 
over the past 35 years~\cite{JL90,KA94,BEPPSBC08} 
are mostly thanks to hardware improvements.
At this pace, another century would be needed before simulations attain  
experimentally-relevant conditions. The glass ceiling thus appears unbreakable.
(ii) There is a fundamental methodological ambiguity 
as to which configurational entropy should be measured in order  
to match theoretical calculations.
Qualitatively, the configurational entropy is defined by subtracting 
vibrational contributions from the total 
entropy~\cite{Kauzmann48,martinez2001thermodynamic,sciortino1999inherent}.
What is specifically meant by ``vibrations'' in amorphous solids, however, 
is ill-defined in general~\cite{BB11} and difficult to measure in 
practice~\cite{Kauzmann48,BC14}. 
Hence consistently determining the configurational entropy 
is in itself a difficult challenge, that may be underestimated
in the literature.

Here, we solve both of these major problems at once. 
First, we take advantage of the flexibility offered by computer simulations 
to dramatically accelerate the equilibrium sampling of 
configuration space~\cite{GP01,BCNO16,PRX17}. 
Namely, we use a system optimized for the nonlocal swap Monte Carlo (MC) 
algorithm, which enables its extremely fast thermalization.
We establish that this approach surpasses any current alternative,
and even experimental protocols.
Second, we measure four proxies for the configurational entropy 
by deploying state-of-the-art computational tools to characterize
{\it in-silico} configurations that are more deeply equilibrated than 
their experimental 
analogs~\cite{angelani2007configurational,BCNO16,BC14,BCY16}
and \res{obtain consistent results that have a clear physical interpretation.} 
By combining these developments for a realistic model glass-former, we shift
computer simulations from lagging eight decades behind 
experiments to exploring novel territory in glass physics. In particular,
our measurements validate Kauzmann's observations that 
the configurational entropy decreases steeply towards the 
glass temperature, and extend these observations 
to a regime previously inaccessible.

\begin{figure}[!t]
\begin{center}
\includegraphics[width=\linewidth]{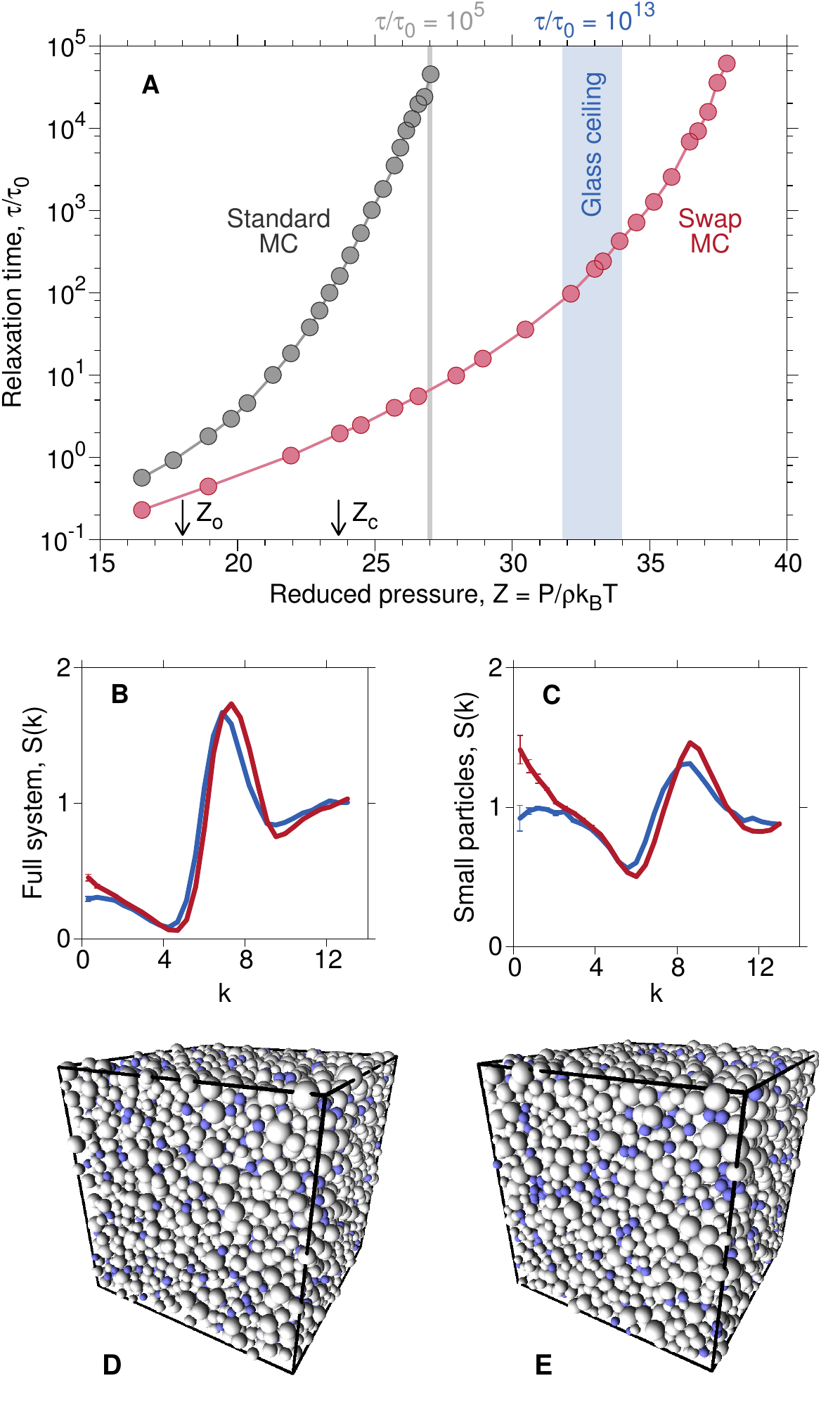}
\end{center}
\caption{
{Breaking the glass ceiling: thermalization beyond the experimental 
glass transition.}
(A) Structural relaxation time, $\tau$, for both standard and accelerated 
swap MC dynamics as a function of the reduced pressure
of a polydisperse hard sphere model, where $\tau$ is obtained from the 
decay of the self-intermediate scattering function at a wavenumber 
$k=5.25$. The onset
of slow dynamics occurs at $Z_0\approx 18$, and the mode-coupling 
crossover at $Z_c\approx 23.5$.
Times are rescaled by $\tau_0=\tau(Z_0)$ for standard MC.
The current limit of colloidal ($\tau/\tau_0 = 10^5$)
and molecular ($\tau/\tau_0 = 10^{13}$) experiments are indicated by vertical 
bands (this uncertainty stems from the extrapolation scheme),
showing that swap MC breaks the glass ceiling.
Static structure factor for $Z=18.8\approx Z_0$ ($\phi=0.568$)
and $Z=33.2$ ($\phi=0.640$) for (B) all particles (C) 
the 40\% particles with the smallest diameter.
(D, E) show typical snapshots for these two state 
points, where the smallest particles are highlighted in 
blue.
}
\label{figure:tau}
\end{figure}

\emph{Results--} 
We simulate a three-dimensional polydisperse mixture of hard spheres, 
as in~\cite{BCNO16}, 
which is a good model for colloids used in 
experiments~\cite{pusey_phase_1986,BEPPSBC08}. We show in 
Appendix~\ref{sec:soft} that our methods and conclusions also
apply to particles with soft and more complex interactions. We control the 
volume fraction $\phi$, and measure pressure $P$ to 
report the (unitless) reduced pressure, $Z = P/(\rho k_B T)$, where  
$\rho$ is the number density, and $k_BT$ the thermal energy. 
This natural control variable for hard spheres plays a role akin to the
inverse temperature in thermal liquids~\cite{berthier_glass_2009}.
Detailed information about the simulations is provided in Appendices~\ref{hard_model},~\ref{hard_methods}, and~\ref{Ap_eos}.
Swap MC complements
standard translational MC moves with nonlocal moves that exchange 
randomly-chosen pairs of particles, ensuring 
equilibrium sampling. Detailed tests 
of thermalization of all glassy degrees of freedom
are reported in Appendix~\ref{equi_make_sure}, see also Ref.~\cite{PRX17}.
We demonstrate the extreme speedup actually achieved by swap MC for 
this model in 
Fig.~\ref{figure:tau}, in which the 
structural relaxation time $\tau$  for both MC sampling methods is reported as the 
system approaches its glass transition.
Note that the rapid increase of $\tau$ in standard MC simulations 
resembles the fragile super-Arrhenius behavior of 
standard glass-formers~\cite{BB11}. We can only 
indirectly assess fragility beyond the reported 
numerical regime, which we estimated to be $m \approx 50$.  
We have fitted several empirical forms to our measurements,
which thermalize up to $Z\approx 27$, 
to estimate the experimental glass transition at $\tau/\tau_0 = 10^{13}$ (see Appendix~\ref{Ap_localization}).
Use of various fits reflects the well-known uncertainties 
associated with the empirical description of data measured over a large 
dynamical range~\cite{stickel95}.
The fits give consistent locations for the glass ceiling,
$Z_\mathrm{g} \approx 32$-34, as highlighted in Fig.~\ref{figure:tau}. 
Remarkably, this dramatic slowdown is completely 
bypassed by swap MC sampling, which thermalizes the system 
up to $Z \approx 38 > Z_\mathrm{g}$. Even most conservative 
extrapolation indicates that we access a dynamical range that is 
broader than in experiments.
Meanwhile, the two-point structure barely 
budges (see Figs.~\ref{figure:tau}B-C), which is a telltale sign of 
glassiness~\cite{BB11} and a confirmation that both 
crystallization and more subtle fractionation effects are absent.
Visual inspection of particle configurations further confirm these conclusions (see, e.g., Figs.~\ref{figure:tau}D-E).
We are therefore in the unique position of studying 
at equilibrium a homogeneous supercooled liquid 
beyond the experimental glass ceiling.

\begin{figure*}[!t]
\centering
\includegraphics[width=0.9\linewidth]{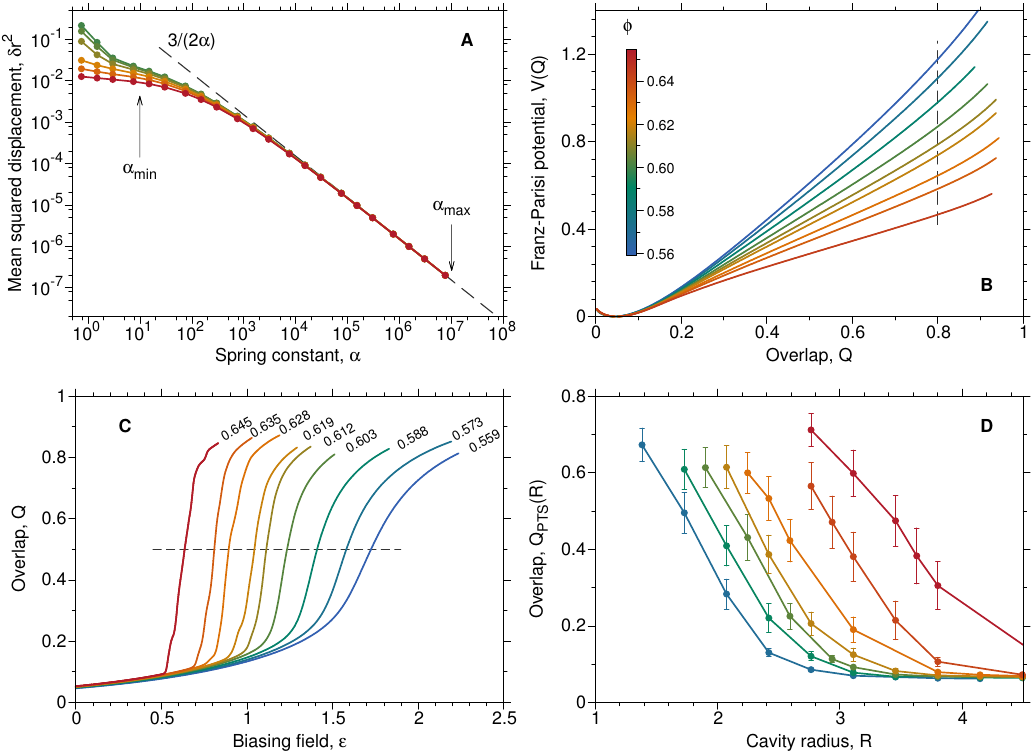}
\caption{
Numerical procedures leading to the four estimates of the 
configurational entropy.
(A) Method 1: 
The Frenkel-Ladd method to obtain the vibrational entropy 
$s_\mathrm{vib}$ performs a thermodynamic integration of the mean-squared 
distance $\delta r^2$ between a reference equilibrium configuration 
and a copy of the system constrained by a harmonic potential
of strength $\alpha$. The integration is carried out from
$\alpha_{\rm max}$, for which the system behaves as an Einstein solid
(indicated by the dashed line $\delta r^2 = 3 / (2 \alpha)$)
to $\alpha_{\rm min}$, for which particles are trapped by their own cages 
on the vibrational time scale.
(B) Method 2: The numerically-determined Franz-Parisi potential
$V(Q)$ is used to 
measure the configurational entropy as 
$s_{\rm conf}= V(Q_{\rm high} = 0.8)-V(Q_{\rm low} \approx 0.05)$. 
(C) Method 3: The evolution of the overlap $Q$ with the biasing
field $\varepsilon$ reveals a first-order jump at a value $\varepsilon^\star$ for which $Q = 1/2$ 
(dashed line). Then, $s_{\rm conf} = \varepsilon^\star (Q_{\rm high} - Q_{\rm low})$.
(D) Method 4: The decay of the cavity overlap correlation function
$Q_{\rm PTS}(R)$
with cavity radius, $R$, defines the point-to-set correlation length 
$\xi_{\mathrm{PTS}}\propto s_{\mathrm{conf}}^{-1/(d-\theta)}$.}
\label{figure:method}
\end{figure*}

We then turn to measuring the configurational 
entropy, $s_{\rm conf}$, in these extremely supercooled configurations.
The numerical procedures leading to the four estimates
of $s_{\rm conf}$ are shown in Fig.~\ref{figure:method}. 
Further details are provided in Appendices~\ref{Ap_method1},~\ref{Ap_method23}, and~\ref{Ap_method4}. In Method 1, we determine the
 configurational entropy from its most conventional definition, 
$s_\mathrm{conf} = s_\mathrm{tot} -s_\mathrm{vib}$, as 
used in many experimental and simulation 
studies~\cite{Kauzmann48,martinez2001thermodynamic,sciortino1999inherent,angelani2007configurational}. 
The total entropy of the equilibrium fluid, 
$s_\mathrm{tot}$, is measured by thermodynamic integration from 
the dilute ideal gas limit to the target volume fraction, while 
the vibrational contribution, $s_\mathrm{vib}$, is measured by Frenkel-Ladd 
thermodynamic integration~\cite{angelani2007configurational,frenkel1984new}. 
The latter 
integration is over the amplitude of 
the Hookean constant, $\alpha$, of a spring that constrains each particle 
to reside close to the position of a quenched reference 
equilibrium configuration.
This requires estimating the mean-squared distance $\delta r^2$ 
between the reference and constrained systems over 
a broad range of $\alpha$ values, 
as illustrated in Fig.~\ref{figure:method}(A). 
In continuously polydisperse systems, special care is 
also needed to account for 
the mixing contribution to the total entropy, 
because this contribution formally diverges~\cite{frenkel2014colloidal,ozawa2017does}.
The mixing entropy is thus determined from 
an independent, additional set of simulations~\cite{ozawa2017does} (see Appendix~\ref{Ap_method1}).
Method 1 is equivalent to partitioning configuration space into
basins of attraction of inherent structures~\cite{stillinger_supercooled_1988}. 
The resulting estimate of the configurational entropy thus counts the number of energy 
minima~\cite{stillinger1985inherent,sciortino1999inherent}, 
which presumably overestimates the number of relevant 
basins in the free energy landscape~\cite{monasson-biroli}. 

Methods 2 and 3 are both based on the Franz-Parisi 
theoretical construction~\cite{FP97},
which expresses the equilibrium free energy of the liquid, $V(Q)$, 
in terms of a global order parameter, the overlap $Q$.
The overlap between two configurations is defined  
as $Q = N^{-1} \sum_{i,j} \theta( a- | {\bf r}_{1,i} - {\bf r}_{2,j} | )$, 
where $\theta(x)$ is the Heaviside function, ${\bf r}_{1,i}$ 
and ${\bf r}_{2,j}$ are the positions of particle $i$ and $j$ 
within configuration 1 and 2, and $a$ is a fraction of the average
particle diameter. By definition,
$Q$ quantifies the similarity between the coarse-grained density profiles
of two configurations.
To compute $V(Q)$, we introduce a coupling between a quenched reference 
equilibrium configuration and a copy of the system through a field
$\varepsilon$ conjugate to $Q$~\cite{FP97,BC14}; 
$\varepsilon$ constrains the 
collective density profile, whereas 
$\alpha$ in Method 1 constrains single-particle displacements.
We define $V(Q) = - \lim_{\varepsilon \to 0} 
\left[ \frac{T}{N} \ln P(Q) \right]$, 
where $P(Q)$ is the equilibrium probability distribution of the overlap
for a given reference configuration, and
brackets denote averaging over these configurations.
In Method 2, we follow \cite{BC14} and use 
the free-energy difference $s_{\rm conf} = V(Q_{\rm high}) - 
V(Q_{\rm low})$ between the global minimum
at $Q_{\rm low} \approx 0.05$ and its value at $Q_{\rm high} = 0.8$
to obtain an estimate of $s_{\rm conf}$ 
that is closest to its theoretical definition, see Fig.~\ref{figure:method}(B). 
Importantly, this estimate only exists for sufficiently supercooled
states, for which $Q_{\rm high}$ can be defined~\cite{BC14}. 
For the present system, this happens
close to the mode-coupling crossover, $Z_c$. 
In Method 3, we determine the value of the biasing $\varepsilon$ 
needed to `tilt' the potential $V(Q)$, so that a first-order phase 
transition, at which $Q$ jumps from $Q_{\rm low}$ 
to $Q_{\rm high}$, takes place as illustrated in 
Fig.~\ref{figure:method}(C). 
\res{We use the maximum variance of the overlap fluctuations to measure
$\varepsilon^\star$ for each volume fraction studied. In practice, this is
equivalent to determining the biasing field at which the
overlap reaches $Q=1/2$, see Fig.~\ref{figure:method}(C).}

Method 4 builds on the physical idea that the 
decrease of the configurational entropy is directly responsible for the 
growth of spatial correlations quantified by the 
point-to-set correlation length, 
$\xi_{\rm PTS}$~\cite{adam1965temperature,BB04,BBCGV08}. 
Following what is becoming common practice~\cite{BBCGV08,BCY16}, 
we measure $\xi_{\rm PTS}$ by pinning the 
position of particles outside a spherical cavity of radius
$R$, equilibrating the liquid 
within it, and measuring the evolution of the 
overlap between interior configurations, $Q_{\rm PTS}(R)$, with 
the cavity radius $R$, as shown in Fig.~\ref{figure:method}(D).
The decay of $Q_{\rm PTS}(R)$ is controlled by
$\xi_{\rm PTS}$, and the variance of the overlap fluctuations 
also presents a maximum~\cite{BCY16} very close to $\xi_{\rm PTS}$. 
Physically, $\xi_{\rm PTS}$ thus represents the 
cavity size above which the system starts to explore a significant
number of distinct states. With minimal hypothesis~\cite{BB04}, 
it can be connected to the configurational entropy through 
$s_{\rm conf}\propto\xi_{\rm PTS}^{-(d-\theta)}$, with
an unknown exponent $\theta \leq (d-1)$.
Various values of $\theta$ have been proposed, including $\theta=2$ from saturating the inequality and $\theta=3/2$ from a wetting argument~\cite{KTW89,reviewlub}.
However, our measurements are consistent with both of these values and thus cannot unambiguously distinguish one proposal from the other. 

\begin{figure}
\begin{center}
\includegraphics[width=\linewidth]{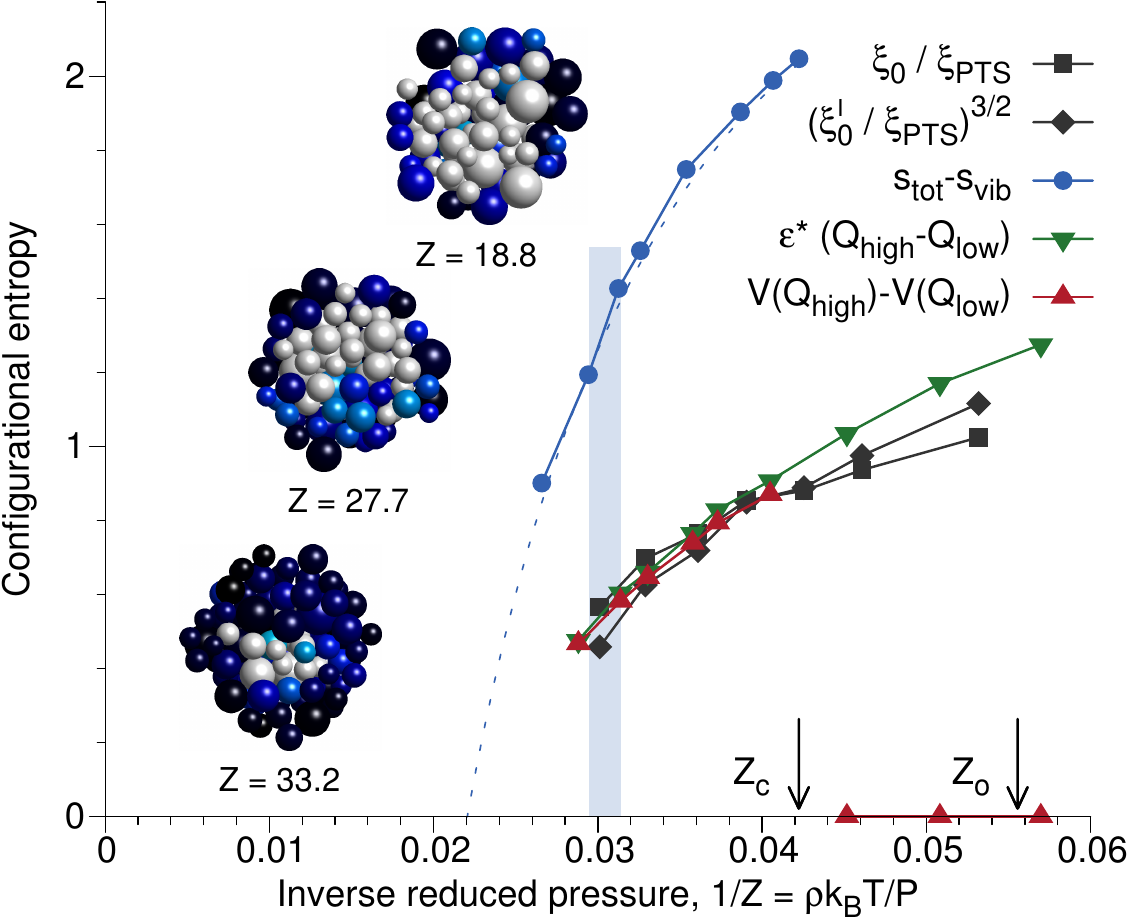}
\end{center}
\caption{
{Convergent measurements of the four estimates of the 
configurational entropy beyond the glass ceiling.}
$s_\mathrm{conf}$ is plotted as a function of 
$1/Z \propto T/P$, which is equivalent to the classic Kauzmann plot.
All measurements indicate a steep decrease
of $s_{\rm conf}$ that continues as the 
experimental glass ceiling is crossed. 
The point-to-set estimates are normalized with $\xi_0=2.0$ for $\theta=2$ and $\xi_0^{\prime}=2.1$ for $\theta=3/2$, respectively, to match the Franz-Parisi estimates at the start of the low-temperature regime, $1/Z=0.04\approx 1/Z_{\rm c}$.
The dashed line is an extrapolation based on 
$s_\mathrm{tot}-s_\mathrm{vib}$ (see Appendix~\ref{Ap_method1}).
Inset: typical overlap profiles measured in a finite cavity of 
radius $R=3.46$, with colors coding for the overlap value 
from low (white) to large (black). Overlap fluctuations 
are uncorrelated around the onset but become strongly correlated 
over the entire cavity at the largest pressure shown.}
\label{figure:entropy}
\end{figure}

We gather the four estimates of the configurational
entropy in Fig.~\ref{figure:entropy} to produce a plot 
akin to the original 1948 Kauzmann representation 
of $s_{\rm conf}(T)$~\cite{Kauzmann48}.
Although in the high-temperature liquid the configurational entropy is 
not sensibly defined~\cite{BC14}, three of the four measures 
can still be estimated.
Note that only this regime was accessible in earlier 
simulations~\cite{sciortino1999inherent,angelani2007configurational,BC14}.
In the more relevant
low-temperature regime, our main finding is that the important
conceptual and technical differences between the four methods
nevertheless result in qualitatively consistent results. 
\res{In particular, the three estimates (Methods 2-4) 
that closely follow the theoretical
definition of the configurational entropy provide numerically indistinguishable results 
at low temperatures. The conventional estimate of the entropy (Method 1) is larger, as expected~\cite{monasson-biroli}, but its
temperature evolution remains qualitatively consistent with 
the other methods.
All our estimates of $s_{\rm conf}$ thus exhibit a 
steep decrease as $Z$ increases towards the glass phase, which 
is consistent with the seemingly fragile behavior of the model
in Fig.~\ref{figure:tau}. 
Although a quantitative extrapolation is hard to control, 
our measurements robustly suggest that $s_{\rm conf}$ may
vanish near $Z \approx 1/0.022 \approx 45$. }
We thus conclude that even for a simple glass-forming system
equilibrated deeper in the landscape than any previously 
studied material, the trend discovered 70 years ago by Kauzmann is confirmed when more precise estimates of $s_{\rm conf}$ are
adopted, and persists even below the experimental glass temperature.

We further show in Appendix~\ref{sec:soft} that similar observations can be performed for a model with a continuous pair 
potential, suggesting our methodological progress and physical 
conclusions are not restricted to hard spheres, and likely apply more generally.
Note that while continuous polydisperse distributions are commonplace in colloidal suspensions, a molecular liquid with a sufficiently large number of components to approximate a continuous size distribution has yet to be considered.

 \emph{Discussion--} 
Our point-to-set measurements go beyond Kauzmann's observation 
by establishing that the decrease in 
$s_{\rm conf}$ is accompanied by an increase of static spatial correlations 
as the glass ceiling is crossed. This result 
reinforces a recent experimental report based on non-linear 
dielectric measurements~\cite{ABMBBLLTWL16}. In absolute value, the 
measured static length scale at the experimental 
glass transition appears somewhat smaller than previous estimates
based on dynamical correlations~\cite{rabo2013,science05}, 
but remains compatible
with the modest growth expected 
from general arguments based on thermally activated 
scaling~\cite{KTW89,BB04,reviewlub} and decorrelation 
between static and dynamical length scales~\cite{sho}.
Our particle-based resolution of such correlations 
further provides a direct visualization of the spatial profile of 
the overlap within a spherical cavity (see insets 
in Fig.~\ref{figure:entropy}). 
In particular, within a cavity comprising about 200 particles, the positions 
of particles freely fluctuate near the onset pressure, but 
become strongly correlated over the entire cavity for the 
largest pressure shown.
The spatial extent of static correlations is thus directly revealed.

The important methodological advances achieved here regarding 
the thermalization of supercooled liquids and the measurement
of configurational entropy therefore support a
thermodynamic view of the glass formation
based on the rarefaction of metastable state accompanied
by growing static correlations that is devoid of the 
experimental ambiguities and that extends to a 
temperature regime that has never been explored before.

\begin{acknowledgments}
The research in Montpellier was supported by funding
from the European Research Council under the European
Union’s Seventh Framework Programme (FP7/2007-
2013) / ERC Grant agreement No 306845 and 
by a grant from the Simons Foundation (\#454933, Ludovic Berthier).
The research at Duke was supported by a grant from the Simons 
Foundation (\#454937, Patrick Charbonneau) and associated computations were carried out through the Duke Compute Cluster.
Data relevant to this work have been archived and can be accessed at http://dx.doi.org/10.7924/G8ZG6Q9T.

\end{acknowledgments}

\clearpage

\appendix
\renewcommand{\thefigure}{S\arabic{figure}}
\setcounter{figure}{0}
\begin{widetext}

\section{Model}\label{hard_model}
We study a three-dimensional hard-sphere model, for which the 
pair interaction is zero for non-overlapping particles 
and infinite otherwise. 
Systems have a continuous size polydispersity, 
with particle diameters $\sigma$ randomly drawn from 
the distribution 
$f(\sigma) = A\sigma^{-3}$, with $\sigma \in [ \sigma_{\rm min}, 
\sigma_{\rm max} ]$ 
with normalization constant $A$. 
Our model is the same as that studied in Ref.~\cite{BCNO16}, with a measure of size polydispersity
$\Delta=\sqrt{\overline{\sigma^2} - \overline{\sigma}^2}/\overline{\sigma}$, 
where $\overline{\cdots}=\int d \sigma f(\sigma)
(\cdots)$,
of $\Delta = 23 \%$, 
and $\sigma_{\rm min} / \sigma_{\rm max} = 0.4492$. The average diameter,
$\overline{\sigma}$, defines the unit of length. 
We simulate systems composed of $N$ particles in a cubic cell 
of volume $V$ under
periodic boundary conditions~\cite{allen1989computer}. 
Depending on the chosen method 
to estimate the configurational entropy (see Appendix~\ref{Ap_method1}),
we simulate systems with either $N=1000$, $8000$ (Method 1) or $N=300$ (Method 2 and 3).
Cavities for Method 4 are carved from 
bulk configurations with $N=8000$.  
The relaxation times shown in Figure~\ref{figure:tau}A 
are obtained from samples with $N=1000$.
Given these parameters, the system is then uniquely characterized by its volume fraction 
$\phi = \pi N \overline{\sigma^3} /(6V)$, and we 
frequently report the data using the
reduced pressure $Z=P/(\rho k_{\rm B}T)$, where $\rho$, 
$k_{\rm B}$, and $T$ are 
the number density, Boltzmann constant and temperature, respectively. Without loss of generality, we set $k_B$ and  $T = 1/\beta$ to unity. 
The pressure $P$ is calculated from the contact value of the pair correlation function properly scaled for a polydisperse system~\cite{Santos05}.

\section{Methods}\label{hard_methods}
To obtain equilibrium fluid configurations deep in the glassy regime, 
we perform Monte-Carlo (MC) simulations with both translational displacements 
and non-local particle 
swaps~\cite{BCNO16,gazzillo1989equation,sindzingre1989calculation,santen2001liquid,pronk2004melting,GP01,FS01,brumer2004numerical,gutierrez2015static}.
The two types of moves are selected randomly: with probability $0.8$ we attempt a translational displacement, and with probability $0.2$ we attempt a swap.
Translational displacements are uniformly drawn over a cube of side 
$0.115$.
For swaps, two randomly selected particles exchange diameter. 
In both cases, proposed moves are accepted if no overlap is created.
Following Ref.~\cite{PRX17}, we also immediately reject swaps between 
particles whose diameters differ by more than $0.2$.

We measure the equilibrium relaxation time $\tau$ 
both with and without the swap moves from the time-decay of the 
self-intermediate scattering function, $F_s(k,\tau)=1/e$, where 
$k=5.25$ is the wavenumber chosen slightly below 
the first maximum of the static structure factor.
Note that the particles' diameters can change during a swap MC simulation, but their trajectories are continuous.
Relaxation times are measured in units of MC sweeps, comprising $N$ MC moves, 
irrespective of their type.

Thermalized systems at each state point are obtained in the same way
for both standard and swap MC dynamics. We measure the
relaxation time $\tau$ and ensure that for each state point simulations 
of a total duration of at least $100\tau$ can be performed. 
We also check for the presence
of aging effects in time correlation functions, and we measure
the static structure factor, the pair correlation function and the equation 
of state over long simulations, paying attention to any temporal drift
that could signal either improper thermalization,
incipient crystallization or demixing of particles with 
distinct sizes. Selected results for the evolution 
of the structure factor with volume fraction are presented in 
Figs.~\ref{figure:tau}B-C. Over the extreme range of densities 
shown here, the static structure evolves very little. Similarly,
a very modest evolution is seen when the partial structure factor 
of the smallest particles is measured. A large increase of the 
low-$k$ value of these quantities, or the emergence of discrete 
peaks would signal that demixing or crystallization is taking place.
In fact, we have found that measuring the relaxation time 
for the swap simulation is the most sensitive test of thermalization, 
because purely static observables may appear thermalized over 
long simulation times, whereas the system is in fact nearly arrested
in a glass state within which sampling is inefficient.

We introduce two dynamical reference states: (i) the onset of slow dynamics at $\phi_0\approx 0.56$ ($Z_0\approx 18$), above which the time decay of correlation functions is non-exponential~\cite{sastry_signatures_1998}, and (ii) the mode-coupling crossover $\phi_{\rm c}\approx 0.598$ ($Z_{\rm c}\approx 23.5$), at which a power-law fit extrapolates a divergence of the relaxation times~\cite{BCNO16}. Note that these particular definitions are not unique~\cite{charbonneau2014hopping}, but are sufficiently accurate for our purposes, where
these values are simply used for qualitative reference.

\section{Equation of state}\label{Ap_eos}

We report in Fig.~\ref{fig:eos} the equilibrium equation of state, $Z=Z(\phi)$ (from Ref.~\cite{BCNO16}), of the system under study, thus enabling a translation of the reported results from $Z$ to $\phi$. 
The specific data points used in the four measurements of the configurational entropy are also presented in Tab.~\ref{tab:eos_N300},~\ref{tab:eos_N1000}, and ~\ref{tab:eos_N8000}. 

\begin{figure}[!htbp]
\begin{center}
\includegraphics[width=\linewidth]{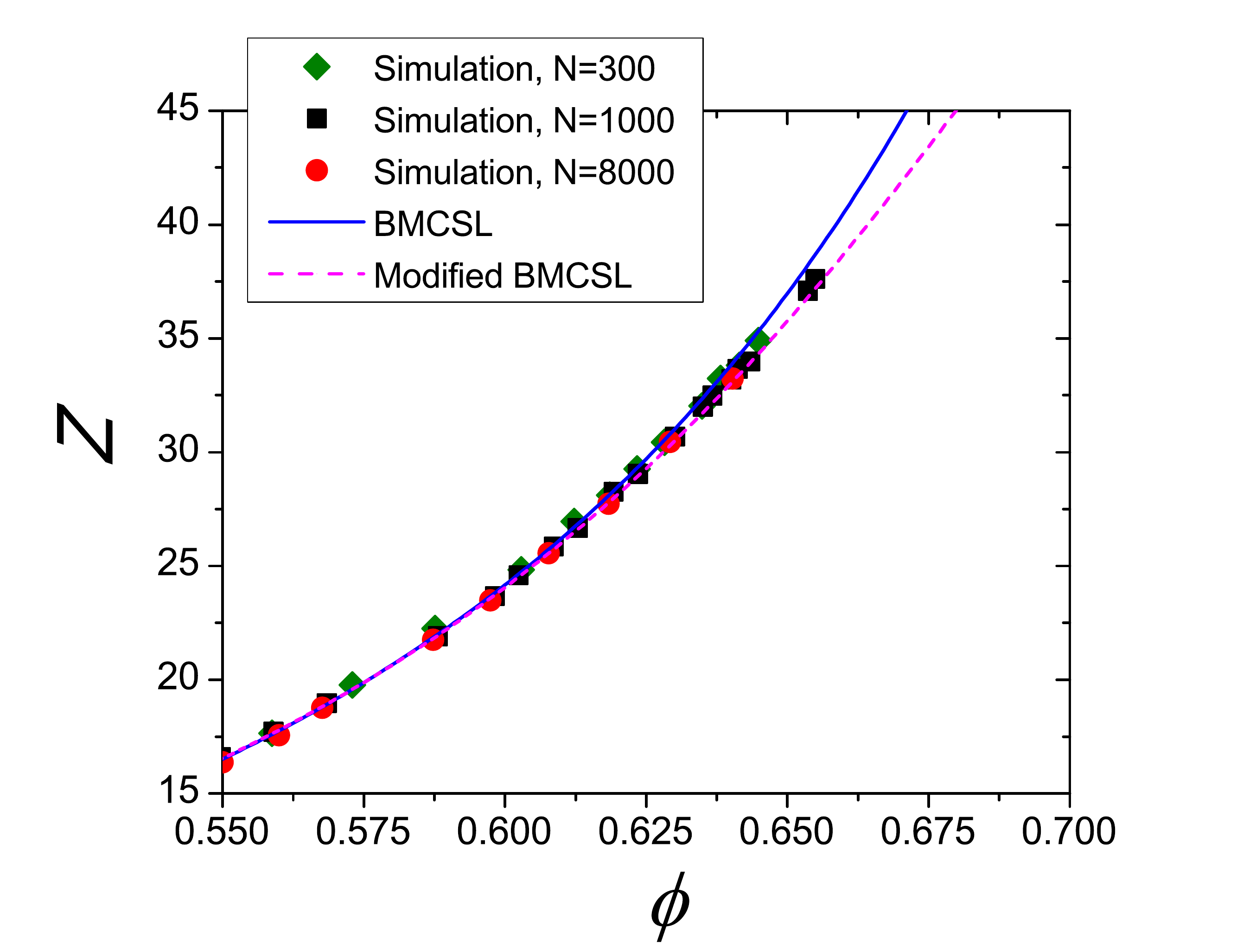}
\caption{Equation of state $Z=Z(\phi)$ for the
hard-sphere system studied in this work. 
The solid line is the empirical expression 
for the equation of state from Refs.~\cite{boublik1970hard,mansoori1971equilibrium}, BMCSL, which describes our data well, except at very large
volume fractions, where it slightly overestimates $Z$.
A modified version of BMCSL~\cite{berthier2016growing} is given by the dashed line.}
\label{fig:eos}
\end{center}
\end{figure}

\begin{table}[!htbp] 
\begin{center} 
\begin{tabular}{c|ccccccccc}
\hline
\hline
\ $Z$ \ & \ 17.6 \ & \ 19.8 \ & \ 22.3 \ & \ 24.8 \ & \ 27.0 \ & \ 28.1 \ & \ 30.4 \ & \ 32.0 \ & \ 34.9 \\
\hline
\ $\phi$ \ & \ 0.559 \ & \ 0.573 \ & \ 0.588 \ & \ 0.603 \ & \ 0.612 \ & \ 0.619 \ & \ 0.628 \ & \ 0.635 \ & \ 0.645 \\
\hline 
\hline
\end{tabular}
\caption{
Data points for $N=300$ used in the Franz-Parisi construction (Methods 2 and 3).
} 
\label{tab:eos_N300}
\end{center}
\end{table}

\begin{table}[!htbp] 
\begin{center} 
\begin{tabular}{c|ccccccccc}
\hline
\hline
\ $Z$ \ & \ 23.7 \ & \ 24.6 \ & \ 25.8 \ & \ 28.3 \ & \ 30.7 \ & \ 32.0 \ & \ 34.0 \ & \ 37.6 \\
\hline
\ $\phi$ \ & \ 0.598 \ & \ 0.602 \ & \ 0.609 \ & \ 0.619 \ & \ 0.630 \ & \ 0.635 \ & \ 0.643 \ & \ 0.655 \\
\hline 
\hline
\end{tabular}
\caption{
Data points for $N=1000$ used in the thermodynamic integrations (Methods 1).
} 
\label{tab:eos_N1000}
\end{center}
\end{table}

\begin{table}[!htbp] 
\begin{center} 
\begin{tabular}{c|ccccccc}
\hline
\hline
\ $Z$ \ & \ 18.8 \ & \ 21.7 \ & \ 23.5 \ & \ 25.6 \ & \ 27.7 \ & \ 30.4 \ & \ 33.2 \\
\hline
\ $\phi$ \ & \ 0.568 \ & \ 0.587 \ & \ 0.597 \ & \ 0.608 \ & \ 0.618 \ & \ 0.629 \ & \ 0.640 \\
\hline 
\hline
\end{tabular}
\caption{
Data points for $N=8000$ used in the thermodynamic integration computations (Methods 1) and the point-to-set correlation (Method 4).
} 
\label{tab:eos_N8000}
\end{center}
\end{table}

\section{Equilibration}\label{equi_make_sure}

\begin{figure}[!htbp]
\begin{center}
\includegraphics[width=0.49\linewidth]{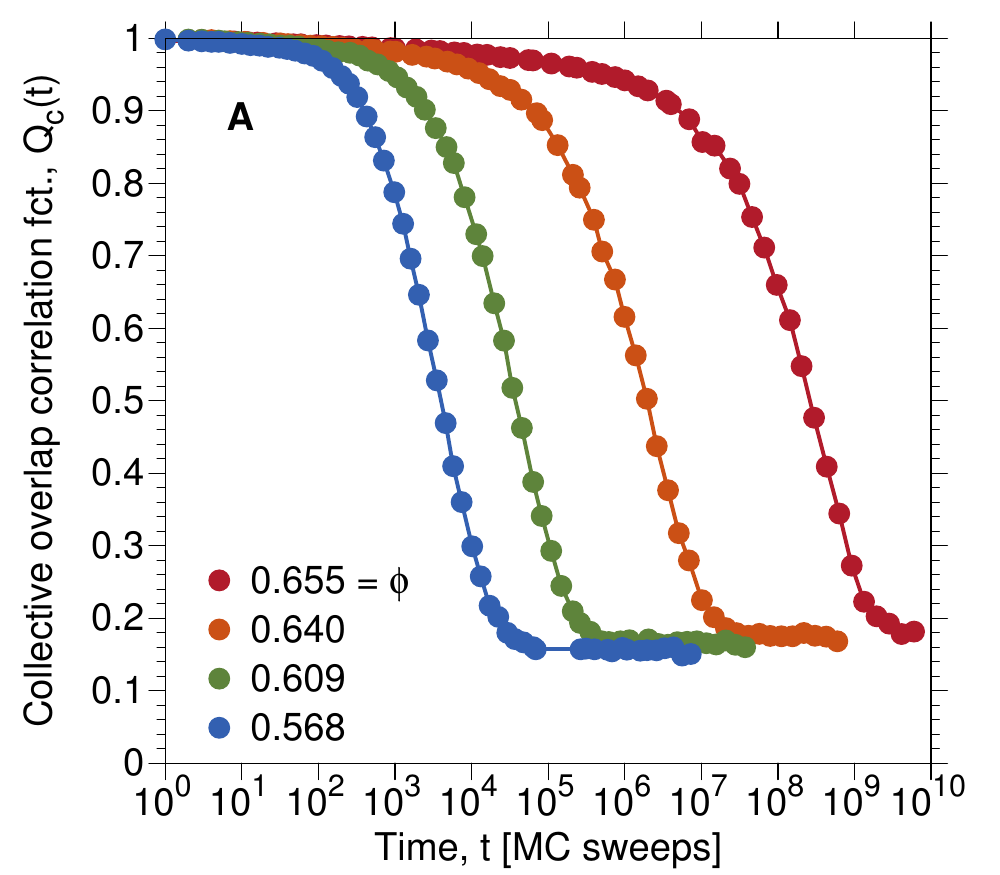}
\includegraphics[width=0.49\linewidth]{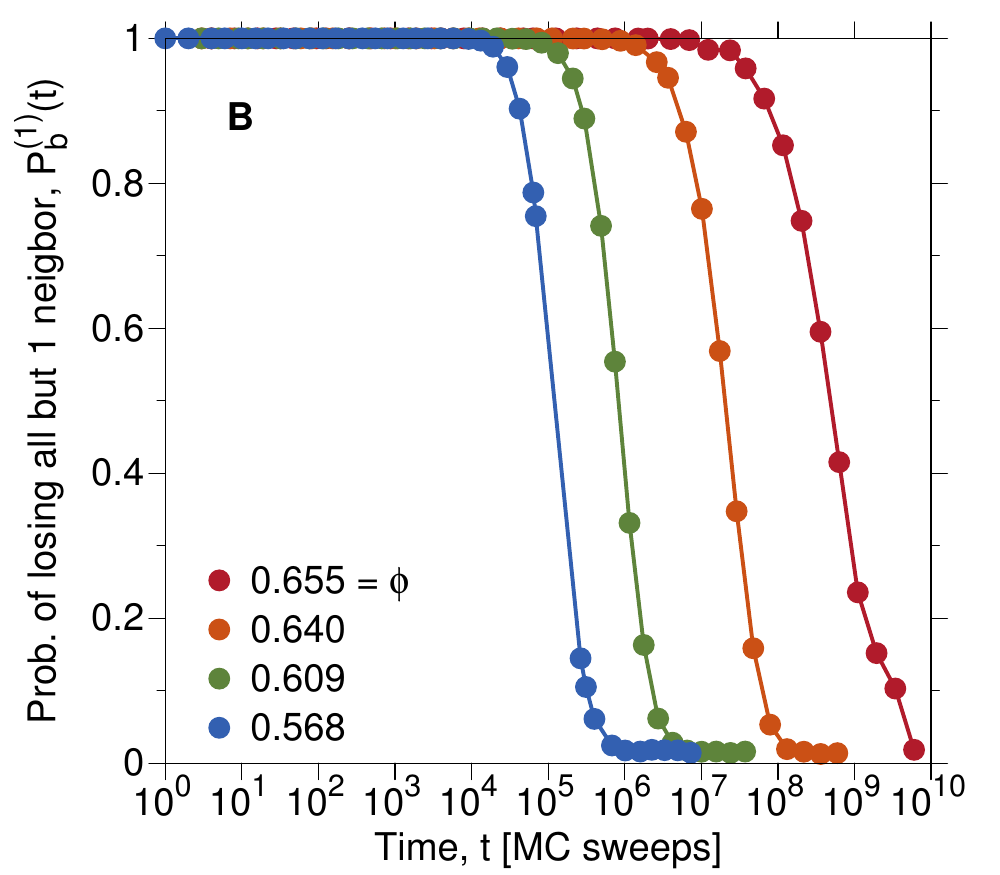}
\caption{Panel A: Collective overlap time correlation function $Q_c(t)$ for several packing fractions (from left to right, $\phi=0.568$, 0.609, 0.640, 0.655). Panel B: Probability of losing all but one neighbor $P_b^{(1)}(t)$ as a function of time for the same packing fractions as in panel A.}
\label{fig:qcoll}
\end{center}
\end{figure}

In this section, we present evidence that strongly supports the equilibration of our samples.
In addition to the tests mentioned in the Methods section of the main text, we also verified the equivalence of derivative and fluctuation expressions for the compressibility, which is a well-known equilibration check~\cite{vashist}.
The validity of this sum rule is, however, not necessarily a strong test of equilibration, because it may appear to hold \textit{within the noise of the data}, even in systems that are not fully equilibrated.
A more stringent test of equilibration is to show that 
all the slow degrees of freedom relax within the observation time.
In a structural glass former, the relevant slow degrees of freedom are those associated with collective density fluctuations and  with cage breaking.
To measure the decorrelation of the former, we introduce the time-dependent collective overlap function
\begin{equation}
Q_c(t) = N^{-1} \sum_i \sum_j \Theta(a - |\vec{r}_i(t)-\vec{r}_j(0)|),
\end{equation}
where $\Theta(x)$ is the Heaviside function and $a=0.346$. The decay of $Q_c(t)$ quantifies the relaxation of density fluctuations up to a coarse-graining length $a$.
In order to analyze cage breaking, here defined as the process 
of a particle losing some of its neighbors within a given 
time span~\cite{kawasaki}, the following procedure is employed.
First, we identify the neighbors of a particle by a radical Voronoi tessellation obtained using the Voro++ package~\cite{voro++}. Note that the average number of neighbors measured for our densest states is about 14.
Second, we measure the probability, $P_b^{(k)}(t)$, that a particle has changed all but $k$ of its initial neighbors after a time $t$.
For small $k$, the decay of $P_b^{(k)}(t)$ provides evidence for the complete restructuring of the cages and ensures that a pseudo-molecular structure is not found in the fluid.

Representative results of $Q_c(t)$ and $P_b^{(1)}(t)$, measured in swap MC simulations of the $N=1000$ system, are shown in Fig.~\ref{fig:qcoll}.
All these functions decay to trivial plateaus within our simulations, with the possible, albeit mild, exception of the data point at the highest packing fraction.
We also note that although the cage breaking probability decays more slowly than $Q_c(t)$, the associated relaxation times grow in a similar way.
Our simulations thus pass this very stringent equilibration test, in addition to those already mentioned in the main text and in Ref.~\cite{BCNO16}.
For the soft polydisperse model studied in Sec.~\ref{sec:soft} of the SM, a similar analysis was performed in Ref.~\cite{ninarello}.

\section{Locating the glass ceiling}\label{Ap_localization}
In the context of this work, we refer to the typical relaxation time 
measured at the laboratory glass transition (at which, conventionally, $\tau/\tau_0 = 10^{13}$~\cite{rossler}) as the glass ceiling, $Z_{\rm g}$.
Because standard MC dynamics can only access relaxation times 
at most of order $\tau/\tau_0 = 10^5$, where $\tau_0=10^4$ MC sweeps is the 
value of $\tau$ at the onset of slow dynamics,
our dynamical data have to be extrapolated to locate this ceiling.
We fit measured relaxation times to various functional forms
up to $\tau/\tau_0 \leq 10^5$, and then extrapolate 
up to the vicinity of the glass ceiling. In an effort to obtain an estimate
as unbiased as possible, we consider a range of possible functional
forms~\cite{berthier_glass_2009,BB11}, as detailed below.  We present in Fig.~\ref{fig:tau_fit} the result of 
this exercise. In particular, note that 
whereas extrapolating the location of a putative divergence of $\tau$ 
is extremely delicate, extrapolating the location of the glass ceiling 
is in fact rather well-controlled.

\begin{figure}[!tbp]
\begin{center}
\includegraphics[width=0.8\linewidth]{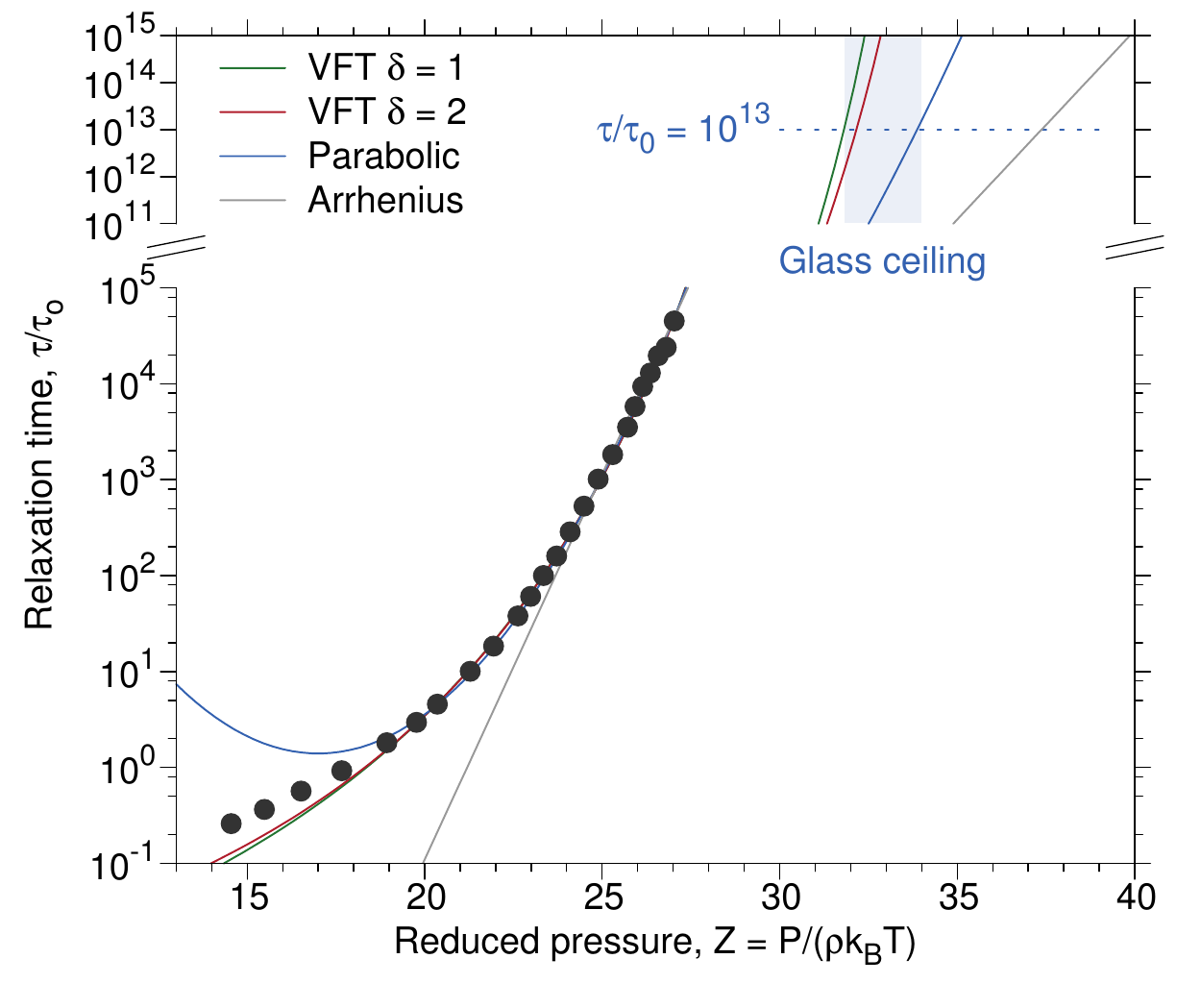}
\caption{The measured relaxation time, $\tau$, (symbols)
is fitted to the VFT form, Eq.~(\ref{fit_vft}), with exponents $\delta=1$ and $\delta=2$, the parabolic law, Eq.~(\ref{fit_parabolic}), and the Arrhenius law, Eq.~(\ref{fit_arrhenius}). The first three fits provide  a good description of the data, and are used to locate the glass ceiling at $\tau/\tau_0 = 10^{13}$ that is used in the main text.} 
\label{fig:tau_fit}
\end{center}
\end{figure}  

The first functional form is the Vogel-Fulcher-Tammann (VFT) expression
\begin{equation}
  \tau  = \tau_\infty \exp \left[ \frac{A}{(Z_{\rm vft}-Z)^\delta} \right],
\label{fit_vft}
\end{equation}
where $\tau_\infty$, $A$, the exponent $\delta$ 
and the critical pressure $Z_{\rm vft}$ are free parameters.
Whereas $\delta=1$ is traditionally used, more recent experimental and 
numerical 
studies~\cite{BEPPSBC08,berthier_glass_2009}
favor $\delta=2$. In Fig.~\ref{fig:tau_fit}
we consider both. As noted 
before~\cite{BCNO16}, 
the fit with $\delta=1$ yields a critical pressure $Z_{\rm vft} \approx 38$ 
that falls within the range in which swap MC sampling 
equilibrates. Although the resulting fit 
is good over the range covered by standard MC dynamics, it 
overestimates the growth of the relaxation time beyond that range, 
which undermines the very logic behind the proposed scaling form. 
When $\delta=2$ is imposed
the fit is still very good, but we now obtain $Z_{\rm vft} \approx 45$,
which is at least beyond the equilibrium range accessible with the swap 
method. Although this fit provides a more consistent description of the 
dynamical data, resolving one form from the other is beyond the scope of the current study.

The second functional form is the parabolic law proposed by 
Elmatad {\it et al.}~\cite{elmatad2009corresponding,isobe2016applicability} 
in the context of facilitated models,
\begin{equation}
  \tau = \tau_\infty \exp\left[ A(Z-Z_0)^2 \right],
\label{fit_parabolic}
\end{equation}
where $Z_0=17$ is around the onset
of slow dynamics and the fit is made over the range $Z>Z_0$.  
In contrast to the VFT law, this parabolic expression 
does not invoke a divergence of the relaxation time at any finite 
pressure; $\tau$ only diverges when $Z$ also diverges.
Note that because $Z \propto 1/T$, this form is equivalent to fitting
the relaxation time of a supercooled liquid without 
any finite-temperature divergence~\cite{berthier_glass_2009}.
As a result this expression necessarily 
provides a less divergent extrapolation 
of the relaxation time at large pressures, but still accounts well for 
the curvature, i.e., the fragility, of the measured dynamical data.
The leading order at large $Z$, 
$\tau \propto \exp [AZ^2]$, indeed grows faster than the Arrhenius law.
Like the VFT form, this fit is quite good over the measured 
range of relaxation times, see Fig.~\ref{fig:tau_fit}.

For completeness, we also include a simple Arrhenius fit to the high-$Z$ 
portion of the equilibrium data
\begin{equation}
  \tau = \tau_\infty \exp[{A}{Z}].
\label{fit_arrhenius}
\end{equation}
We find this fit not to be very good
as it does not capture the 
fragility of the system. 
It is also inconsistent with the thermodynamic 
behavior of the system reported in the main text, because the
configurational entropy typically does not vary much in 
glass-formers with an Arrhenius-like behavior. 
This fit is thus most probably incorrect in the sense that
it underestimates considerably the evolution of the
relaxation time of the system. 

We conclude that
the first two families of expressions account well for the non-Arrhenius 
dependence of the relaxation time data observed in 
Fig.~\ref{fig:tau_fit} and yield reasonable descriptions 
of the available equilibrium data over several orders of magnitude.
Whereas the VFT law with $\delta=1$ can be logically ruled out by the swap MC 
measurements, the other two fits cannot be excluded on the basis of any further
measurement we could perform. As is common in glass simulations,
it is thus difficult to discriminate between fitting forms \emph{with} or \emph{without} a 
finite temperature singularity~\cite{BB11}.
These forms nonetheless allow us to locate the glass ceiling $Z_{\rm g}$, for which the relaxation 
time is $10^{13}$ larger than its value at $\phi_0$.
Fits to the relevant expressions yield estimates ranging from $Z_{\rm g} \approx 32$ (VFT 
law with $\delta=1$) to $Z_{\rm g}=34$ (parabolic law).
These values are used as boundaries of the glass ceiling box in 
Fig.~1 of the main text. 

As discussed above, there are several reasons for which we do not expect 
the Arrhenius expression to describe accurately the relaxation data 
at high $Z$. If we nonetheless considered it, the 
equilibrium swap MC data would still be able to equilibrate the liquid
at pressures higher than the 
 glass transition it predicts. Hence, this very 
conservative extrapolation allows us to confidently state 
that we have successfully broken the glass ceiling.

\section{Method 1: Conventional definition}\label{Ap_method1}

In order to compute the configurational entropy, $s_{\rm conf}$, we define 
\begin{equation}
s_{\rm conf}=s_{\rm tot}-s_{\rm vib} \quad {\rm( Method \ 1 )},
\end{equation}
where $s_{\rm tot}$ and $s_{\rm vib}$ are the total and vibrational entropy, respectively.
Note that this definition is common in experimental and computational studies~\cite{Kauzmann48,richert_dynamics_1998,martinez2001thermodynamic,sciortino1999inherent,Sastry01,angelani2007configurational}.

\subsection{Total entropy}

The total entropy is obtained by thermodynamic integration from the ideal gas limit ($\phi \to 0$) up to the target volume fraction $\phi$,
\begin{equation}
s_{\rm tot}(\phi) = \frac{5}{2} -\ln \left( \frac{6 \phi}{\pi M_3} \right) - \ln \Lambda^3 - \int_0^{\phi} {\rm d} \phi' \frac{\left(Z(\phi') -1 \right)}{\phi'} + s_{\rm mix},
\end{equation}
where $M_k = \overline{\sigma^k}$ ($k=1,2,3, \cdots$) are the $k$-th moments and $\Lambda=\sqrt{2\pi \beta \hbar^2/m}$ is the thermal de Broglie wavelength. Without loss of generality, we here set $\Lambda=1$. 
For continuous polydisperse systems one needs to pay special attention to the mixing entropy, $s_{\rm mix}$~\cite{salacuse1982polydisperse,frenkel2014colloidal}. We get back to this point in Sec.~\ref{sec:mixing}. 
For now, we report $s_{\rm tot}-s_{\rm mix}$ for $N=1000$ and $8000$ in Fig.~\ref{fig:s_total_vib}(A).

In order to validate the numerical thermodynamic integration, we also consider an analytical approximation of the equation of state (EOS). The polydisperse version of the Carnahan-Starling EOS, i.e., the so-called BMCSL EOS~\cite{boublik1970hard,mansoori1971equilibrium}, 
\begin{equation}
Z_{\rm BMCSL}(\phi) = \frac{1}{1-\phi} + \frac{3 M_1 M_2}{M_3} \frac{\phi}{(1-\phi)^2} + \frac{M_2^3}{M_3^2} \frac{(3-\phi)\phi^2}{(1-\phi)^3},
\end{equation}
is known to describe experiments and simulations of polydisperse hard-sphere systems rather well in the liquid regime. We also consider a modified version of $Z_{\rm BMCSL}$, 
\begin{equation}
Z_{\rm modBMCSL} (\phi) = 1 + h(\phi) (Z_{\rm BMCSL}(\phi)-1),
\end{equation}
where $h(\phi)=0.005-\tanh(14(\phi-0.79))$~\cite{berthier2016growing}.
Both EOSs trace our simulation data very well, as can be seen in Figs.~\ref{fig:eos} and \ref{fig:s_total_vib}(A). 
These EOSs can thus also be used to extrapolate the configurational entropy toward very high volume fraction (Sec.~\ref{sec:sconf}).

\begin{figure}[htbp]
\includegraphics[width=0.48\columnwidth]{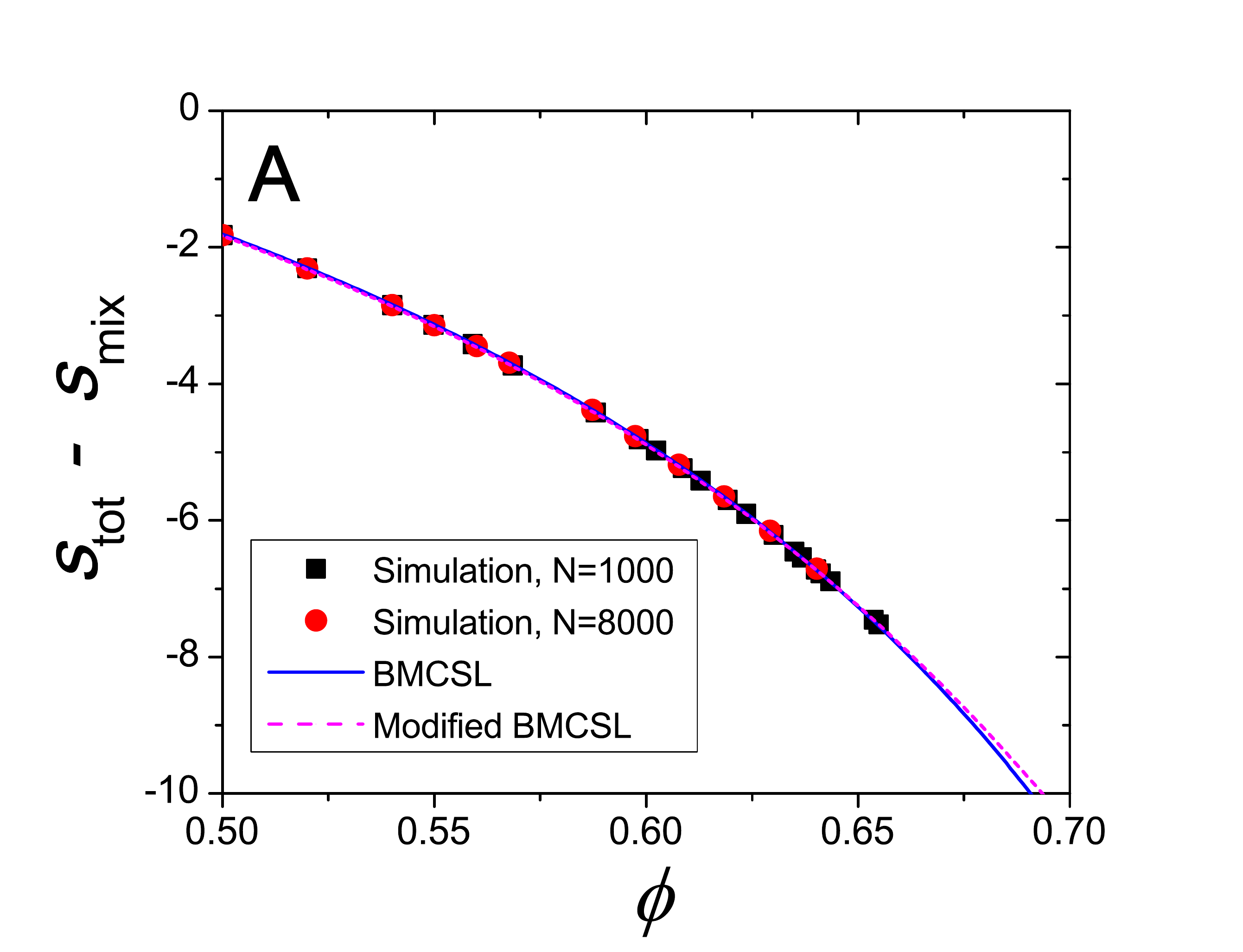}
\includegraphics[width=0.48\columnwidth]{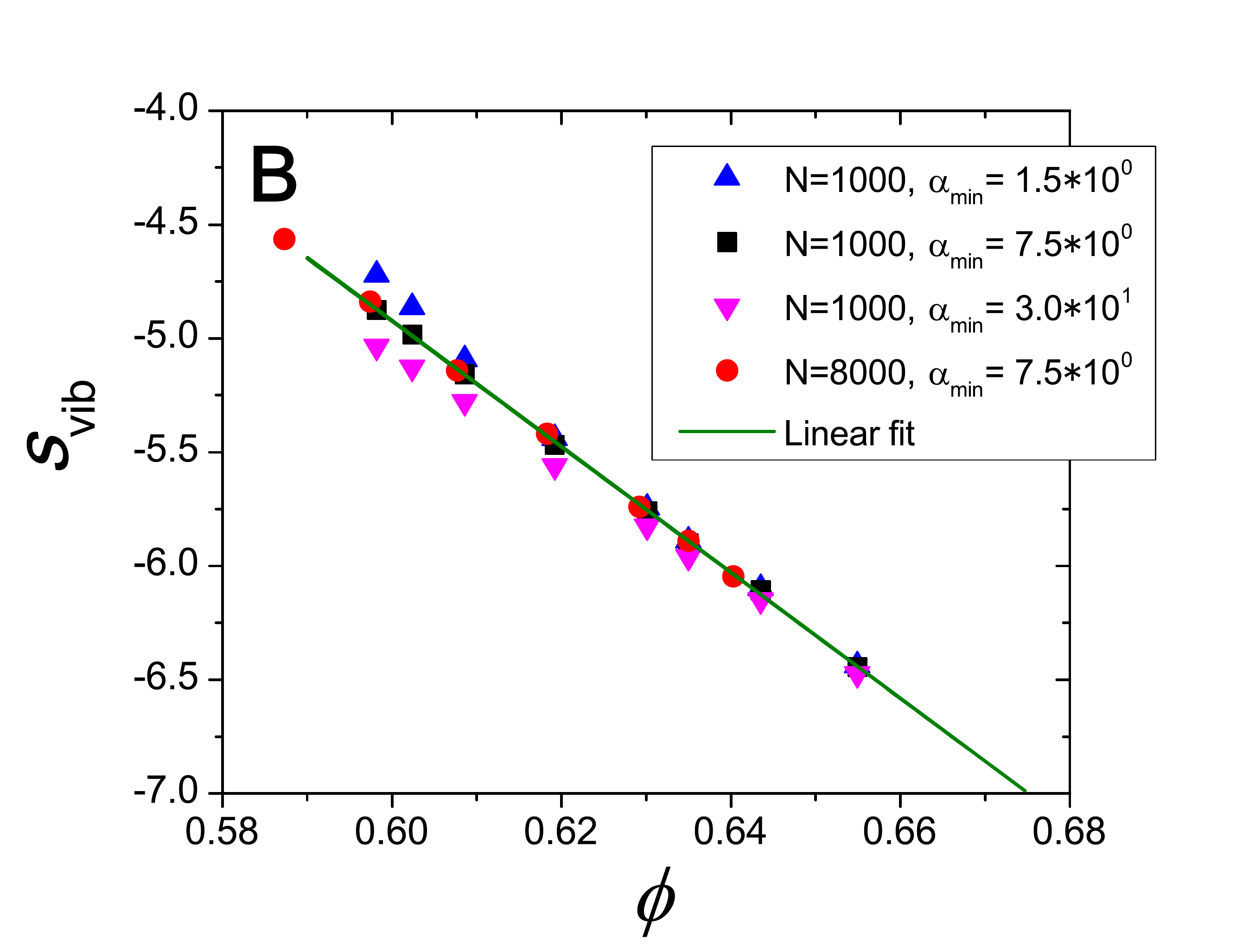}
\caption{
(A) The total entropy, $s_{\rm tot}$, minus the mixing contribution, $s_{\rm mix}$, from simulations and from the (modified) BMCSL EOS.
(B) The vibrational entropy, $s_{\rm vib}$, as a function of $\phi$ for several $\alpha_{\rm min}$.
The solid straight line is fitted to the $N=1000$ results with $\alpha_{\rm min}=7.5 \times 10^0$ in the high $\phi$ region.}
\label{fig:s_total_vib}
\end{figure}

\subsection{Vibrational entropy}

The vibrational entropy, $s_{\rm vib}$, is obtained by Frenkel-Ladd (FL) thermodynamic integration~\cite{FS01,angelani2007configurational} by performing MC simulations of a constrained system with Hamiltonian
\begin{equation}
\beta H(\alpha) = \beta H(0) + \alpha \sum_{i=1}^{N}({\bf r}_i - {\bf r}_{0 i})^2,
\label{eq:hamiltonian}
\end{equation}
for a template configuration, $\{ {\bf r}_{0 i} \}$, obtained from an equilibrium target system under $H(0)$.
In short, the FL method integrates from a large $\alpha_{\rm max} \gg 1$, at which particles experience a nearly pure harmonic oscillator, down to a very weak $\alpha_{\rm min} \ll 1$, at which particles vibrate within cages.
The vibrational entropy is then obtained by 
\begin{eqnarray}
s_{\rm vib} &=& \frac{3}{2} -  \ln \Lambda^3 - \frac{3}{2} \ln \left( \frac{\alpha_{\rm max}}{\pi} \right) + \int_{\alpha_{\rm min}}^{\alpha_{\rm max}} {\rm d} \alpha  \delta r^2(\alpha) + \alpha_{\rm min} \delta r^2(\alpha_{\rm min}), \label{eq:final_expression} \\
\delta r^2(\alpha) &=& \left[ \left\langle \frac{1}{N} \sum_{i=1}^{N}({\bf r}_i - {\bf r}_{0 i})^2 \right\rangle_{\alpha} \right], \label{eq:msd}
\end{eqnarray}
where $\langle (\cdots) \rangle_{\alpha}$ and $[(\cdots)]$ denote the thermal average with Hamiltonian $H(\alpha)$ and averaging over template configurations, $\{ {\bf r}_{0 i} \}$, respectively, and $\delta r^2$ in Eq.~(\ref{eq:msd}) is the mean-squared displacement shown in Fig.~2(A) of the main text.

Numerical integration of $\delta r^2$ is performed from $\alpha_{\rm max}$ to $\alpha_{\rm min}$.
We set $\alpha_{\rm max}=7.5 \times 10^6$, which is well into the harmonic-oscillator scaling regime [see Fig.~2(A) of the main text],
\begin{equation}
\delta r^2(\alpha_{\rm max}) = \frac{3}{2 \alpha_{\rm max}}.
\end{equation}
The choice of $\alpha_{\rm min}$ is such that $\delta r^2$ of the constrained system is comparable to the mean-squared displacement of the target system without constraint.
The last term in Eq.~(\ref{eq:final_expression}) corresponds to assuming a constant $\delta r^2$ for $\alpha\in[0,\alpha_{\rm min}]$ in the thermodynamic integration.
In Fig.~\ref{fig:s_total_vib}(B), we show $s_{\rm vib}$ for $N=1000$ and $8000$, with $\alpha_{\min}= 1.5 \times 10^0$, $7.5 \times 10^0$, and  $3.0 \times 10^1$.
At high volume fraction $\phi$, the results are insensitive to the choice of $\alpha_{\rm min}$, as expected. We can thus confidently set $\alpha_{\rm min}=7.5 \times 10^0$. Also, we empirically observe a linear relation between $s_{\rm vib}$ and $\phi$, which allows us to linearly extrapolate the fit and thus the configurational entropy (see Sec.~\ref{sec:sconf}).

\subsection{Mixing entropy}
\label{sec:mixing}
The mixing entropy of the continuous polydisperse system formally diverges in the thermodynamic limit because each particle in the system then belongs to one of an infinite number of different components~\cite{salacuse1982polydisperse,frenkel2014colloidal}.
One, however, can subtract from this quantity a physically relevant contribution, which we call the effective mixing entropy, $s_{\rm mix}^*$. 
The main idea is that a continuous polydisperse system can be regarded as an effective $M^*$-component system (see Ref.~\cite{sollich2001predicting} and references therein). 
We then assume that the effective $M^*$-component system shares physical properties, in particular a same (free-)energy basin, with the original continuous polydisperse system.
Here, a practical description of how to estimate $s_{\rm mix}^*$ from this scheme in our system is provided; a full explanation is provided in Ref.~\cite{ozawa2017does}.

To estimate $M^*$, we first decompose the distribution $f(\sigma)$ into $M$ species, as shown in the inset of Fig.~\ref{fig:s_mix}(B). We define $M$ species by dividing $f(\sigma)$ into equal intervals $\Delta \sigma=(\sigma_{\rm max}-\sigma_{\rm min})/M$, such that each species occupies the same fraction of the total volume, $A \rho \pi \Delta \sigma /6 = C$, where $C$ is a constant.
Note that $M$ is an integer and that $M \to \infty$ corresponds to the original continuous polydisperse system.
We then perform a quench of the discretized system from the original configuration, and determine whether or not it remains in the same glassy basin as the original system by measuring the mean-squared displacement before and after the quench, 
\begin{equation}
\Delta_{M}= \left[ \frac{1}{N} \sum_{i=1}^{N} | {\bf r}_{Mi}^{\rm IS} - {\bf r}_{0 i} |^2 \right],
\end{equation}
where $\{ {\bf r}_{Mi}^{\rm IS} \}$ is the inherent structure configuration of a $M$-discretized system.
If $M$ is large, the discretized system is almost identical with the original continuous polydisperse system and stays within the same basin, hence $\Delta_{M} \simeq \Delta_{M \to \infty}$.
If $M$ is small, however, discretization destroys the glassy basin of the original system. The system thus structurally rearranges into a different glassy basin, and $\Delta_{M} \gg \Delta_{M \to \infty}$. This determination is done here by considering inherent structures, which for hard-sphere systems correspond to an out-of-equilibrium compression up to jamming~\cite{stillinger1985inherent}. 
We determine $M^*$ as the crossover between these two limit cases.

More precisely, we follow the following algorithm. 
\begin{enumerate}
\item[1)] Obtain an equilibrium configuration of the original continuous polydisperse system, $\{ {\bf r}_{0 i} \}$, for the initial configuration.
\item[2)] Discretize the diameters $\sigma$ of the original system into $M$ species, keeping $\phi$ fixed.
\item[3)] Quench the system to its inherent structure,  $\{ {\bf r}_{M i}^{\rm IS} \}$, using the algorithm described in Refs.~\cite{xu2005random,desmond2009random}. 
\item[4)] Repeat 1) - 3) for a range of $M$ and over different initial configurations. 
\item[5)] Determine the crossover $M^*$ from $\Delta_{M}$ as a function of $M$.
\end{enumerate}

In Fig.~\ref{fig:s_mix}(A), we show how $\Delta_{M}$ evolves with $M$.
At large $M$, $\Delta_{M}$ is flat and near $\Delta_{M \to \infty}$. The discretized system stays in its original basin.
Upon decreasing $M$, however, $\Delta_{M}$ deviates from $\Delta_{M \to \infty}$, indicating that the discretized system escapes its basin.
We estimate $M^*$ as an onset of this deviation by fitting the two linear lines from large and small $M$ regions.
We find a weak $\phi$ dependence with $M^*\approx9-10$ for all $\phi$ considered. We use linear fits to more precisely locate the 
crossover at $M^*=9.13$, see Fig.~\ref{fig:s_mix}(A).

We map $M^*$ onto the effective mixing entropy $s_{\rm mix}^*$ by computing $s_{\rm mix}$ for different $M$,
\begin{eqnarray}
s_{\rm mix}(M) &=& - \sum_{m=1}^{M} X_m \ln X_m, \label{eq:s_mix} \\
X_m &=& \int_{ \sigma_{\rm min} + (m-1) \Delta \sigma}^{\sigma_{\rm min} + m\Delta \sigma} \mathrm{d} \sigma f(\sigma), \label{eq:X_m}
\end{eqnarray}
where $X_m$ is the concentration of species $m$.
Fig.~\ref{fig:s_mix}(B) shows the resulting $s_{\rm mix}(M)$. From the mapping indicated by the arrows, we obtain $s_{\rm mix}^* = 1.98$. Note that $s_{\rm mix}^*$
in Fig.~\ref{fig:s_mix}(B)
varies relatively weakly with $M^*$ and so a more precise 
estimate of the value of $M^*$ is not needed for our purposes.

\begin{figure}[htbp]
\includegraphics[width=0.48\columnwidth]{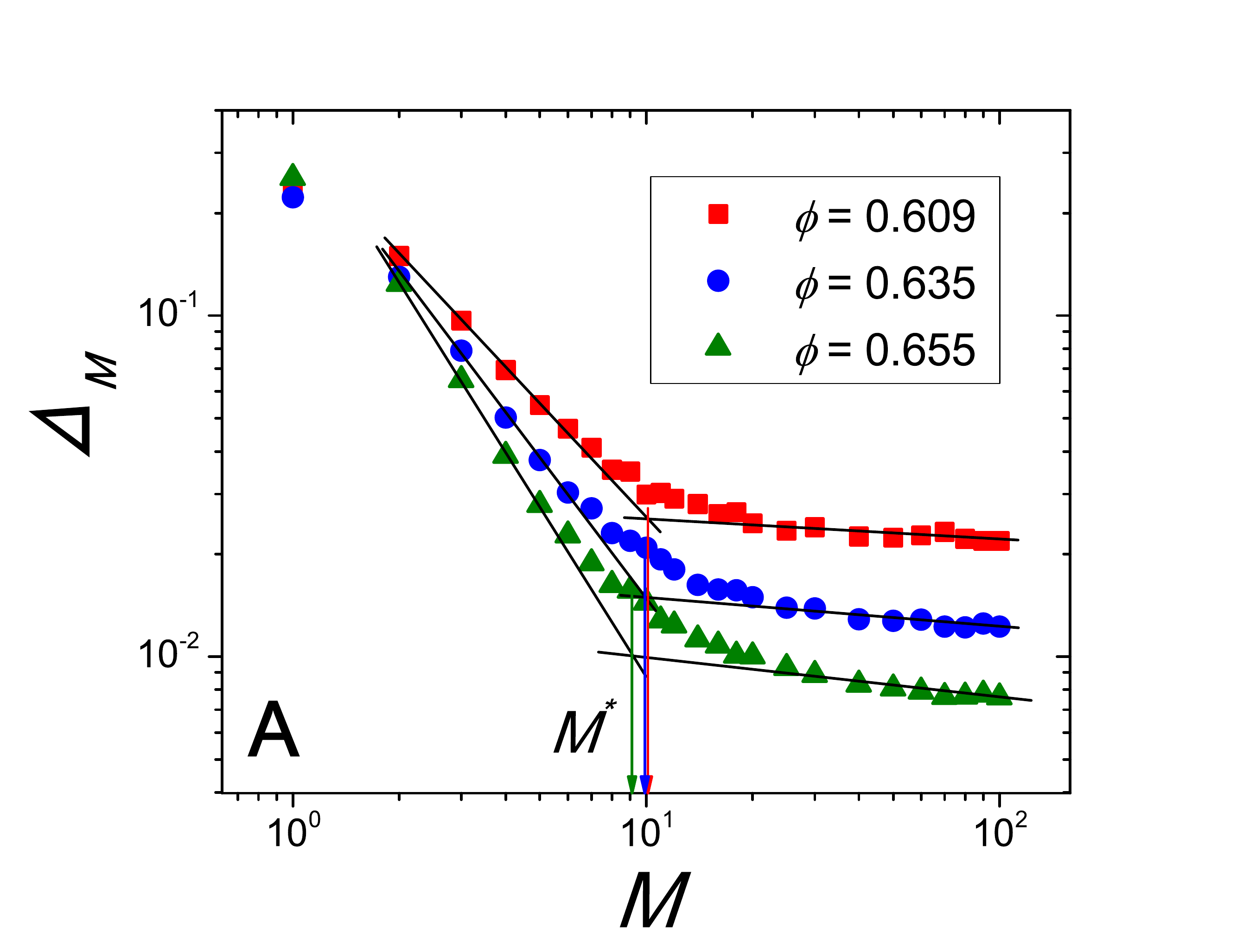}
\includegraphics[width=0.48\columnwidth]{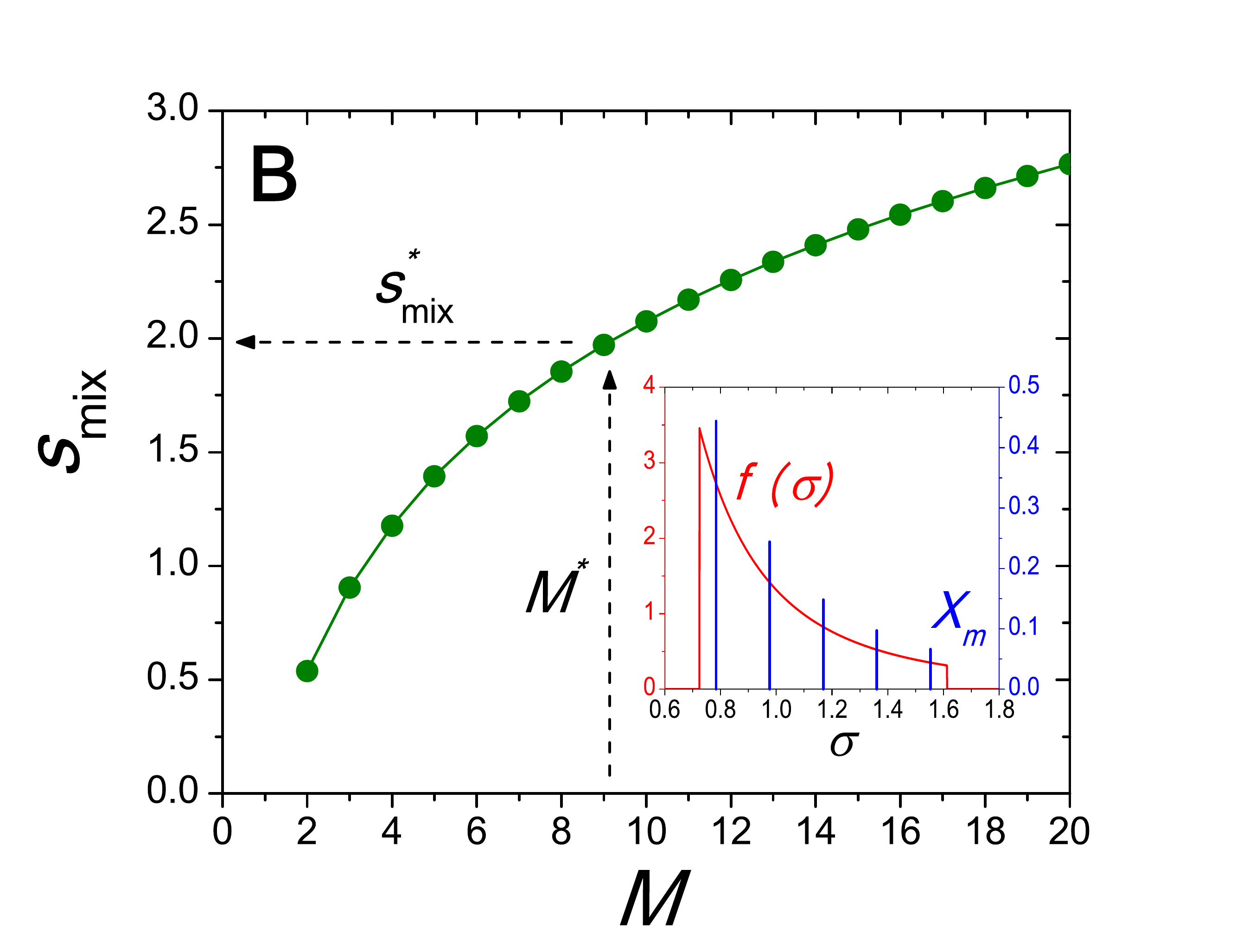}
\caption{
(A) Mean-squared displacement $\Delta_{M}$ between configurations before and after the quench.
The crossover, $M^*$, is determined from the intersection of the two linear regimes at small and large $M$.
(B) The mixing entropy $s_{\rm mix}$ as a function of $M$ from Eqs.~(\ref{eq:s_mix}) and (\ref{eq:X_m}).
The dashed arrows map $M^*$ onto $s_{\rm mix}^*$.
The inset shows the continuous polydispersity distribution, $f(\sigma)$, and the weights of delta functions for the discretized distribution with $M=5$, $\left\{X_m\right\}_{m=1,\ldots,5}$, as an example.}
\label{fig:s_mix}
\end{figure}  

\subsection{Configurational entropy}
\label{sec:sconf}
Using the above results, we obtain the configurational entropy from $s_{\rm tot}-s_{\rm vib}$ 
as a function of $\phi$. As can be seen in Fig.~\ref{fig:s_config},
no significant difference is observed between $N=1000$ and $8000$.
We also find that the combination of $s_{\rm tot}$ from the (modified) BMCSL EOS in Fig.~\ref{fig:s_total_vib}(A) and $s_{\rm vib}$ from the linear fit in Fig.~\ref{fig:s_total_vib}(B) gives a reasonable extrapolation toward $s_{\rm tot}-s_{\rm vib}=0$, and thus an estimate for the Kauzmann transition, $\phi_{\rm K} \approx 0.68$ ($Z_{\rm K} \approx 45$).
In Fig. 3 from the main text we use the curve extrapolated by the modified BMCSL EOS.

\begin{figure}[htbp]
\begin{center}
\includegraphics[width=0.6\columnwidth]{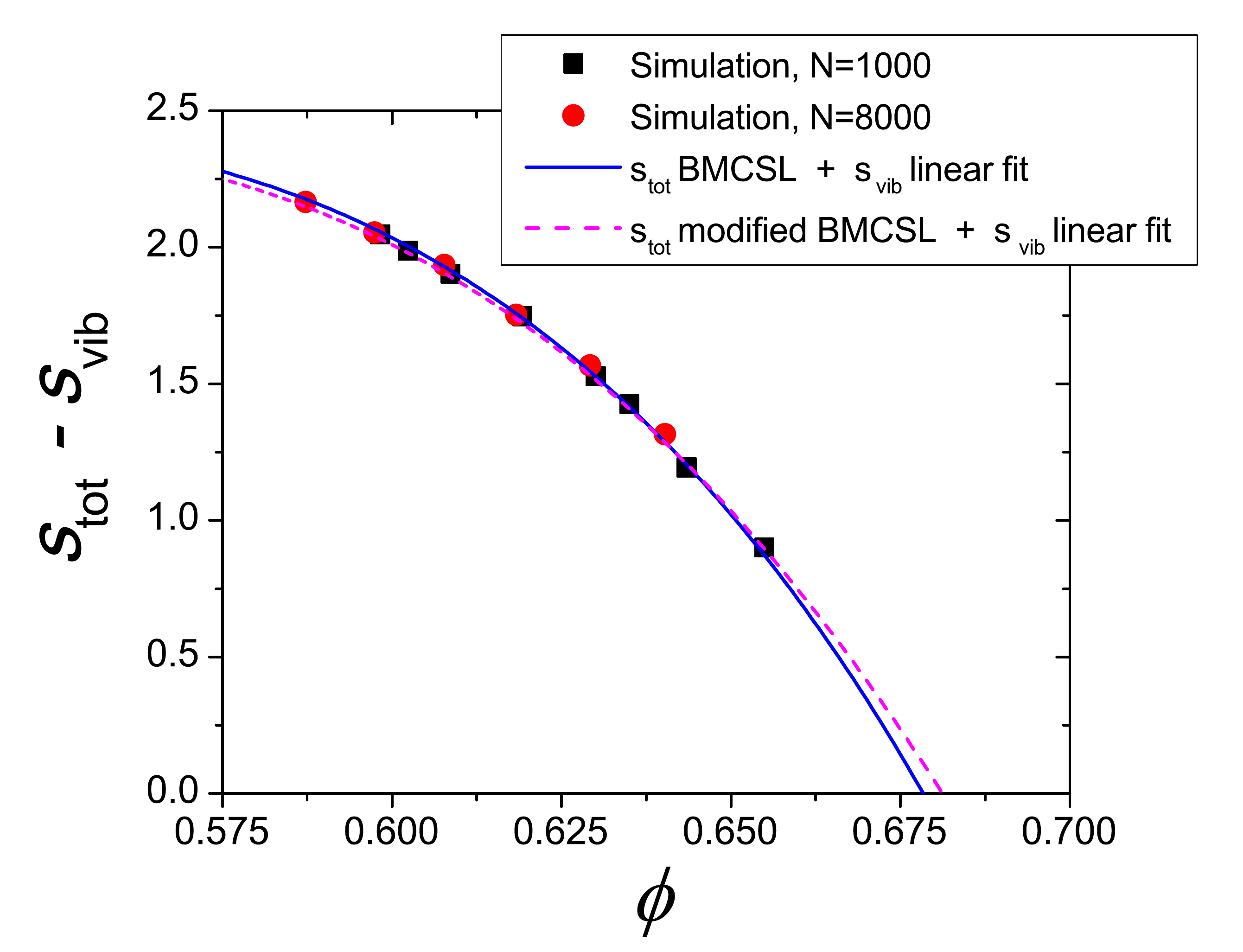}
\caption{
Configurational entropy $s_{\rm tot}-s_{\rm vib}$ as a function of volume fraction $\phi$, including the effective mixing entropy $s_{\rm mix} = s_{\rm mix}^*=1.98$.
The solid (dashed) curve is a combination of $s_{\rm tot}$ from the (modified) BMCSL EOS and $s_{\rm vib}$ from the linear fit.
}
\label{fig:s_config}
\end{center}
\end{figure}

\section{Methods 2 and 3: The Franz-Parisi free energy}\label{Ap_method23}

In this section we detail the Franz-Parisi construction~\cite{FP97}.
We describe the order parameter of the glass transition, 
the umbrella sampling, as well as the parallel tempering 
and histogram reweighting techniques used to measure 
numerically the configurational entropy defined by Methods 2 and 3.
Because the presentation closely follows earlier reports
in which the Franz-Parisi free energy was determined
numerically~\cite{Berthier13,BC14,BJ15}, we only provide a brief summary of the computational
methodology used.

\subsection{Order parameter}

The order parameter of the 
Franz-Parisi construction is the overlap $Q$ (or similarity) 
between two disordered configurations, defined as
\begin{equation}
Q = \frac{1}{N} \sum_{i,j} \theta( a- | {\bf r}_{1,i} - {\bf r}_{2,j} | ),
\end{equation} 
where $\theta(x)$ is the Heaviside function, and ${\bf r}_{1,i}$ and 
${\bf r}_{2,j}$ are the positions of particle $i$ and $j$ within configurations 1 
and 2, respectively. By definition, $Q$ is large when the 
density fields in configurations 1 and 2 are very close, and
$Q$ is close to zero when the density profiles are uncorrelated.   
The length, $a=0.23$, accounts for small-amplitude
thermal vibrations, so that density profiles that only differ by local
vibrations caused by thermal fluctuations
have $Q$ close to unity.

\subsection{Umbrella sampling}

First, a reference configuration, $\{ {\bf r}_2 \}$, 
is chosen at random from the equilibrium states of the system at a fixed volume 
fraction.
In order to access the probability distribution function of the overlap $Q_{12}$ 
between configurations $\{ {\bf r}_1 \}$ and $\{ {\bf r}_2 \}$, we conduct $n$ distinct simulations.
In each simulation, $\ell = 1, \cdots, n$, system 1 
evolves according to the Hamiltonian
\begin{equation}
H_\ell = H ( \{ {\bf r}_1 \})  - \varepsilon Q_{12} +  W_\ell(Q_{12}), 
\label{hi}
\end{equation}
where $H (\{ {\bf r} \} )$ is the original hard-sphere Hamiltonian and 
$\varepsilon$ is the thermodynamic field conjugate to the overlap $Q_{12}$.
The biasing potential $W_\ell(Q)$ is taken to have the form
\begin{equation}
W_\ell(Q) = k_\ell (Q - Q_\ell)^2,
\label{wi}
\end{equation} 
with parameters $(k_\ell, Q_\ell)$ chosen to constrain the 
overlap $Q_{12}$ to explore values away from its average equilibrium value. 
We then perform local displacements and swap MC moves,
which are accepted using a Metropolis acceptance rate given by the
Hamiltonian (\ref{hi}). 

Provided that the system is properly thermalized,
our measurements yield the equilibrium probability distribution 
function of the overlap, 
\begin{equation}
P_\ell(Q,\varepsilon,\phi) =  \langle \delta(Q - Q_{12}) \rangle_\ell ,
\end{equation} 
where $\langle \cdots \rangle_\ell$ denotes the thermal average with Hamiltonian $H_\ell$ in Eq.~(\ref{hi}) at fixed 
reference configuration 2. 
Because we perform a quenched average to compute the overlap 
distribution, however, a second averaging,
over the reference configuration 2,  is needed
to determine the Franz-Parisi free energy.

The idea behind the biasing potentials $W_\ell(Q)$ in Eqs.~(\ref{hi}),
and (\ref{wi})
is that the fluctuations of the overlap in each simulation can be
tailored to explore a narrow region centered around $Q_\ell$. 
Each simulation thus explores but a small range of 
overlap values, and it becomes unnecessary to wait for 
very rare overlap fluctuations to take place. 
Umbrella sampling enables the
efficient measure of the atypical overlap fluctuations
needed to reconstruct the Franz-Parisi free energy.

\subsection{Parallel tempering}

While the umbrella sampling technique described above considerably accelerates 
the measurement of overlap fluctuations, we have observed 
that when $\phi$ is high, $N$ is large, and/or $Q_\ell$ is large, the particle 
dynamics still slows down considerably. It then becomes
difficult to perform an accurate sampling of the overlap 
fluctuations imposed by the Hamiltonian (\ref{hi}).

To accurately sample the density regime explored in the present work, 
we have implemented parallel tempering MC moves~\cite{Hukushima96}. 
We conduct the $n$ simulations needed for the umbrella sampling 
at volume fraction $\phi$ in parallel, and propose MC exchange moves 
between neighboring simulations, as characterized by nearby sets of parameters,
say $(k_\ell, Q_\ell)$ and $(k_{\ell+1}, Q_{\ell+1})$. An exchange between 
simulations $\ell$ and $\ell+1$ is proposed 
with a low frequency (typically every $10^3$ MC sweeps)
and they are accepted with a Metropolis acceptance rate 
given by the Hamiltonians $H_\ell$ and $H_{\ell+1}$, ensuring 
detailed balance. 

Because each simulation now performs a random walk in the 
parameter space defined by $\{ (k_\ell, Q_\ell), \, \ell=1, \cdots, n \}$, 
the sampling of overlap fluctuations is greatly enhanced.
For the method to be efficient, however, we need to adjust the 
biasing potentials $W_\ell(Q)$ such that the distributions $P_\ell(Q)$
of neighboring simulations overlap significantly.
We have used up to 
$n=20$ to gather data. Thermalization was carefully checked 
by running long simulations (up to $10^{10}$ MC sweeps per simulation), in order
to make sure that each state point was visited several times 
by all simulations via the replica exchange. 
This represents a significant numerical effort. 
 
\subsection{Histogram reweighting}

Having obtained thermalized results from $n$ biased simulations
run in parallel, we process the simulation outcome 
using multi-histogram reweighting 
methods to reconstruct the unbiased probability $P(Q)$ 
from the independently measured $P_\ell(Q)$, 
\be
P(Q,\varepsilon,\phi) 
= \frac{\sum_{\ell=1}^n P_\ell(Q,\varepsilon,\phi) }{ \sum_{\ell=1}^n e^{-\beta W_\ell} / Z_\ell },
\label{pq}
\ee 
where $Z_\ell$ are defined self-consistently as 
\be
Z_\ell = \int_0^1 {\rm d}Q' \frac{ \sum_{m=1}^n P_j(Q',\varepsilon,\phi) }{ 
\sum_{m=1}^n e^{\beta(W_\ell-W_m) }/Z_m }.
\ee 

Note that the actual value of $\varepsilon$ used in simulations plays no 
conceptual role because reweighting 
directly provides $P(Q,\varepsilon',\phi)$ from $P(Q,\varepsilon,\phi)$
for distinct field values $\varepsilon$ and $\varepsilon'$:
\begin{equation}
P(Q,\varepsilon',\phi) = \frac{P(Q,\varepsilon,\phi) e^{-\beta Q 
(\varepsilon'-\varepsilon)}}{\int_0^1 {\rm d}Q' P(Q',\varepsilon,\phi) e^{-\beta Q' 
(\varepsilon'-\varepsilon)} } . 
\end{equation}
Two values of the field $\varepsilon$ are nonetheless particularly 
relevant to determining the configurational entropy, as described in the following two subsections.

\subsection{Method 2}

First, the Franz-Parisi free energy $V(Q)$ is obtained as
\begin{equation}
V(Q) = - \lim_{\varepsilon \to 0} \left[ \frac{T}{N} \ln P(Q, \varepsilon, 
\phi) \right],
\end{equation}
where $[ \cdots ]$ denotes averaging over the quenched 
reference configuration 2. We used 60 independent 
reference configurations for each value of the volume
fraction. 
Because $V(Q)$ is only defined up to an additive constant, 
we arbitrarily adjust it, such that $V (Q_{\rm low}) = 0$, where
$Q_{\rm low} \approx 0.05$ is defined as the location of the global minimum in 
$V(Q)$. This additive constant is irrelevant because only the free energy difference, 
\begin{equation}
s_{\rm conf} =V(Q_{\rm high})-V(Q_{\rm low}) \quad {\rm(Method \ 2)}
\end{equation}
is needed to determine the configurational entropy.
In this work, we set $Q_{\rm high}=0.8$ as motivated in the following subsection.

\subsection{Method 3}

Second, we obtain the critical field value 
$\varepsilon = \varepsilon^\star$ 
needed to induce a phase coexistence between 
low and high overlap states~\cite{FP97,BJ15}.
In practice, we use the strength of the reweighting 
method to finely explore a range of $\varepsilon$ values, 
and define $\varepsilon^\star$ as the field strength for which 
the overlap shows a value intermediate 
between $Q_{\rm low}$ and $Q_{\rm high}$,
where the distribution $P(Q,\varepsilon^\star,\phi)$ shows two peaks 
of equal amplitude, and where the variance of the 
overlap fluctuations (the susceptibility) is maximal. 
The position of the second peak of $P(Q,\varepsilon^\star,\phi)$
at coexistence  sets $Q_{\rm high}$. Because its volume fraction 
dependence is very weak, we fix it close to its average 
value and use $Q_{\rm high} = 0.8$ for all $\phi$. 
Note that a secondary minimum cannot exist in the 
large system size limit of finite-dimensional simulations~\cite{BC14}, which
is why we resort to the above definition of $Q_{\rm high}$.

\res{The field $\varepsilon^\star$ has a simple graphical 
interpretation. It represents the amplitude of the field needed to `tilt' the 
potential $V(Q)$ towards coexistence [see Fig.~2(B) of the main text].
Because the relation
$V(Q_{\rm high}) \approx \varepsilon^\star (Q_{\rm high} - Q_{\rm low})$,
holds to a good approximation, $\varepsilon^\star$ provides an estimate of the 
free energy difference $V(Q_{\rm high})-V(Q_{\rm low})$.
We can thus define the configurational entropy as
\begin{equation}
s_{\rm conf} = \varepsilon^\star (Q_{\rm high} - Q_{\rm low}) \quad {\rm( Method \ 3)}.
\end{equation}}

\section{Method 4: The point-to-set correlation}\label{Ap_method4}

\subsection{Definition of cavity core overlap and point-to-set correlations}
The similarity between two configurations, ${\bf X}=\le\{{\bf x}_i\ri\}$ and ${\bf Y}=\le\{{\bf y}_i\ri\}$, is characterized by using an overlap field, $q_{\bf X,Y}\le({\bf r}\ri)$, computed as in Ref.~\cite{BCY16}. 
For each particle ${\bf x}_i$, we find the nearest particle ${\bf y}_{i_{\rm nn}}$ and assign an overlap value $q_{{\bf X; Y}} \le({\bf x}_i\ri)\equiv w\le(\big|{\bf x}_i-{\bf y}_{i_{\rm nn}}\big|\ri)$, where
\be
w(z)\equiv {\rm exp}\le[-\le(\frac{z}{b}\ri)^2\ri],
\ee
with $b=0.23$.
This function defines overlap values $q_{\bf X; Y} \le({\bf x}_i\ri)$ at scattered points $\le\{{\bf x}_i\ri\}$.
We then define a continuous function passing through these points, $q_{\bf X; Y} \le({\bf r}\ri)$.
Specifically, we first perform a Delaunay tessellation of space and, to a point ${\bf r}$ within a simplex spanned by four points $\le\{{\bf x}_i\ri\}_{i=i_1,i_2,i_3,i_4}$, associate a linearly interpolated value
\be
q_{\bf X; Y} \le({\bf r}\ri)=\sum_{i=i_1,i_2,i_3,i_4}c_i q_{\bf X} \le({\bf x}_i\ri),
\ee
where $\le\{c_i\ri\}_{i=i_1,i_2,i_3,i_4}$ satisfies ${\bf r}=\sum_{i=i_1,i_2,i_3,i_4}c_i {\bf x}_i$ with the constraint $\sum_{i=i_1,i_2,i_3,i_4}c_i=1$.
We can similarly obtain $q_{\bf Y; X} \le({\bf r}\ri)$, allowing us to define the overlap field
\be q_{\bf X,Y}\le({\bf r}\ri)\equiv \frac{1}{2}\le\{q_{\bf X; Y}\le({\bf r}\ri)+q_{\bf Y; X}\le({\bf r}\ri)\ri\}\, .
\ee

In order to capture the similarity between two configurations near the center of the cavity, we also define the cavity core overlap
\begin{equation}
\qc\equiv\frac{3}{4\pi r_{\rm c}^3}\int_{|{\bf r'}|<r_{\rm c}}{\rm d}{\bf r'}\ q_{\bf X,Y}\le({\bf r'}\ri),
\end{equation}
where $r_{\rm c}=0.576$ and ${\bf r'}={\bf 0}$ is the cavity center. This integral is numerically evaluated by Monte Carlo integration using $10^4$ points.

For each volume fraction $\PF$ and cavity radius $R$, the point-to-set correlation function, $Q_{\rm PTS}(R;\PF)$, is evaluated by disorder-averaging over $40$ cavity centers and, within each cavity, thermal-averaging over $s_{\rm{prod}}$ pairs of equilibrated configurations (see Tables \ref{sample1}-\ref{sample7})~\cite{BCY16}.
We extract the point-to-set correlation length through the compressed exponential fit, 
\begin{equation}
Q_{\rm PTS}(R;\PF)=A\exp[-\le\{R/\xi_{\rm PTS}(\PF)\ri\}^{\gamma}]+Q_{\rm PTS}^{\rm bulk}(\PF),
\end{equation}
with the compressed exponent $\gamma=4$. Here the bulk value, $Q_{\rm PTS}^{\rm bulk}$, corresponds to the value at $R=\infty$ and is evaluated by taking $10^5$ pairs of independent configurations in bulk samples.

At the point-to-set length scale, we expect the probability distribution function of core overlaps to display broad fluctuations.
Figure~\ref{bimodal} bears out this expectation.
In particular, as anticipated in Ref.~\cite{BCY16}, we observe that the full disorder-averaged distribution becomes bimodal at high volume fractions, showing that we have indeed entered a deeply glassy regime that remains currently inaccessible for a standard binary Lennard-Jones liquid (see also \cite{ozawa2017exploring}).

\begin{figure*}[htbp]
\includegraphics[width=0.48\textwidth]{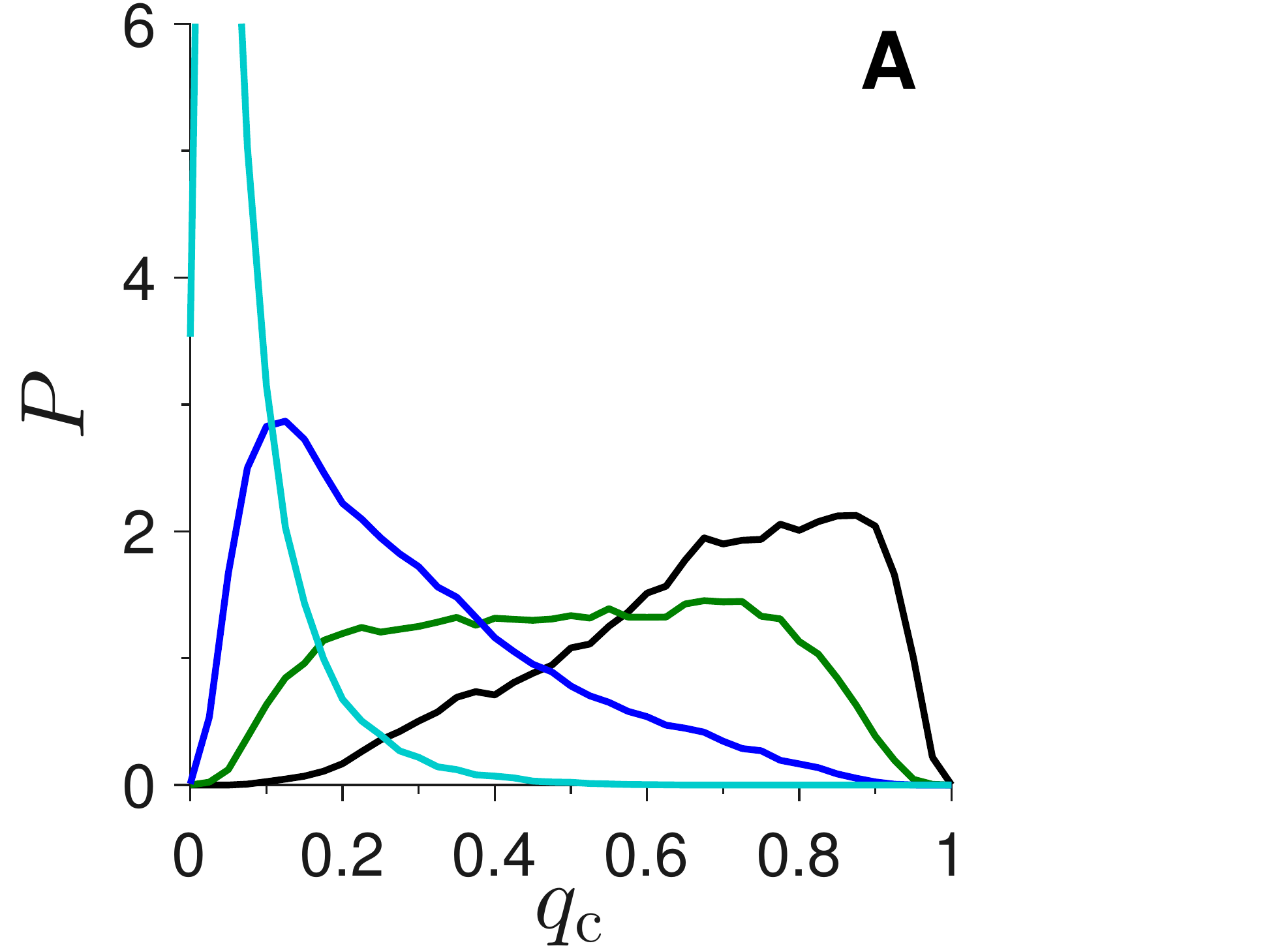}
\includegraphics[width=0.48\textwidth]{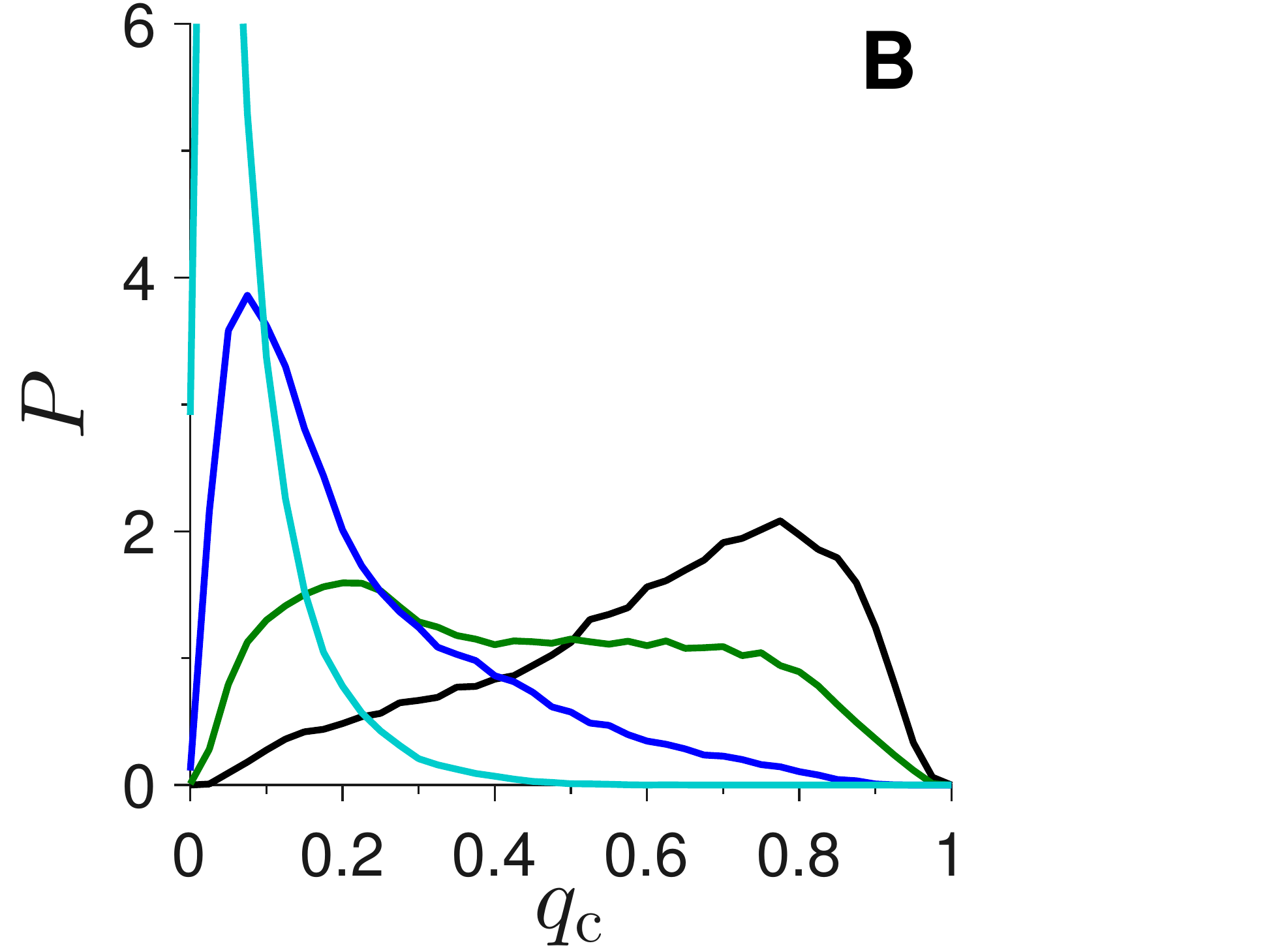}
\includegraphics[width=0.48\textwidth]{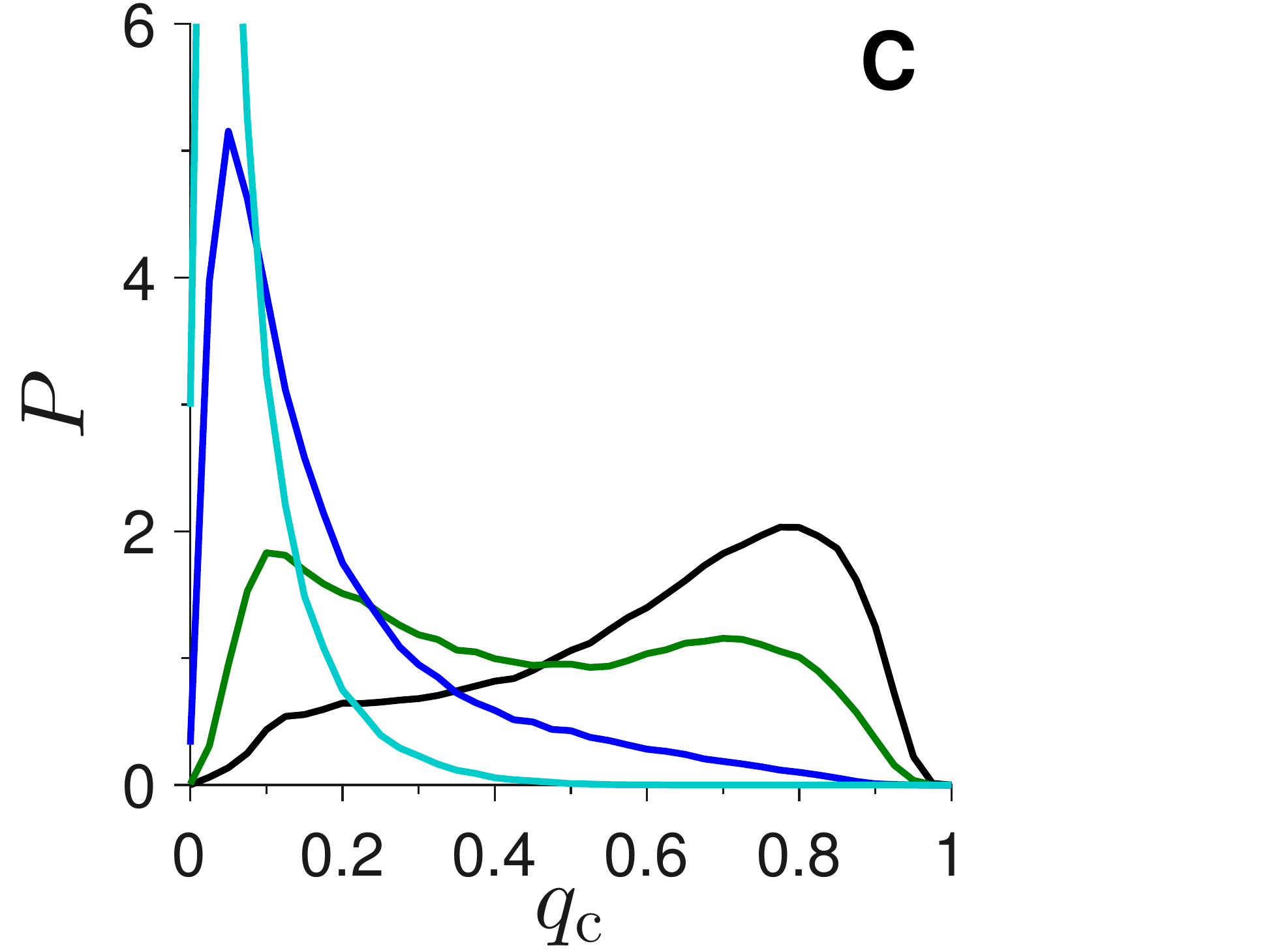}
\includegraphics[width=0.48\textwidth]{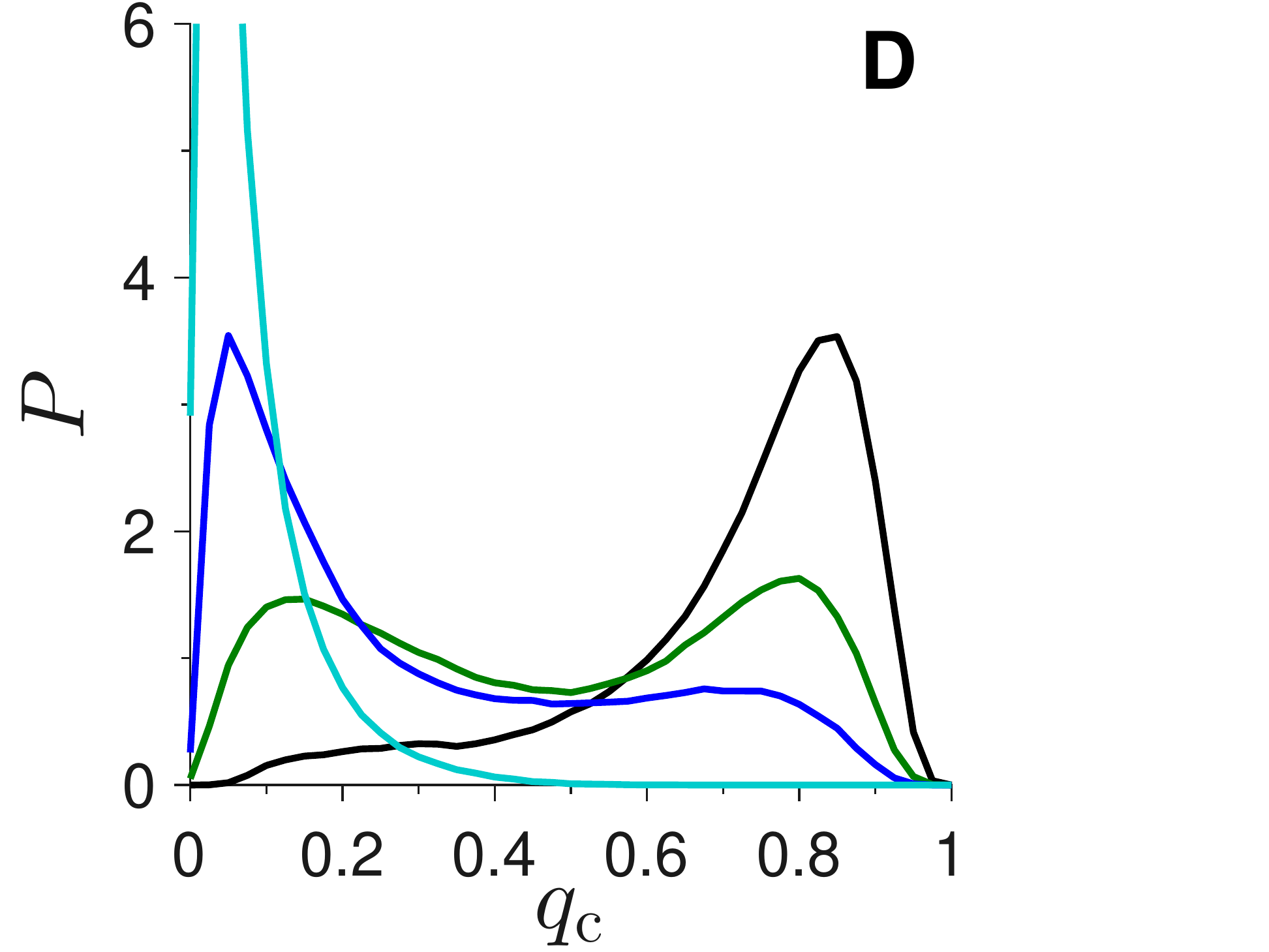}
\caption{Disorder-averaged probability distribution function of core overlap $P(\qc)$, at volume fractions $\PF = 0.568$ (A), $0.597$ (B), $0.618$ (C), and $0.640$ (D) with varying radius $R=R_1$ (black),  $R_2$ (green), $R_3$ (blue), and $R_4$ (cyan). We chose 
$(R_1, R_2,R_3,R_4)|_{\PF=0.568}=(1.38,1.73,2.07,3.11)$; 
$(R_1, R_2,R_3,R_4)|_{\PF=0.597}=(1.90,2.25,2.59,3.46)$; 
$(R_1, R_2,R_3,R_4)|_{\PF=0.618}=(2.25,2.59,3.11,4.15)$; 
and $(R_1, R_2,R_3,R_4)|_{\PF=0.640}=(2.77,3.46,3.80,5.18)$.}
\label{bimodal}
\end{figure*}

\subsection{Parallel-tempering sampling with fuzzy-ensemble replicas}
Denote positions of mobile particles inside the cavity $\mathbf{x}^{\rm M}_{i}$ with $i=1,\ldots,N_{\rm cav}$ and their associated diameters $\sigma^{\rm M}_i$.
Similarly, denote positions and diameters of pinned particles outside the cavity $\mathbf{x}^{\rm P}_{j}$ and $\sigma^{\rm P}_j$, respectively.
For cavity point-to-set measurements, one samples cavity configurations $\le\{\mathbf{x}^{\rm M}_{i}\ri\}$ with probability
\bea\label{oriE}
&&P\le(\le\{\mathbf{x}^{\rm M}_{i}\ri\} \big| \le\{\sigma^{\rm M}_{i}\ri\}, \le\{\mathbf{x}^{\rm P}_{j}\ri\}, \le\{\sigma^{\rm P}_{j}\ri\}\ri)\\
&\simeq &{\rm exp}\le[-\beta U\le(\le\{\mathbf{x}^{\rm M}_{i}\ri\}, \le\{\sigma^{\rm M}_{i}\ri\}, \le\{\mathbf{x}^{\rm P}_{j}\ri\},  \le\{\sigma^{\rm P}_{j}\ri\}\ri)\ri]\nonumber,
\eea
where $\simeq$ denotes equality up to a normalization constant and a hard wall constraint $\big|{\bf x}^{\rm M}_{i}\big|<R$ at the edge of the cavity of size $R$ is imposed. Note that for the standard hard-sphere potential, $U$, this expression does not depend on the inverse temperature $\beta\in\le(0,\infty\ri)$.
Properly sampling from this weight can be computationally demanding, because the equilibration time grows rapidly both as the cavity size decreases and as the volume fraction increases (see Fig.~\ref{versus} and Sec.~\ref{swapisnotenough}).
In these systems, phase space can actually become fully disconnected, which makes proper sampling impossible through (semi-)local Monte Carlo moves.

We here overcome this difficulty through a parallel-tempering algorithm~\cite{FS01,FWT02}. The simplest scheme, analogous to that employed in Ref.~\cite{BCY16} for a thermal binary Lennard-Jones liquid, would be to couple the above ensemble with replicas governed by a probability
\bea
&&P_{\lambda}\le(\le\{\mathbf{x}^{\rm M}_{i}\ri\} \big| \le\{\sigma^{\rm M}_{i}\ri\}, \le\{\mathbf{x}^{\rm P}_{j}\ri\}, \le\{\sigma^{\rm P}_{j}\ri\}\ri)\\
&\simeq&{\rm exp}\le[-\beta U\le(\le\{\mathbf{x}^{\rm M}_{i}\ri\},\le\{ \lambda\sigma^{\rm M}_{i}\ri\}, \le\{\mathbf{x}^{\rm P}_{j}\ri\},  \le\{\sigma^{\rm P}_{j}\ri\}\ri)\ri]\, ,\nonumber
\eea
where the diameter of the particles inside the cavity is uniformly shrunk by $\lambda$.
This approach, however, is highly inefficient for hard interactions. Replica-swap attempts are indeed very likely to be rejected, because even small $\lambda$ increments can result in hard overlaps. More precisely, in order to have appreciable replica-swapping acceptance rates, the spacing between consecutive $\lambda$'s must scale as $O(1/N_{\rm cav})$, and thus $O(N_{\rm cav})$ replicas are needed, as opposed to $O(\sqrt{N_{\rm cav}})$ for thermal systems~\cite{FS01}. 

We therefore introduce a \textit{fuzzy} ensemble governed by probability weights controlled by two parameters, $\alpha$ and $\hat{\lambda}$,
\bea
&&P_{\alpha,\hat{\lambda}}\le(\le\{\mathbf{x}^{\rm M}_{i}\ri\}, \lambda\ \big| \le\{\sigma^{\rm M}_{i}\ri\}, \le\{\mathbf{x}^{\rm P}_{j}\ri\}, \le\{\sigma^{\rm P}_{j}\ri\}\ri)\\
&\simeq&{\rm exp}\le\{-\frac{\alpha^2N_{\rm cav}}{2}\le(\lambda-\hat{\lambda}\ri)^2\ri\}\, \nonumber\\
&&\times {\rm exp}\le[-\beta U\le(\le\{\mathbf{x}^{\rm M}_{i}\ri\},\le\{ \lambda\sigma^{\rm M}_{i}\ri\}, \le\{\mathbf{x}^{\rm P}_{j}\ri\},  \le\{\sigma^{\rm P}_{j}\ri\}\ri)\ri]\, ,\nonumber
\eea
hence the shrinking factor $\lambda$ is allowed to fluctuate.
With this modification, a $O(\sqrt{N_{\rm cav}})$ scaling for the number of replicas needed is recovered.
One caveat is that one of the replicas must obey the original sharp distribution defined in Eq.~(\ref{oriE})--formally corresponding to the limit $\alpha\rightarrow\infty$ with $\hat{\lambda}=1$--and the aforementioned challenge imposed by hard interactions still persists for that case.
This problem is here surmounted by using a large number of replica-swap attempts near the original replica--$O(1000N_{\rm cav})$ more frequent than others--as detailed in Sec.~\ref{PTdetail}.

\subsection{Details of Monte Carlo moves}
This section details the implementation of the Monte Carlo scheme for the cavity simulations.

\subsubsection{Basic moves within a replica}
Within each replica, we perform Monte Carlo moves consisting of local displacements, particle identity swaps, and, for fuzzy ensembles, diameter fluctuations.
\begin{itemize}
\item Local displacements consist of: (i) randomly choosing a particle $i$ from the $N_{\rm cav}$ mobile particles within the cavity; (ii) displacing particle $i$ by $\Delta {\bf x}=l \hat{\bf{n}}$ with uniform $l\in[0,0.1]$ and uniform $\hat{\bf{n}}$ on the unit sphere $S^2$;  and (iii) accepting displacement only if no hard overlaps are created. Note that an additional hard spherical wall at the edge of the cavity guarantees that mobile particles cannot leave the cavity--a rare instance, even at the lowest volume fraction studied.
\item Particle identity swaps consist of: (i) choosing two distinct particles $i$ and $j$ within a cavity; and (ii) swapping their diameters if it results in no hard overlaps.
(For the highest volume fraction $\PF=0.640$ with the cavity size $R\geq3.11$, in order to accelerate runs, we attempt these moves only for pairs with diameter difference $\lambda|\sigma_i-\sigma_j|<0.086$.)
\item Diameter fluctuations consist of: (i) uniformly drawing a shift $\Delta\lambda\in\le[-\frac{0.2}{\alpha\sqrt{N_{\rm cav}}},+\frac{0.2}{\alpha\sqrt{N_{\rm cav}}}\ri]$,
and (ii) accepting the shift with probability $p={\rm min}\{p_{\Lambda},1\}$, where
\be
p_{\Lambda}={\rm exp}\le[-\frac{\alpha^2N_{\rm cav}}{2}\le\{\le(\lambda+\Delta\lambda-\hat{\lambda}\ri)^2-\le(\lambda-\hat{\lambda}\ri)^2\ri\}\ri]\, ,
\ee
if no hard overlaps are created.
\end{itemize}
One MC sweep consists of $N_{\rm cav}$ MC trial moves of the above kinds for each replica. For large cavities with $R > \xi_{\rm PTS} + 1$,
each MC sweep consists of $N_{\rm cav}$ MC trial moves with $70\%$ ($80\%$) local
displacements and $30\%$ ($20\%$) particles identity swaps (parenthesis are for $\phi=0.640$) alone. For small cavities, however, proper sampling requires replica exchange and diameter fluctuations  (see Fig.~\ref{versus}). 

\subsubsection{Parallel tempering}
\label{PTdetail}
For parallel tempering, we label replicas $a=1,\ldots,n$, where $a=1$ corresponds to the original ensemble with $\hat{\lambda}_1=1$ and $\alpha_1=\infty$. For the other replicas, we choose $\alpha_{a\geq2}=20$, and $\{\hat{\lambda}_a\}$ are tuned to enable appreciable replica-swap rate (see below). For $a\geq3$, each MC sweep consists of $N_{\rm cav}$ MC trial moves with $60\%$ ($70\%$) local displacements, $30\%$ ($20\%$) particles identity swaps, and $10\%$ diameter fluctuations (parenthesis are for $\PF=0.640$ with $R\geq3.11$).
For $a=1$ and $a=2$, a different scheme is employed. The two cavities are run as a pair with a large number of replica-swap attempts. Each MC sweep consists of $4N_{\rm cav}$ attempts, $50\%$ replica-swap between $\le\{\mathbf{x}^{\rm M}_{i}\ri\}_{1}$ and $\le\{\mathbf{x}^{\rm M}_{i}\ri\}_{2}$ --accepted only when swapping $\lambda_1=1$ and $\lambda_2$ does not result in hard overlaps--, $15\%$  ($17.5\%$) local displacements respectively for $a=1$ and $a=2$, $7.5\%$ ($5\%$) particles identity swaps respectively for $a=1$ and $a=2$, and $5\%$ diameter fluctuations for $a=2$. 

For each $a\geq2$, a replica-identity swap between $\le(\le\{\mathbf{x}^{\rm M}_{i}\ri\}_{a}, \lambda_{a}\ri)$ and $\le(\le\{\mathbf{x}^{\rm M}_{i}\ri\}_{a+1}, \lambda_{a+1}\ri)$ is attempted every $1000$ MC sweeps on average, with acceptance probability $p={\rm min}\{p_{\rm RS},1\}$, where
\be
p_{\rm RS}={\rm exp}\le\{-\alpha^2N_{\rm cav}\le(\hat{\lambda}_{a+1}-\hat{\lambda}_{a}\ri)\le(\lambda_{a+1}-\lambda_{a}\ri)\ri\}\, .
\ee

The replica parameters, $\{\hat{\lambda}_a\}_{a\geq2}$, are tuned to ensure sufficient replica-swap rates. In order to achieve this sampling, we first define the average of the fluctuating $\lambda$
\be
\langle\lambda\rangle_a=\int {\rm d}\lambda {\rm d}\mathbf{x}^{\rm M}_{i}\ \lambda\ P_{\alpha_a,\hat{\lambda}_a}\le(\le\{\mathbf{x}^{\rm M}_{i}\ri\}, \lambda\ \big| \le\{\sigma^{\rm M}_{i}\ri\}, \le\{\mathbf{x}^{\rm P}_{j}\ri\}, \le\{\sigma^{\rm P}_{j}\ri\}\ri)\, ,
\ee
For the replica $a=2$, we ensure that $\langle\lambda\rangle_2=1+O\le(\sqrt{\langle\lambda^2\rangle_2-\langle\lambda\rangle_2^2}\ri)$, which typically requires $\hat{\lambda}_2>1$, because of the relatively high system pressure.
Replicas are then added one by one, with $\hat{\lambda}_2>\hat{\lambda}_3>\ldots>\hat{\lambda}_n$, each time targeting a replica-swap acceptance rate of $\sim20\%$.
This process is stopped upon reaching $\hat{\lambda}_n$, such that $\langle\lambda\rangle_n<\lambda_{\rm dec}$ (see Tables \ref{sample1}-\ref{sample7}).
Although this linear approach does not attain a globally uniform replica-swap acceptance rate (see~\cite{KTHT06,VP15} for more systematic approaches), the resulting scheme suffices to ensure equilibration and convergence, as defined in the next subsection.

The concerted use of replica exchange, fuzzy ensembles, and specialized sampling scheme around the original ensemble provides a sufficient number of independent cavity configurations from the desired ensemble defined in Eq.~(\ref{oriE}).

\subsubsection{Convergence criterion}
The quality of equilibration within each cavity is evaluated by monitoring the convergence of two initialization schemes~\cite{BICtest12,BCY16}: (i) from the original configuration, and (ii) from a randomized configuration prepared by first running $10^6$ MC sweeps with shrunk cavity particles at $(\hat{\lambda},\alpha)=(0.5,20)$ and then slowly regrowing particles back to $\lambda=1$. We then record configurations every $t_{\rm rec}$ MC sweeps (see Tables \ref{sample1}-\ref{sample7}), and monitor the core overlap between new configurations and the original configuration, $\qc^{\rm on}\le(t\ri)$, as a function of the number of MC sweeps, $t$.
The first $s_{\rm eq}$ configurations are discarded, and the average overlap for the following $s_{\rm prod}$ configurations, 
\be\label{qon}
\langle \qc^{\rm on}\rangle\equiv\frac{1}{s_{\rm prod}}\sum_{s=s_{\rm eq}+1}^{s_{\rm eq}+s_{\rm prod}}\qc^{\rm on}\le(t_{\rm rec}s\ri)\, ,
\ee
is computed. Convergence is deemed achieved when the results of both approaches lie within $\pm0.1$ of each other for each cavity. 
This criterion also allows us to estimate the convergence time for 
a given value of $R$ and $\phi$. 
Replica parameters as well as $s_{\rm eq}$ and $s_{\rm prod}$ (see Tables \ref{sample1}-\ref{sample7}) are chosen, such that at least $95\%$ out of $40$ cavities pass this convergence test, except for the most challenging data point, $(\PF,R)=(0.640,3.80)$, where a $92.5\%$ rate was tolerated. Because the difference between the two approaches is not systematic, however, averaging over $40$ cavities results in a rather close agreement between the two schemes, \textit{i.e.}, they converge within $\pm0.01$.

\subsection{Swap is not enough}\label{swapisnotenough}
\begin{figure}
\begin{center}
\includegraphics[width=0.6\textwidth]{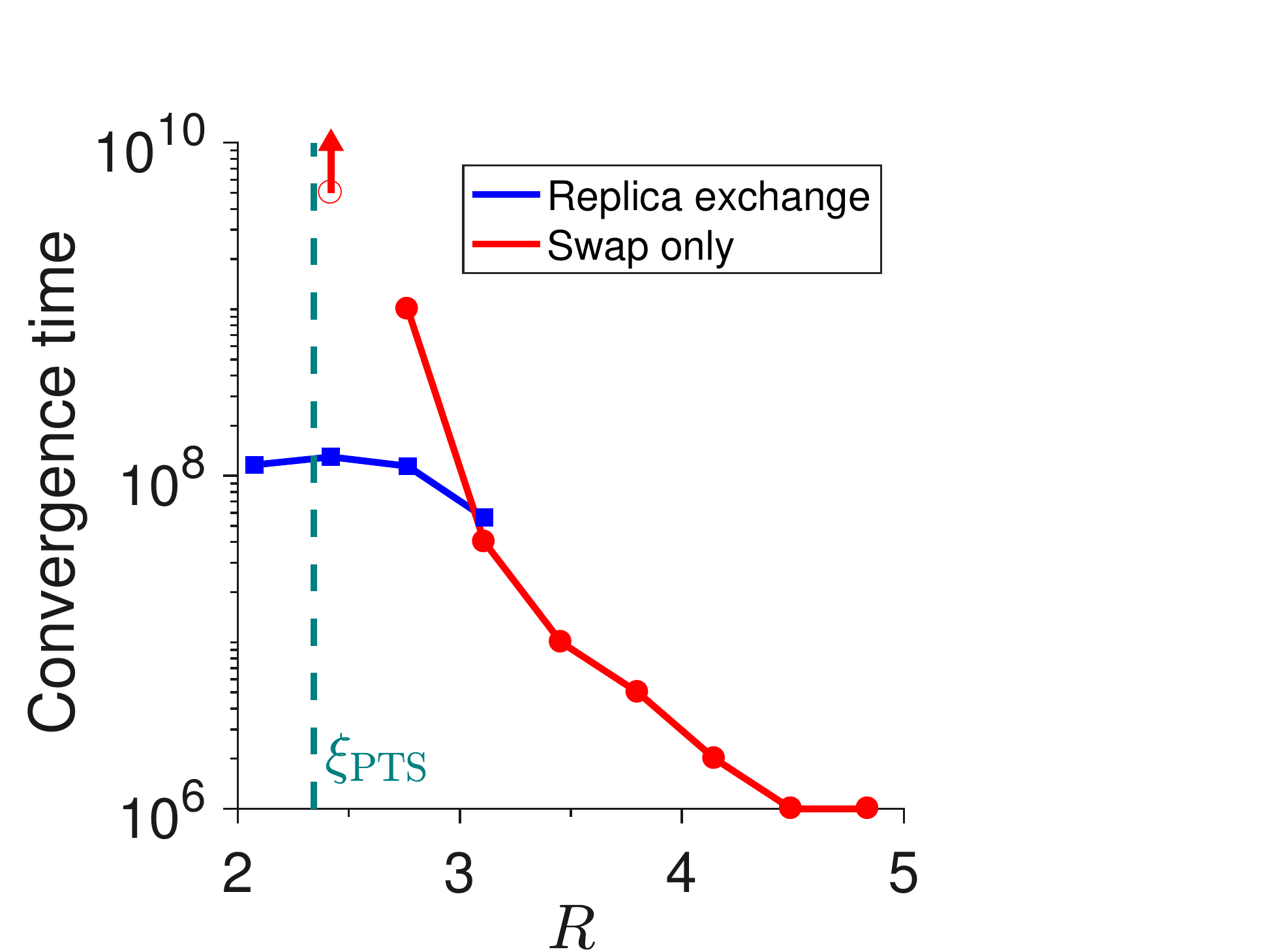}
\caption{Characteristic convergence time (in unit of MC sweeps) multiplied by the number of replicas with (blue-square) and without (red-circle) replica exchange, as a function of cavity size $R$.
Note that the result without replica exchange at $R=2.42$ is an underestimate, because particle swaps alone did not allow the results to converge even after $5\cdot10^{9}$ MC sweeps.
Convergence is also not achieved for $R=1.73$.
}
\label{versus}
\end{center}
\end{figure}
It has been suggested that particle identity swaps by themselves suffice to thermalize cavities, especially small ones with 
$R < \xi_{\rm PTS}$~\cite{BBCGV08,BICtest12}.
We here test this hypothesis for our polydisperse system, for which particle swaps are extremely effective at sampling bulk configurations.
 Figure~\ref{versus} contrasts the computational time needed to equilibrate cavity configurations at $\PF=0.608$, with and without parallel tempering.
Without replica exchange, the equilibration time rapidly grows as the cavity size decreases, which is in line with the suggestion that confinement enhances the breaking of ergodicity~\cite{BCY16}. We observe that when $R$ 
approaches the point-to-set length from above, however, the convergence time 
becomes too long to be measured. Tests performed with even smaller $R$ values
confirm that trend. Convergence is never achieved for cavities of the order
of the point-to-set length or smaller. By contrast, using
replica exchange with swap dynamics keeps convergence time within computational reach even for small cavities. 

This result seems to contrast with the findings of Ref.~\cite{BICtest12}
that the dynamics actually speeds up inside small cavities. We have no 
explanation for this discrepancy 
as we never observed such speed-up in any of our simulations, but we note that the timescale studied in 
Ref.~\cite{BICtest12} is different from the one shown in Fig.~\ref{versus}.
Specifically, whereas we report a convergence time for the overlap inside the 
cavity, Cavagna {\it et al.}~study the time decay of the 
overlap-\textit{fluctuation} auto-correlation function in equilibrium. 
We have not tried to systematically measure this latter auto-correlation because it is already very clear that, according to the former convergence criterion, we cannot properly thermalize small cavities in simulations that use identity swaps alone.
Therefore, we would not access an equilibrium correlation timescale
but a nonequilibrium one.
In addition, no quantitative analysis of the convergence time is provided in Refs.~\cite{BBCGV08,BICtest12}, which makes a direct comparison with our results impossible.

We conclude that one must generally employ a parallel-tempering scheme, as developed in Ref.~\cite{BCY16},
in order to properly sample small cavities and measure point-to-set correlations.

\begin{table}[htbp]
\begin{tabular}{| l | c | c | c | c | c | c | c | c | c | c |}
\hline
\ \ \ $R$ &&\ $1.38$ &\ $1.73$  &\ $2.07$ &\ $2.42$ &\ $2.77$ &\ $3.11$ &\ $3.46$ &\ $3.80$ &\ $4.15$\\
\hline
\hline
\ \ \ $n_{\rm ave}$ && 10 & 8 & 7 & 7 & 6 & 1 & 1 & 1 & 1 \\
\hline
\ \ \ $\lambda_{\rm dec}$ && 0.850 & 0.920 & 0.960 & 0.970 & 0.980 & NA & NA & NA & NA \\
\hline
\ \ \ $t_{\rm rec}$ && 4000 & 4000  & 4000  & 4000  & 4000  & 1000 & 400 & 200 & 200 \\
\hline
\ \ \ $s_{\rm eq}$ && 500 & 500 & 500 & 500 & 500 & 500 & 500 & 500 & 500  \\
\hline
\ \ \ $s_{\rm prod}$ && 2000 & 2000 & 2000 & 2000 & 2000 & 2000 & 2000 & 2000 & 2000 \\
\hline
\end{tabular}
\caption{Cavity PTS measurement parameters $\PF=0.568$. Runs without parallel tempering have $n_{\rm ave}=1$.}
\label{sample1}
\end{table}

\begin{table}[htbp]
\begin{tabular}{| l | c | c | c | c | c | c | c | c | c | c |}
\hline
\ \ \ $R$ &&\ $1.73$  &\ $2.07$ &\ $2.42$ &\ $2.77$ &\ $3.11$ &\ $3.46$ &\ $3.80$ &\ $4.15$ &\ $4.49$\\
\hline
\hline
\ \ \ $n_{\rm ave}$ && 10 & 10 & 10 & 8 & 7 & 1 & 1 & 1 & 1 \\
\hline
\ \ \ $\lambda_{\rm dec}$ && 0.900 & 0.930 & 0.950 & 0.970 & 0.980 & NA & NA & NA & NA \\
\hline
\ \ \ $t_{\rm rec}$ && $2\cdot10^4$ & $2\cdot10^4$ & $2\cdot10^4$ & $10^4$ & 4000 & 2000 & 1000 & 400 & 400 \\
\hline
\ \ \ $s_{\rm eq}$ && 500 & 500 & 500 & 500 & 500 & 500 & 500 & 500 & 500  \\
\hline
\ \ \ $s_{\rm prod}$ && 2000 & 2000 & 2000 & 2000 & 2000 & 2000 & 2000 & 2000 & 2000 \\
\hline
\end{tabular}
\caption{Cavity PTS measurement parameters $\PF=0.587$. Runs without parallel tempering have $n_{\rm ave}=1$.}
\label{sample2}
\end{table}

\begin{table}[htbp]
\begin{tabular}{| l | c | c | c | c | c | c | c | c | c | c |}
\hline
\ \ \ $R$ &&\ $1.90$ &\ $2.25$ &\ $2.59$ &\ $2.94$ &\ $3.11$ &\ $3.46$ &\ $3.80$ &\ $4.15$ &\ $4.49$\\
\hline
\hline
\ \ \ $n_{\rm ave}$ && 12 & 13 & 12 & 11 & 9 & 1 & 1 & 1 & 1 \\
\hline
\ \ \ $\lambda_{\rm dec}$ && 0.900 & 0.920 & 0.940 & 0.960 & 0.970 & NA & NA & NA & NA \\
\hline
\ \ \ $t_{\rm rec}$ && $2\cdot10^4$ & $2\cdot10^4$ & $2\cdot10^4$ & $10^4$ & $10^4$ & $10^4$ & 4000 & 2000 & 1000 \\
\hline
\ \ \ $s_{\rm eq}$ && 500 & 500 & 500 & 500 & 500 & 500 & 500 & 500 & 500  \\
\hline
\ \ \ $s_{\rm prod}$ && 2000 & 2000 & 2000 & 2000 & 2000 & 2000 & 2000 & 2000 & 2000 \\
\hline
\end{tabular}
\caption{Cavity PTS measurement parameters $\PF=0.597$. Runs without parallel tempering have $n_{\rm ave}=1$.}
\label{sample3}
\end{table}

 \begin{table}[htbp]
\begin{tabular}{| l | c | c | c | c | c | c | c | c | c | c |}
\hline
\ \ \ $R$ &&\ $2.07$ &\ $2.42$ &\ $2.77$ &\ $3.11$ &\ $3.46$ &\ $3.80$ &\ $4.15$ &\ $4.49$ &\ $4.84$\\
\hline
\hline
\ \ \ $n_{\rm ave}$ && 13 & 14 & 12 & 12 & 1 & 1 & 1 & 1 & 1 \\
\hline
\ \ \ $\lambda_{\rm dec}$ && 0.910 & 0.920 & 0.950 & 0.960 & NA & NA & NA & NA & NA \\
\hline
\ \ \ $t_{\rm rec}$ && $2\cdot10^4$ & $2\cdot10^4$ & $2\cdot10^4$ & $10^4$ & $2\cdot10^4$ & $10^4$ & 4000 & 2000 & 2000 \\
\hline
\ \ \ $s_{\rm eq}$ && 500 & 500 & 500 & 500 & 500 & 500 & 500 & 500 & 500  \\
\hline
\ \ \ $s_{\rm prod}$ && 2000 & 2000 & 2000 & 2000 & 2000 & 2000 & 2000 & 2000 & 2000 \\
\hline
\end{tabular}
\caption{Cavity PTS measurement parameters $\PF=0.608$. Runs without parallel tempering have $n_{\rm ave}=1$.}
\label{sample4}
\end{table}

\begin{table}[htbp]
\begin{tabular}{| l | c | c | c | c | c | c | c | c | c | c |}
\hline
\ \ \ $R$ &&\ $2.25$ &\ $2.42$ &\ $2.59$  &\ $3.11$  &\ $3.80$ &\ $4.15$ &\ $4.49$ &\ $4.84$\\
\hline
\hline
\ \ \ $n_{\rm ave}$ && 15 & 16 & 18  & 17  & 1 & 1 & 1 & 1 \\
\hline
\ \ \ $\lambda_{\rm dec}$ && 0.900 & 0.910 & 0.910  & 0.940  & NA & NA & NA & NA \\
\hline
\ \ \ $t_{\rm rec}$ && $10^4$ & $10^4$ & $10^4$  & $10^4$  & $5\cdot10^4$ & $3\cdot10^4$ & $10^4$  & 5000  \\
\hline
\ \ \ $s_{\rm eq}$ && 2000 & 2000 & 2000   & 1000  & 500 & 500 & 500 & 500  \\
\hline
\ \ \ $s_{\rm prod}$ && 8000 & 8000 & 8000   & 4000 & 2000 & 2000 & 2000 & 2000 \\
\hline
\end{tabular}
\caption{Cavity PTS measurement parameters $\PF=0.618$. Runs without parallel tempering have $n_{\rm ave}=1$.}
\label{sample5}
\end{table}

\begin{table}[htbp]
\begin{tabular}{| l | c | c | c | c | c | c | c | c |  }
\hline
\ \ \ $R$ &&\  $2.77$ &\ $2.94$ &\ $3.11$ &\ $3.46$ &\ $3.80$ &\ $4.49$ &\ $4.84$\\
\hline
\hline
\ \ \ $n_{\rm ave}$ && 17 & 19 & 18 & 17 & 16 &  1 & 1 \\
\hline
\ \ \ $\lambda_{\rm dec}$ && 0.930 & 0.930 &  0.940 & 0.955 & 0.965 &  NA & NA \\
\hline
\ \ \ $t_{\rm rec}$ && $10^4$ & $10^4$ & $10^4$ & $10^4$ & $10^4$  &  $10^5$ & $4\cdot10^4$ \\
\hline
\ \ \ $s_{\rm eq}$ && 1500 & 3000 & 4000 & 6000 & 8000  & 500 & 500   \\
\hline
\ \ \ $s_{\rm prod}$ && 6000 & 12000 & 16000 & 9000 & 12000  & 2000 & 2000 \\
\hline
\end{tabular}
\caption{Cavity PTS measurement parameters $\PF=0.629$. Runs without parallel tempering have $n_{\rm ave}=1$.}
\label{sample6}
\end{table}
 
\begin{table}[htbp]
\begin{tabular}{| l | c | c | c | c | c | c | c | c |}
\hline
\ \ \ $R$ &&\  $2.77$ &\ $3.11$  &\ $3.46$ &\ $3.63$ &\ $3.80$ &\ $4.84$ &\ $5.18$ \\
\hline
\hline
\ \ \ $n_{\rm ave}$ && 21 & 21  & 22 & 23 & 22 & 1 & 1 \\
\hline
\ \ \ $\lambda_{\rm dec}$ && 0.910 & 0.930  & 0.940 & 0.940 & 0.950 & NA & NA  \\
\hline
\ \ \ $t_{\rm rec}$ && $10^4$ & $10^4$ & $10^4$ & $10^4$ & $10^4$ & $2\times 10^5$ & $10^5$ \\
\hline
\ \ \ $s_{\rm eq}$ && 2000 & 6000 & 8000 & 12000 & 12000 & 1000 & 1000   \\
\hline
\ \ \ $s_{\rm prod}$ && 8000 & 9000  & 12000 & 18000 & 18000 & 4000 & 4000 \\
\hline
\end{tabular}
\caption{Cavity PTS measurement parameters $\PF=0.640$. Runs without parallel tempering have $n_{\rm
ave}=1$. }
\label{sample7}
\end{table}

\section{Method 1 for soft spheres}
\label{sec:soft}
We compute the configurational entropy of a continuous polydisperse 
soft sphere system by the standard definition, 
\begin{equation}
s_{\rm conf}=s_{\rm tot} - s_{\rm vib},
\end{equation}
where $s_{\rm tot}$ and $s_{\rm vib}$ are the total and vibrational entropies~\cite{sciortino1999inherent,Sastry01}.
The mixing entropy is also treated by the strategy used in Ref.~\cite{ozawa2017does};
we compute $s_{\rm tot}$ and $s_{\rm vib}$ as if the system were monodisperse, and then add the effective mixing entropy $s_{\rm mix}^*$ independently determined in simulations~\cite{ozawa2017does}.

\subsection{Model}

We use a continuous size polydispersity, 
where the particle diameter $\sigma$ of each particle is randomly drawn from 
the following particle size distribution: 
$f(\sigma) = A\sigma^{-3}$, for $\sigma \in [ \sigma_{\rm min}, 
\sigma_{\rm max} ]$, 
where $A$ is a normalization constant. 
We define the size polydispersity 
as $\Delta=\sqrt{\overline{\sigma^2} - \overline{\sigma}^2}/\overline{\sigma}$, where $\overline{\cdots}=\int \mathrm{d} \sigma f(\sigma) (\cdots)$.
We use $\Delta = 0.23$, choosing $\sigma_{\rm min} / \sigma_{\rm max} = 0.4492$, and set $\overline{\sigma}$ as the unit length. 
We simulate systems composed of $N$ particles in a cubic 
cell of volume $V$ with periodic boundary conditions in three dimensions, $d=3$. 
We use the following pairwise potential for the soft sphere model,
\begin{eqnarray}
v_{ij}(r) &=& v_0 \left( \frac{\sigma_{ij}}{r} \right)^n + c_0 + c_1 \left( \frac{r}{\sigma_{ij}} \right)^2 + c_2 \left( \frac{r}{\sigma_{ij}} \right)^4, \label{eq:soft_v} \\
\sigma_{ij} &=& \frac{(\sigma_i + \sigma_j)}{2} (1-\epsilon |\sigma_i - \sigma_j|), \label{eq:non_additive}
\end{eqnarray}
where $v_0$ is the unit of energy, and $\epsilon$  
quantifies the degree of non-additivity of the particle 
diameters. Non-additivity conveniently prevents crystallization
and thus enhances the glass-forming ability of the model.
The constants, $c_0$, $c_1$ and $c_2$, are chosen so that the first and 
second derivatives of $v_{ij}(r)$ vanish at the interaction cut-off 
$r_{\rm cut}=1.25 \sigma_{ij}$.
We employ the non-additive soft sphere model 
with parameters $n=12$ and $\epsilon=0.2$, and set the number density to $\rho=N/V=1.0$ for $N=1500$ particles.

Equilibrium configurations are produced by swap MC simulations after high temperature configurations are instantaneously quenched to the target temperature. Equilibration is ensured by the fact that particles lose memory of their initial positions and dynamical observables do not present aging. The absence of crystalline nuclei is also verified. We additionally perform standard MC simulations to obtain relaxation times down to the mode coupling crossover temperature $T_{\rm c}$. Following the procedure described in the Materials and methods section of the main text, we carry out three kinds of fits on these dynamical data. First we employ a power-law fit to extrapolate $T_{\rm c}=0.104$. We then perform a VFT fit with $\delta=1$ and a parabolic fit to estimate the glass-ceiling temperature $T_{\rm g}=0.0720-0.0817$. 

\subsection{Total entropy}

\begin{figure}[htbp]
  \centering
  \includegraphics[width=0.45\linewidth]{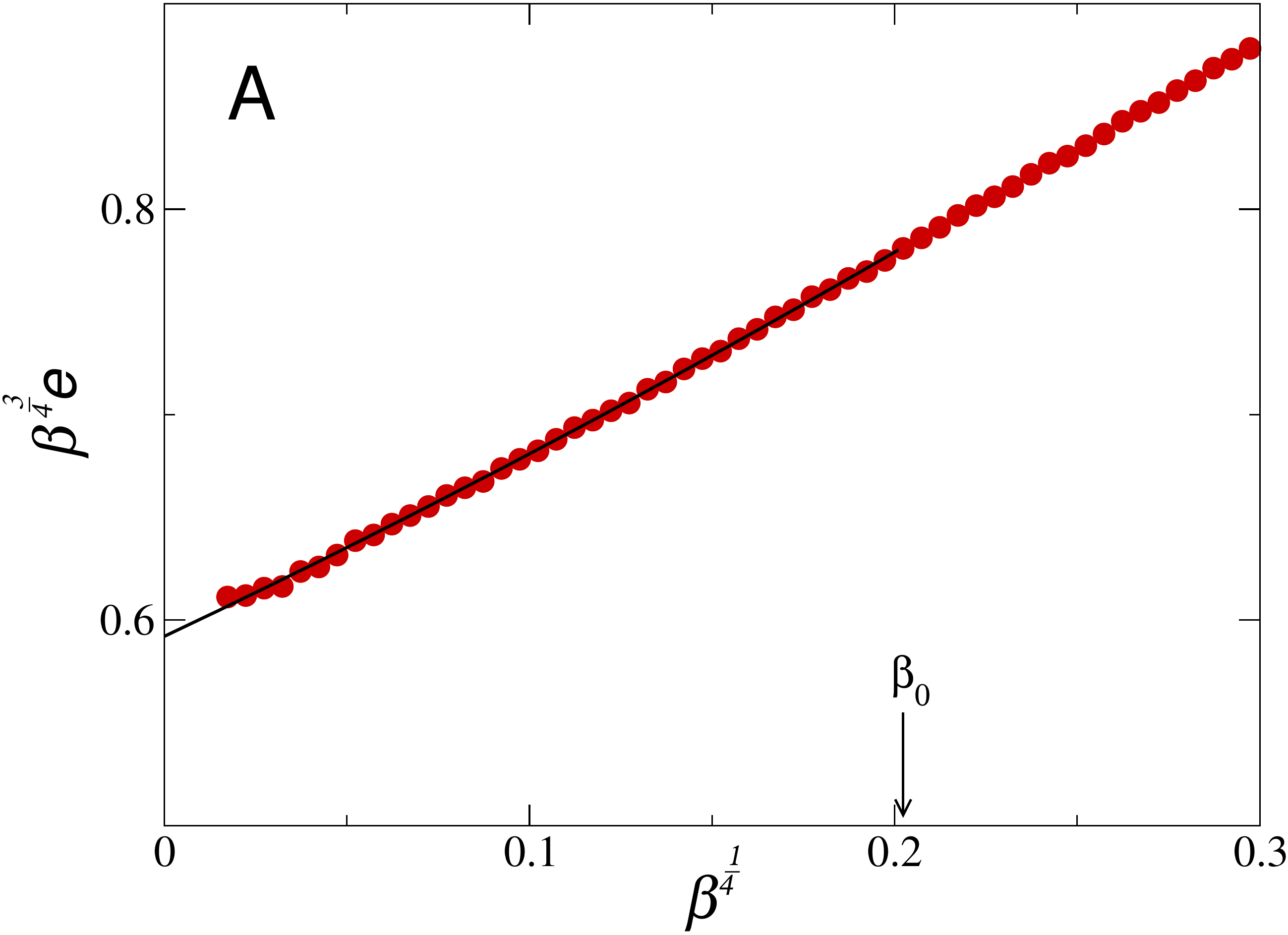}\hspace{0.02\linewidth}
  \includegraphics[width=0.44\linewidth]{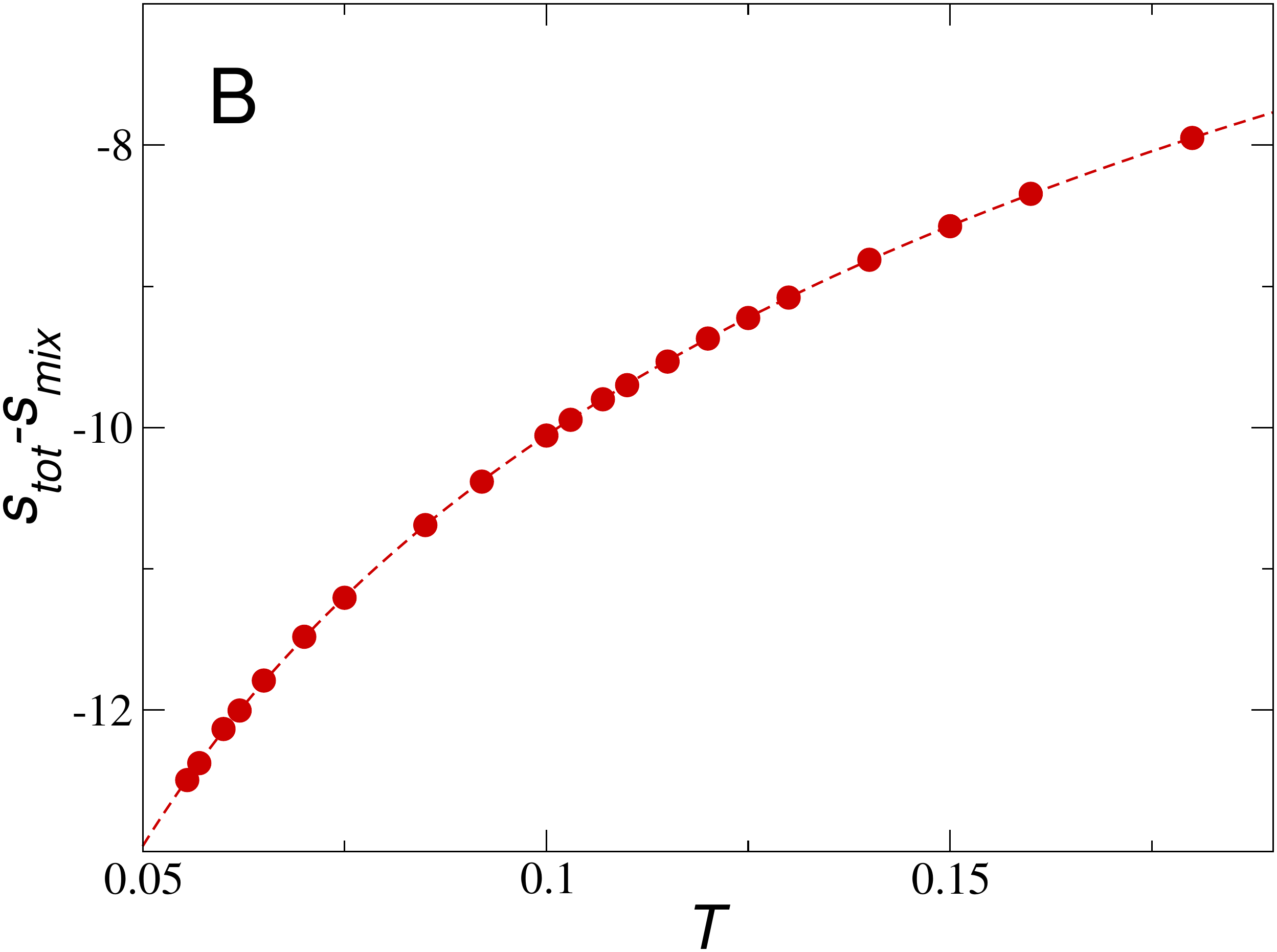}
\caption{
(A) Plot of $\beta^{3/4} e(\beta)$ vs.~$\beta^{1/4}$ to extract the fit parameters, $A$, $B$, and $C$, in Eq.~(\ref{eq:fitting_function}).
The solid line is the resulting fitting curve.
The vertical arrow indicates the position of $\beta_0$.
(B) $s_{\rm tot}-s_{\rm mix}$ as a function of $T$.
The dashed curve is the Rosenfeld-Tarazona expression, $s_{\rm tot}-s_{\rm mix}+3\ln\Lambda \propto T^{-2/5}$, which is well confirmed by our data.
}
\label{fig:s_tot}
\end{figure}  

We perform the thermodynamic integration over the inverse temperature $\beta'$ from the ideal gas limit $\beta' \to 0$ to the target temperature $\beta$,
\begin{equation}
s_{\rm tot}(\beta) = \frac{5}{2} - \ln \rho - 3 \ln \Lambda + \beta e(\beta) - \int_{0}^{\beta} \mathrm{d}\beta' e(\beta') + s_{\rm mix},
\label{eq:s_tot}
\end{equation} 
where $e(\beta)$ is the potential energy of the system and for this system we set $m=1$ and $\hbar=1$ such that the thermal de Broglie wavelength $\Lambda=\sqrt{2\pi \beta}$.

Special care is needed to compute the integral in Eq.~(\ref{eq:s_tot}) because the potential energy $e(\beta)$ diverges in the high temperature limit~\cite{coluzzi2000lennard,ozawa2015equilibrium}.
To accurately calculate the integral, we thus decompose the integration range into a very high temperature regime, $\beta' \in [0, \beta_0]$, and an intermediate regime, 
$\beta' \in (\beta_0, \beta]$, 
with $\beta_0$ the boundary between the two regimes.
Therefore, the integral in Eq.~(\ref{eq:s_tot}) can be decomposed as
\begin{eqnarray}
I &=& \int_{0}^{\beta} \mathrm{d}\beta' e(\beta') = \int_{0}^{\beta_0} \mathrm{d}\beta' e(\beta') + \int_{\beta_0}^{\beta} \mathrm{d}\beta' e(\beta') \\
&=& I_{\rm F} + I_{\rm N},
\label{eq:def_integral}
\end{eqnarray}
where $I_{\rm F}$ and $I_{\rm N}$ are integrals over the very high 
and the intermediate temperature regimes, respectively.
We set $\beta_0=1.68\times 10^{-3}$ in this work. The integral
$I_{\rm N}$ can be performed by usual numerical integration.
To obtain $I_{\rm F}$ we fit the potential energy data to a polynomial function, then analytically
integrate the function, which enables us to avoid the numerical integration of the diverging $e(\beta \to 0)$~\cite{coluzzi2000lennard,ozawa2015equilibrium}.
In a three-dimensional system of particles interacting via $v(r) \propto r^{-12}$, the high temperature expansion of the potential energy reads:
\begin{equation}
e(\beta)=A \beta^{-3/4} + B \beta^{-1/2} + C \beta^{-1/4} + O\left(\beta^0 \right)\, ,
\label{eq:highT_expansion}
\end{equation}
where $A$, $B$, and $C$ are constants.
Using Eqs.~(\ref{eq:def_integral}) and (\ref{eq:highT_expansion}), we get 
\begin{equation}
I_{\rm F} = 4 A \beta_0^{1/4} + 2 B \beta_0^{1/2} + (4/3)C \beta_0^{3/4} +  O\left(\beta_0 \right).
\end{equation}
Therefore, we can compute $I_{\rm F}$ from 
the fitting parameters, $A$, $B$, and $C$.

In order to obtain $A$, $B$, and $C$ by fitting, we rewrite Eq.~(\ref{eq:highT_expansion}) as
\begin{equation}
\beta^{3/4} e(\beta) = A + B \left( \beta^{1/4} \right) + C \left( \beta^{1/4} \right)^2 + O\left( \beta^{3/4} \right),
\label{eq:fitting_function}
\end{equation}
and use a quadratic fit as shown in Fig.~\ref{fig:s_tot}(A).

Figure~\ref{fig:s_tot}(B) shows the results for $s_{\rm tot}-s_{\rm mix}$.
The dashed line corresponds to the Rosenfeld-Tarazona (RT) expression~\cite{rosenfeld1998density} $e(T) \propto T^{3/5}$ or $s_{\rm tot} +3\ln\Lambda \propto T^{-2/5}$, which allows us to extrapolate to lower temperatures. 
Figure~\ref{fig:s_tot}(B) confirms that the RT expression works very well over our simulation range. 

\subsection{Vibrational entropy}

\begin{figure}[htbp]
\begin{center}
  \includegraphics[width=0.432\linewidth]{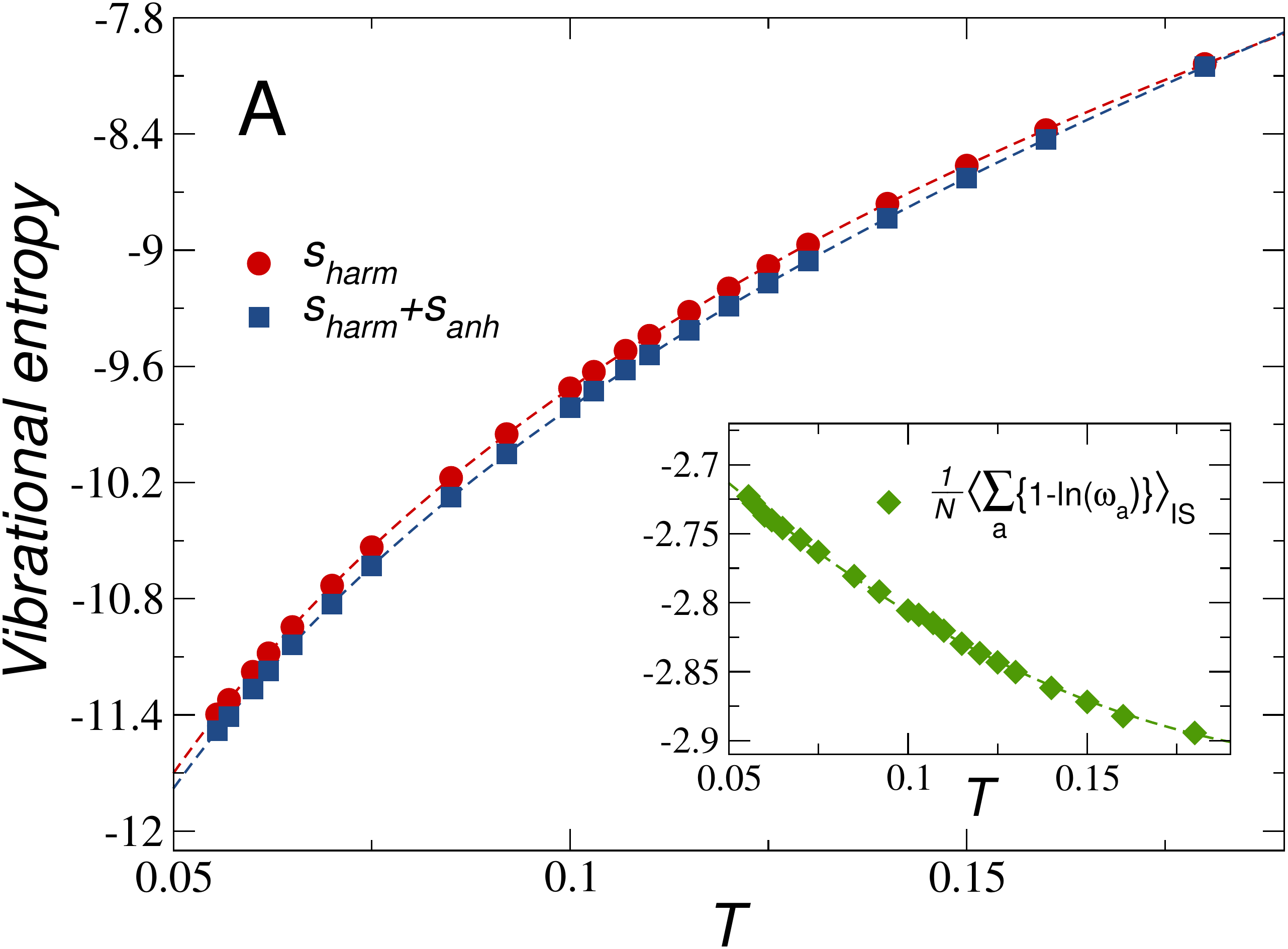}\hspace{0.03\linewidth}
  \includegraphics[width=0.45\linewidth]{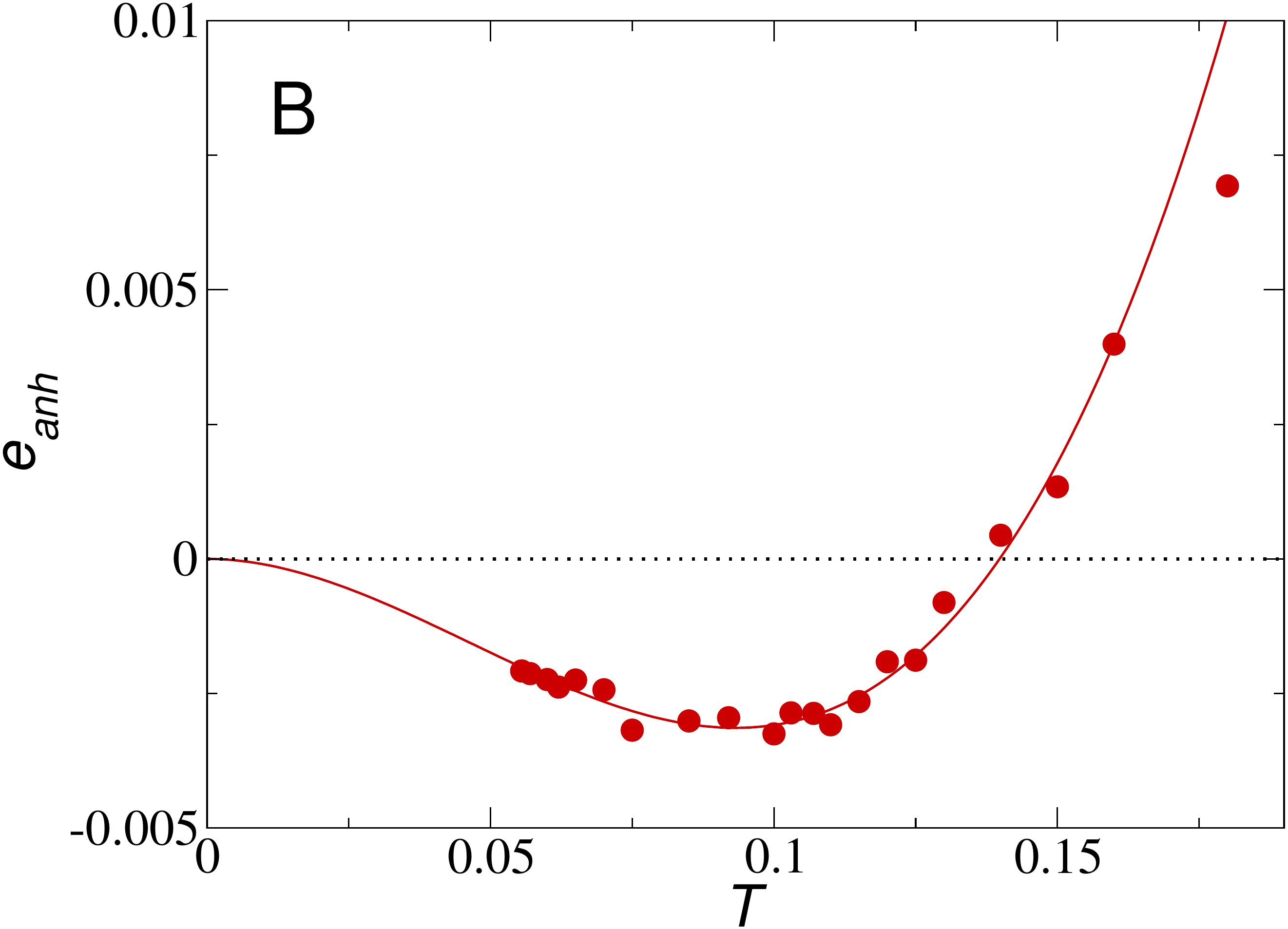}
\caption{
(A) The harmonic vibrational entropy $s_{\rm harm}$ and anharmonic correction $s_{\rm anh}$.
Dashed curves are obtained by quadratic fit to $(1/N)\sum_{a=1}^{3N} \{ 1 - \ln (\omega_a) \}$.
(B) The anharmonic contribution of the potential energy $e_{\rm anh}(T)$.
The solid line is obtained by fitting over a range $T \in [0.0555, 0.125]$
}
\label{fig:s_vib}
\end{center}
\end{figure}  

We compute the vibrational entropy $s_{\rm vib}$ by 
\begin{equation}
s_{\rm vib}=s_{\rm harm} + s_{\rm anh},
\end{equation}
where $s_{\rm harm}$ and $s_{\rm anh}$ are the harmonic vibrational entropy and the anharmonic correction, respectively~\cite{sciortino2005potential}.
We compute $s_{\rm harm}$~\cite{sciortino1999inherent,Sastry01} using
\begin{equation}
s_{\rm harm}(\beta) = \frac{1}{N} \left\langle \sum_{a=1}^{3N} \{1 - \ln (\beta \hbar \omega_a) \} \right\rangle_{\rm IS},
\end{equation}
where $\langle \cdots \rangle_{\rm IS}$ denotes an average over the inherent structure obtained by the conjugate gradient method, and $\omega_a=\sqrt{\lambda_a/m}$, and $\lambda_a$ is the eigenvalue of the Hessian.
In Fig.~\ref{fig:s_vib}(A), we show $s_{\rm harm}$ as a function of $T$.
In order to extrapolate $s_{\rm harm}$ to lower temperatures, we follow the common scheme of fitting $(1/N) \langle \sum_{a=1}^{3N} \{1 - \ln ( \omega_a) \} \rangle_{\rm IS}$ to a second-degree polynomial in $T$~\cite{sciortino1999inherent,flenner2006hybrid}.

We also evaluate the anharmonic contribution $s_{\rm anh}$~\cite{sciortino2005potential} to the potential energy,
\begin{equation}
e_{\rm anh}(T) = e(T) - e_{\rm IS}(T) - \frac{3}{2} T,
\label{eq:U_anh}
\end{equation}
where $e_{\rm IS}$ is the inherent structure energy.
The last term is the energy of the harmonic vibration.
Using $e_{\rm anh}(T)$, $s_{\rm anh}(T)$ is given by
\begin{equation}
s_{\rm anh}(T) = \int_0^T \mathrm{d}T' \frac{1}{T'} \frac{\partial e_{\rm anh}(T')}{\partial T'}.
\label{eq:S_anh}
\end{equation}
when assuming that there is no anharmonic contribution at $T=0$, {\it i.e.}, $s_{\rm anh}(T=0)=0$.
Expanding $e_{\rm anh}(T)$ around zero temperature then gives
\begin{equation}
e_{\rm anh}(T) = \sum_{k=2} a_k  T^k,
\label{eq:U_anh_expand}
\end{equation}
where $a_k $ is a $T$ independent coefficient.
We also assume the linear term in Eq.~(\ref{eq:U_anh_expand}) is zero, $a_1=0$, which means that the anharmonic contribution to the specific heat vanishes at $T=0$.
Substituting Eq.~(\ref{eq:U_anh_expand}) into Eq.~(\ref{eq:S_anh}), we obtain
\begin{equation}
s_{\rm anh}(T) = \sum_{k=2} \frac{k}{k-1} a_k T^{k-1}.
\label{eq:S_anh2}
\end{equation}

In Fig.~\ref{fig:s_vib}(B), we show a fit of $e_{\rm anh}$ using $a_2$ and $a_3$, and in Fig.~\ref{fig:s_vib}(A) we show $s_{\rm harm} + s_{\rm anh}$.
We find that the anharmonic contribution is small,  $|s_{\rm anh}| < 0.08$, over our range of interest, and thus that the details of the above procedure 
have little influence on the final estimate of $s_{\rm vib}$.

\subsection{Mixing entropy}

\begin{figure}[htbp]
\begin{center}
  \includegraphics[width=0.6\linewidth]{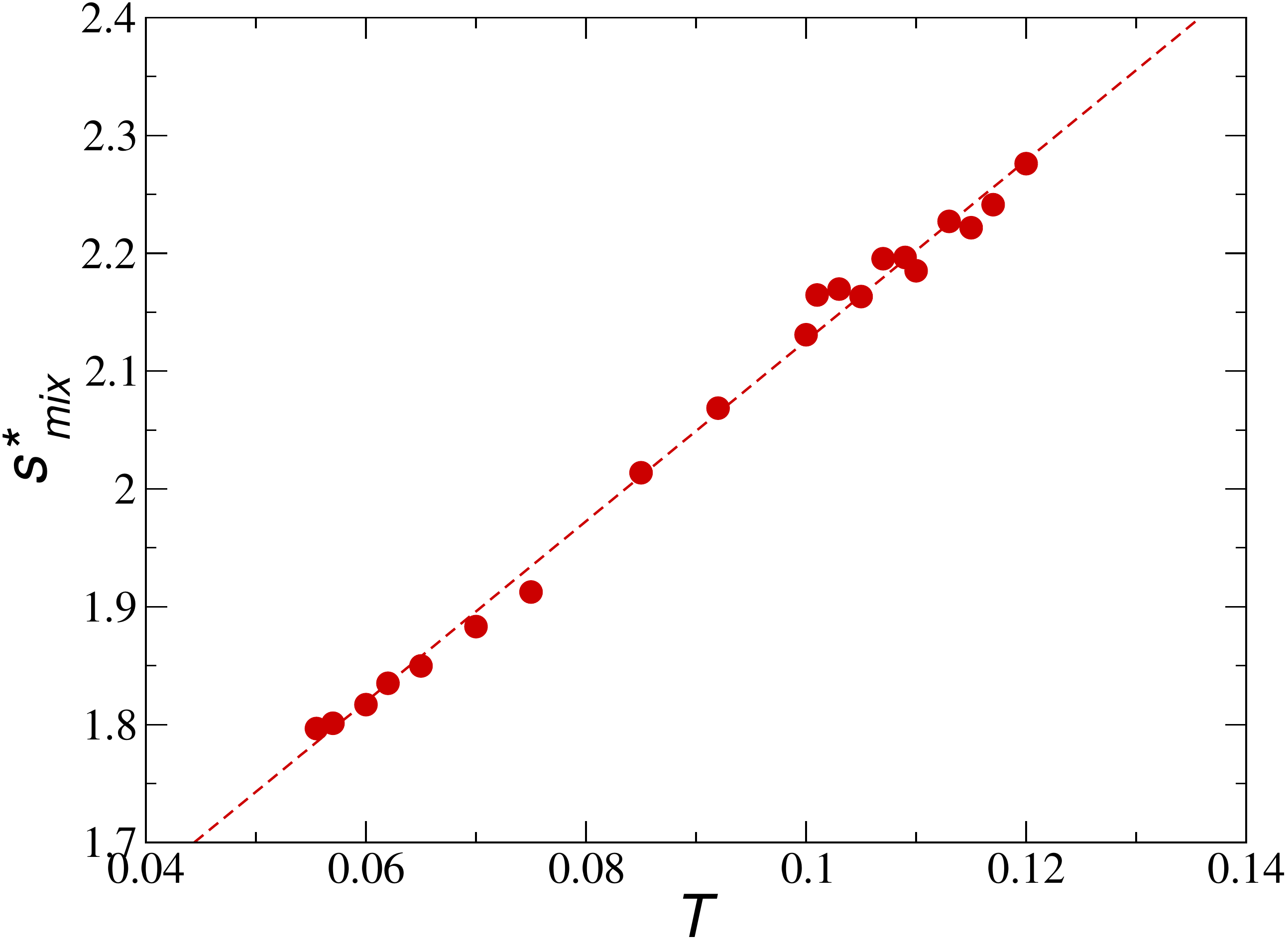}
\caption{
The effective mixing entropy $s_{\rm mix}^*$.
The dashed line is an empirical linear fit.}
\label{fig:s_mix2}
\end{center}
\end{figure}  

We include the effective mixing entropy $s_{\rm mix}^*$ obtained by an independent set of simulations in Ref.~\cite{ozawa2017does}.
In contrast to the hard sphere system, $s_{\rm mix}^*$ of the soft sphere system has a weakly linear temperature dependence on the control parameter as shown in Fig.~\ref{fig:s_mix2}.
A linear fit, $s_{\rm mix}^* = b_0 + b_1 T$, gives $b_0=1.3601$ and $b_1=7.6565$.

\subsection{Configurational entropy}

\begin{figure}[htbp]
\begin{center}
  \includegraphics[width=0.6\linewidth]{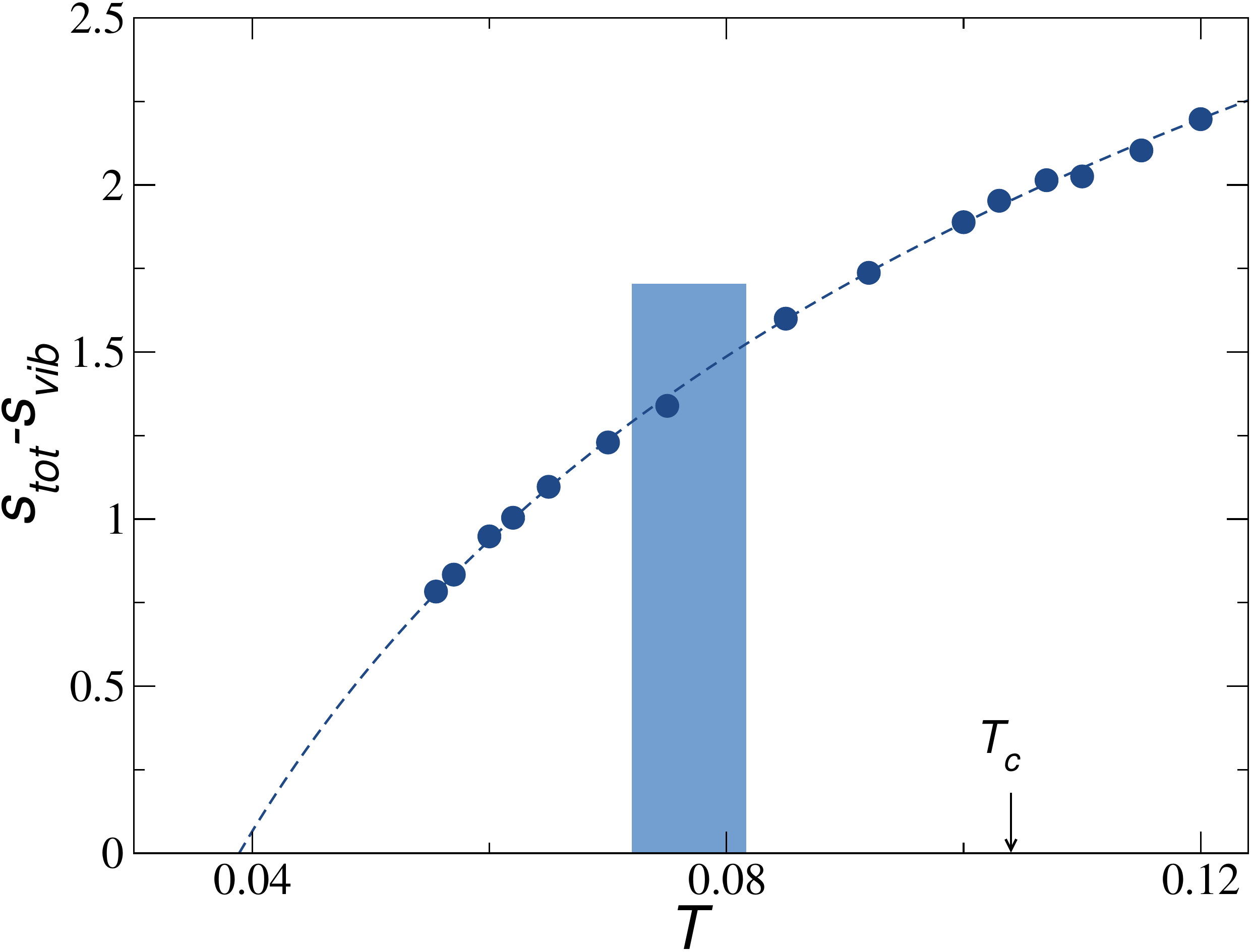}
\caption{The configurational entropy of soft spheres, $s_{\rm conf}=s_{\rm tot}-s_{\rm vib}$. The glass ceiling is indicated in the blue region and the dashed curve is an extrapolation based on separately fitting the individual terms.
The mode-coupling crossover temperature $T_{\rm c}$ is denoted by a vertical arrow.} 
\label{fig:s_conf}
\end{center}
\end{figure}  

The resulting configurational entropy $s_{\rm conf}$ is shown in Fig.~\ref{fig:s_conf}.
The data demonstrates that $s_{\rm conf}$ decreases 
further below the experimental glass transition, and does not 
show any sign of a crossover, bending, or saturation. 
Also, combining the extrapolations for $s_{\rm tot}$, $s_{\rm harm}$, $s_{\rm anh}$, and $s_{\rm mix}^*$, we extrapolate $s_{\rm conf}$ down to zero, and estimate the Kauzmann transition to be around $T_{\rm K} \simeq 0.04$.
We conclude from this analysis that the 
results reported in the main text for hard spheres
are not specific to this interaction potential. Our methods thus apply equally well to 
models of supercooled liquids characterized by continuous pair 
potentials.

\end{widetext}

\bibliography{paper.bib}

\begin{thebibliography}{79}%
\makeatletter
\providecommand \@ifxundefined [1]{%
 \@ifx{#1\undefined}
}%
\providecommand \@ifnum [1]{%
 \ifnum #1\expandafter \@firstoftwo
 \else \expandafter \@secondoftwo
 \fi
}%
\providecommand \@ifx [1]{%
 \ifx #1\expandafter \@firstoftwo
 \else \expandafter \@secondoftwo
 \fi
}%
\providecommand \natexlab [1]{#1}%
\providecommand \enquote  [1]{``#1''}%
\providecommand \bibnamefont  [1]{#1}%
\providecommand \bibfnamefont [1]{#1}%
\providecommand \citenamefont [1]{#1}%
\providecommand \href@noop [0]{\@secondoftwo}%
\providecommand \href [0]{\begingroup \@sanitize@url \@href}%
\providecommand \@href[1]{\@@startlink{#1}\@@href}%
\providecommand \@@href[1]{\endgroup#1\@@endlink}%
\providecommand \@sanitize@url [0]{\catcode `\\12\catcode `\$12\catcode
  `\&12\catcode `\#12\catcode `\^12\catcode `\_12\catcode `\%12\relax}%
\providecommand \@@startlink[1]{}%
\providecommand \@@endlink[0]{}%
\providecommand \url  [0]{\begingroup\@sanitize@url \@url }%
\providecommand \@url [1]{\endgroup\@href {#1}{\urlprefix }}%
\providecommand \urlprefix  [0]{URL }%
\providecommand \Eprint [0]{\href }%
\providecommand \doibase [0]{http://dx.doi.org/}%
\providecommand \selectlanguage [0]{\@gobble}%
\providecommand \bibinfo  [0]{\@secondoftwo}%
\providecommand \bibfield  [0]{\@secondoftwo}%
\providecommand \translation [1]{[#1]}%
\providecommand \BibitemOpen [0]{}%
\providecommand \bibitemStop [0]{}%
\providecommand \bibitemNoStop [0]{.\EOS\space}%
\providecommand \EOS [0]{\spacefactor3000\relax}%
\providecommand \BibitemShut  [1]{\csname bibitem#1\endcsname}%
\let\auto@bib@innerbib\@empty
\bibitem [{\citenamefont {Kauzmann}(1948)}]{Kauzmann48}%
  \BibitemOpen
  \bibfield  {author} {\bibinfo {author} {\bibfnamefont {W.}~\bibnamefont
  {Kauzmann}},\ }\href@noop {} {\bibfield  {journal} {\bibinfo  {journal}
  {Chem. Rev.}\ }\textbf {\bibinfo {volume} {43}},\ \bibinfo {pages} {219}
  (\bibinfo {year} {1948})}\BibitemShut {NoStop}%
\bibitem [{\citenamefont {Kirkpatrick}\ and\ \citenamefont
  {Thirumalai}(1987)}]{KT87}%
  \BibitemOpen
  \bibfield  {author} {\bibinfo {author} {\bibfnamefont {T.~R.}\ \bibnamefont
  {Kirkpatrick}}\ and\ \bibinfo {author} {\bibfnamefont {D.}~\bibnamefont
  {Thirumalai}},\ }\href@noop {} {\bibfield  {journal} {\bibinfo  {journal}
  {Phys. Rev. B}\ }\textbf {\bibinfo {volume} {36}},\ \bibinfo {pages} {5388}
  (\bibinfo {year} {1987})}\BibitemShut {NoStop}%
\bibitem [{\citenamefont {Kirkpatrick}\ and\ \citenamefont
  {Wolynes}(1987)}]{KW87}%
  \BibitemOpen
  \bibfield  {author} {\bibinfo {author} {\bibfnamefont {T.~R.}\ \bibnamefont
  {Kirkpatrick}}\ and\ \bibinfo {author} {\bibfnamefont {P.~G.}\ \bibnamefont
  {Wolynes}},\ }\href@noop {} {\bibfield  {journal} {\bibinfo  {journal} {Phys.
  Rev. B}\ }\textbf {\bibinfo {volume} {36}},\ \bibinfo {pages} {8552}
  (\bibinfo {year} {1987})}\BibitemShut {NoStop}%
\bibitem [{\citenamefont {Charbonneau}\ \emph
  {et~al.}(2014{\natexlab{a}})\citenamefont {Charbonneau}, \citenamefont
  {Kurchan}, \citenamefont {Parisi}, \citenamefont {Urbani},\ and\
  \citenamefont {Zamponi}}]{CKPUZ14}%
  \BibitemOpen
  \bibfield  {author} {\bibinfo {author} {\bibfnamefont {P.}~\bibnamefont
  {Charbonneau}}, \bibinfo {author} {\bibfnamefont {J.}~\bibnamefont
  {Kurchan}}, \bibinfo {author} {\bibfnamefont {G.}~\bibnamefont {Parisi}},
  \bibinfo {author} {\bibfnamefont {P.}~\bibnamefont {Urbani}}, \ and\ \bibinfo
  {author} {\bibfnamefont {F.}~\bibnamefont {Zamponi}},\ }\href@noop {}
  {\bibfield  {journal} {\bibinfo  {journal} {Nat. Commun.}\ }\textbf {\bibinfo
  {volume} {5}},\ \bibinfo {pages} {3725} (\bibinfo {year}
  {2014}{\natexlab{a}})}\BibitemShut {NoStop}%
\bibitem [{\citenamefont {Berthier}\ and\ \citenamefont {Biroli}(2011)}]{BB11}%
  \BibitemOpen
  \bibfield  {author} {\bibinfo {author} {\bibfnamefont {L.}~\bibnamefont
  {Berthier}}\ and\ \bibinfo {author} {\bibfnamefont {G.}~\bibnamefont
  {Biroli}},\ }\href@noop {} {\bibfield  {journal} {\bibinfo  {journal} {Rev.
  Mod. Phys.}\ }\textbf {\bibinfo {volume} {83}},\ \bibinfo {pages} {587}
  (\bibinfo {year} {2011})}\BibitemShut {NoStop}%
\bibitem [{\citenamefont {Kirkpatrick}\ \emph {et~al.}(1989)\citenamefont
  {Kirkpatrick}, \citenamefont {Thirumalai},\ and\ \citenamefont
  {Wolynes}}]{KTW89}%
  \BibitemOpen
  \bibfield  {author} {\bibinfo {author} {\bibfnamefont {T.~R.}\ \bibnamefont
  {Kirkpatrick}}, \bibinfo {author} {\bibfnamefont {D.}~\bibnamefont
  {Thirumalai}}, \ and\ \bibinfo {author} {\bibfnamefont {P.~G.}\ \bibnamefont
  {Wolynes}},\ }\href@noop {} {\bibfield  {journal} {\bibinfo  {journal} {Phys.
  Rev. A}\ }\textbf {\bibinfo {volume} {40}},\ \bibinfo {pages} {1045}
  (\bibinfo {year} {1989})}\BibitemShut {NoStop}%
\bibitem [{\citenamefont {Stillinger}(1988)}]{stillinger_supercooled_1988}%
  \BibitemOpen
  \bibfield  {author} {\bibinfo {author} {\bibfnamefont {F.~H.}\ \bibnamefont
  {Stillinger}},\ }\href@noop {} {\bibfield  {journal} {\bibinfo  {journal} {J.
  Chem. Phys.}\ }\textbf {\bibinfo {volume} {88}},\ \bibinfo {pages} {7818}
  (\bibinfo {year} {1988})}\BibitemShut {NoStop}%
\bibitem [{\citenamefont {Garrahan}\ and\ \citenamefont
  {Chandler}(2003)}]{GC03}%
  \BibitemOpen
  \bibfield  {author} {\bibinfo {author} {\bibfnamefont {J.~P.}\ \bibnamefont
  {Garrahan}}\ and\ \bibinfo {author} {\bibfnamefont {D.}~\bibnamefont
  {Chandler}},\ }\href@noop {} {\bibfield  {journal} {\bibinfo  {journal}
  {Proc. Natl. Acad. Sci. U.S.A.}\ }\textbf {\bibinfo {volume} {100}},\
  \bibinfo {pages} {9710} (\bibinfo {year} {2003})}\BibitemShut {NoStop}%
\bibitem [{\citenamefont {Dyre}(2006)}]{Dyre}%
  \BibitemOpen
  \bibfield  {author} {\bibinfo {author} {\bibfnamefont {J.}~\bibnamefont
  {Dyre}},\ }\href@noop {} {\bibfield  {journal} {\bibinfo  {journal} {Rev.
  Mod. Phys.}\ }\textbf {\bibinfo {volume} {78}},\ \bibinfo {pages} {953}
  (\bibinfo {year} {2006})}\BibitemShut {NoStop}%
\bibitem [{\citenamefont {Wyart}(2010)}]{Wyart}%
  \BibitemOpen
  \bibfield  {author} {\bibinfo {author} {\bibfnamefont {M.}~\bibnamefont
  {Wyart}},\ }\href@noop {} {\bibfield  {journal} {\bibinfo  {journal} {Phys.
  Rev. Lett.}\ }\textbf {\bibinfo {volume} {104}},\ \bibinfo {pages} {095901}
  (\bibinfo {year} {2010})}\BibitemShut {NoStop}%
\bibitem [{\citenamefont {Rabochiy}\ \emph
  {et~al.}(2013{\natexlab{a}})\citenamefont {Rabochiy}, \citenamefont
  {Wolynes},\ and\ \citenamefont {Lubchenko}}]{Wolynes}%
  \BibitemOpen
  \bibfield  {author} {\bibinfo {author} {\bibfnamefont {P.}~\bibnamefont
  {Rabochiy}}, \bibinfo {author} {\bibfnamefont {P.~G.}\ \bibnamefont
  {Wolynes}}, \ and\ \bibinfo {author} {\bibfnamefont {V.}~\bibnamefont
  {Lubchenko}},\ }\href@noop {} {\bibfield  {journal} {\bibinfo  {journal} {J.
  Phys. Chem. B}\ }\textbf {\bibinfo {volume} {117}},\ \bibinfo {pages} {15204}
  (\bibinfo {year} {2013}{\natexlab{a}})}\BibitemShut {NoStop}%
\bibitem [{\citenamefont {Sciortino}\ \emph {et~al.}(1999)\citenamefont
  {Sciortino}, \citenamefont {Kob},\ and\ \citenamefont
  {Tartaglia}}]{sciortino1999inherent}%
  \BibitemOpen
  \bibfield  {author} {\bibinfo {author} {\bibfnamefont {F.}~\bibnamefont
  {Sciortino}}, \bibinfo {author} {\bibfnamefont {W.}~\bibnamefont {Kob}}, \
  and\ \bibinfo {author} {\bibfnamefont {P.}~\bibnamefont {Tartaglia}},\
  }\href@noop {} {\bibfield  {journal} {\bibinfo  {journal} {Phys. Rev. Lett.}\
  }\textbf {\bibinfo {volume} {83}},\ \bibinfo {pages} {3214} (\bibinfo {year}
  {1999})}\BibitemShut {NoStop}%
\bibitem [{\citenamefont {Berthier}\ and\ \citenamefont
  {Coslovich}(2014)}]{BC14}%
  \BibitemOpen
  \bibfield  {author} {\bibinfo {author} {\bibfnamefont {L.}~\bibnamefont
  {Berthier}}\ and\ \bibinfo {author} {\bibfnamefont {D.}~\bibnamefont
  {Coslovich}},\ }\href@noop {} {\bibfield  {journal} {\bibinfo  {journal}
  {Proc. Natl. Acad. Sci. U.S.A.}\ }\textbf {\bibinfo {volume} {111}},\
  \bibinfo {pages} {11668} (\bibinfo {year} {2014})}\BibitemShut {NoStop}%
\bibitem [{\citenamefont {Petzold}\ \emph {et~al.}(2013)\citenamefont
  {Petzold}, \citenamefont {Schmidtke}, \citenamefont {Kahlau}, \citenamefont
  {Bock}, \citenamefont {Meier}, \citenamefont {Micko}, \citenamefont {Kruk},\
  and\ \citenamefont {R{\"o}ssler}}]{rossler}%
  \BibitemOpen
  \bibfield  {author} {\bibinfo {author} {\bibfnamefont {N.}~\bibnamefont
  {Petzold}}, \bibinfo {author} {\bibfnamefont {B.}~\bibnamefont {Schmidtke}},
  \bibinfo {author} {\bibfnamefont {R.}~\bibnamefont {Kahlau}}, \bibinfo
  {author} {\bibfnamefont {D.}~\bibnamefont {Bock}}, \bibinfo {author}
  {\bibfnamefont {R.}~\bibnamefont {Meier}}, \bibinfo {author} {\bibfnamefont
  {B.}~\bibnamefont {Micko}}, \bibinfo {author} {\bibfnamefont
  {D.}~\bibnamefont {Kruk}}, \ and\ \bibinfo {author} {\bibfnamefont
  {E.}~\bibnamefont {R{\"o}ssler}},\ }\href@noop {} {\bibfield  {journal}
  {\bibinfo  {journal} {J. Chem. Phys.}\ }\textbf {\bibinfo {volume} {138}},\
  \bibinfo {pages} {12A510} (\bibinfo {year} {2013})}\BibitemShut {NoStop}%
\bibitem [{\citenamefont {Brambilla}\ \emph {et~al.}(2009)\citenamefont
  {Brambilla}, \citenamefont {El~Masri}, \citenamefont {Pierno}, \citenamefont
  {Berthier}, \citenamefont {Cipelletti}, \citenamefont {Petekidis},\ and\
  \citenamefont {Schofield}}]{BEPPSBC08}%
  \BibitemOpen
  \bibfield  {author} {\bibinfo {author} {\bibfnamefont {G.}~\bibnamefont
  {Brambilla}}, \bibinfo {author} {\bibfnamefont {D.}~\bibnamefont {El~Masri}},
  \bibinfo {author} {\bibfnamefont {M.}~\bibnamefont {Pierno}}, \bibinfo
  {author} {\bibfnamefont {L.}~\bibnamefont {Berthier}}, \bibinfo {author}
  {\bibfnamefont {L.}~\bibnamefont {Cipelletti}}, \bibinfo {author}
  {\bibfnamefont {G.}~\bibnamefont {Petekidis}}, \ and\ \bibinfo {author}
  {\bibfnamefont {A.~B.}\ \bibnamefont {Schofield}},\ }\href@noop {} {\bibfield
   {journal} {\bibinfo  {journal} {Phys. Rev. Lett.}\ }\textbf {\bibinfo
  {volume} {102}},\ \bibinfo {pages} {085703} (\bibinfo {year}
  {2009})}\BibitemShut {NoStop}%
\bibitem [{\citenamefont {Barrat}\ \emph {et~al.}(1990)\citenamefont {Barrat},
  \citenamefont {Roux},\ and\ \citenamefont {Hansen}}]{JL90}%
  \BibitemOpen
  \bibfield  {author} {\bibinfo {author} {\bibfnamefont {J.-L.}\ \bibnamefont
  {Barrat}}, \bibinfo {author} {\bibfnamefont {J.-N.}\ \bibnamefont {Roux}}, \
  and\ \bibinfo {author} {\bibfnamefont {J.-P.}\ \bibnamefont {Hansen}},\
  }\href@noop {} {\bibfield  {journal} {\bibinfo  {journal} {Chem. Phys.}\
  }\textbf {\bibinfo {volume} {149}},\ \bibinfo {pages} {197} (\bibinfo {year}
  {1990})}\BibitemShut {NoStop}%
\bibitem [{\citenamefont {Kob}\ and\ \citenamefont {Andersen}(1994)}]{KA94}%
  \BibitemOpen
  \bibfield  {author} {\bibinfo {author} {\bibfnamefont {W.}~\bibnamefont
  {Kob}}\ and\ \bibinfo {author} {\bibfnamefont {H.~C.}\ \bibnamefont
  {Andersen}},\ }\href@noop {} {\bibfield  {journal} {\bibinfo  {journal}
  {Phys. Rev. Lett.}\ }\textbf {\bibinfo {volume} {73}},\ \bibinfo {pages}
  {1376} (\bibinfo {year} {1994})}\BibitemShut {NoStop}%
\bibitem [{\citenamefont {Martinez}\ and\ \citenamefont
  {Angell}(2001)}]{martinez2001thermodynamic}%
  \BibitemOpen
  \bibfield  {author} {\bibinfo {author} {\bibfnamefont {L.-M.}\ \bibnamefont
  {Martinez}}\ and\ \bibinfo {author} {\bibfnamefont {C.}~\bibnamefont
  {Angell}},\ }\href@noop {} {\bibfield  {journal} {\bibinfo  {journal}
  {Nature}\ }\textbf {\bibinfo {volume} {410}},\ \bibinfo {pages} {663}
  (\bibinfo {year} {2001})}\BibitemShut {NoStop}%
\bibitem [{\citenamefont {Grigera}\ and\ \citenamefont {Parisi}(2001)}]{GP01}%
  \BibitemOpen
  \bibfield  {author} {\bibinfo {author} {\bibfnamefont {T.~S.}\ \bibnamefont
  {Grigera}}\ and\ \bibinfo {author} {\bibfnamefont {G.}~\bibnamefont
  {Parisi}},\ }\href@noop {} {\bibfield  {journal} {\bibinfo  {journal} {Phys.
  Rev. E}\ }\textbf {\bibinfo {volume} {63}},\ \bibinfo {pages} {045102}
  (\bibinfo {year} {2001})}\BibitemShut {NoStop}%
\bibitem [{\citenamefont {Berthier}\ \emph
  {et~al.}(2016{\natexlab{a}})\citenamefont {Berthier}, \citenamefont
  {Coslovich}, \citenamefont {Ninarello},\ and\ \citenamefont
  {Ozawa}}]{BCNO16}%
  \BibitemOpen
  \bibfield  {author} {\bibinfo {author} {\bibfnamefont {L.}~\bibnamefont
  {Berthier}}, \bibinfo {author} {\bibfnamefont {D.}~\bibnamefont {Coslovich}},
  \bibinfo {author} {\bibfnamefont {A.}~\bibnamefont {Ninarello}}, \ and\
  \bibinfo {author} {\bibfnamefont {M.}~\bibnamefont {Ozawa}},\ }\href@noop {}
  {\bibfield  {journal} {\bibinfo  {journal} {Phys. Rev. Lett.}\ }\textbf
  {\bibinfo {volume} {116}},\ \bibinfo {pages} {238002} (\bibinfo {year}
  {2016}{\natexlab{a}})}\BibitemShut {NoStop}%
\bibitem [{\citenamefont {Ninarello}\ \emph
  {et~al.}(2017{\natexlab{a}})\citenamefont {Ninarello}, \citenamefont
  {Berthier},\ and\ \citenamefont {Coslovich}}]{PRX17}%
  \BibitemOpen
  \bibfield  {author} {\bibinfo {author} {\bibfnamefont {A.}~\bibnamefont
  {Ninarello}}, \bibinfo {author} {\bibfnamefont {L.}~\bibnamefont {Berthier}},
  \ and\ \bibinfo {author} {\bibfnamefont {D.}~\bibnamefont {Coslovich}},\
  }\href@noop {} {\bibfield  {journal} {\bibinfo  {journal} {Phys. Rev. X}\
  }\textbf {\bibinfo {volume} {7}},\ \bibinfo {pages} {021039} (\bibinfo {year}
  {2017}{\natexlab{a}})}\BibitemShut {NoStop}%
\bibitem [{\citenamefont {Angelani}\ and\ \citenamefont
  {Foffi}(2007)}]{angelani2007configurational}%
  \BibitemOpen
  \bibfield  {author} {\bibinfo {author} {\bibfnamefont {L.}~\bibnamefont
  {Angelani}}\ and\ \bibinfo {author} {\bibfnamefont {G.}~\bibnamefont
  {Foffi}},\ }\href@noop {} {\bibfield  {journal} {\bibinfo  {journal} {J.
  Phys. Cond. Matter}\ }\textbf {\bibinfo {volume} {19}},\ \bibinfo {pages}
  {256207} (\bibinfo {year} {2007})}\BibitemShut {NoStop}%
\bibitem [{\citenamefont {Berthier}\ \emph
  {et~al.}(2016{\natexlab{b}})\citenamefont {Berthier}, \citenamefont
  {Charbonneau},\ and\ \citenamefont {Yaida}}]{BCY16}%
  \BibitemOpen
  \bibfield  {author} {\bibinfo {author} {\bibfnamefont {L.}~\bibnamefont
  {Berthier}}, \bibinfo {author} {\bibfnamefont {P.}~\bibnamefont
  {Charbonneau}}, \ and\ \bibinfo {author} {\bibfnamefont {S.}~\bibnamefont
  {Yaida}},\ }\href@noop {} {\bibfield  {journal} {\bibinfo  {journal} {J.
  Chem. Phys.}\ }\textbf {\bibinfo {volume} {144}},\ \bibinfo {pages} {024501}
  (\bibinfo {year} {2016}{\natexlab{b}})}\BibitemShut {NoStop}%
\bibitem [{\citenamefont {Pusey}\ and\ \citenamefont
  {Van~Megen}(1986)}]{pusey_phase_1986}%
  \BibitemOpen
  \bibfield  {author} {\bibinfo {author} {\bibfnamefont {P.}~\bibnamefont
  {Pusey}}\ and\ \bibinfo {author} {\bibfnamefont {W.}~\bibnamefont
  {Van~Megen}},\ }\href@noop {} {\bibfield  {journal} {\bibinfo  {journal}
  {Nature}\ }\textbf {\bibinfo {volume} {320}},\ \bibinfo {pages} {340}
  (\bibinfo {year} {1986})}\BibitemShut {NoStop}%
\bibitem [{\citenamefont {Berthier}\ and\ \citenamefont
  {Witten}(2009)}]{berthier_glass_2009}%
  \BibitemOpen
  \bibfield  {author} {\bibinfo {author} {\bibfnamefont {L.}~\bibnamefont
  {Berthier}}\ and\ \bibinfo {author} {\bibfnamefont {T.~A.}\ \bibnamefont
  {Witten}},\ }\href@noop {} {\bibfield  {journal} {\bibinfo  {journal} {Phys.
  Rev. E}\ }\textbf {\bibinfo {volume} {80}},\ \bibinfo {pages} {021502}
  (\bibinfo {year} {2009})}\BibitemShut {NoStop}%
\bibitem [{\citenamefont {Stickel}\ \emph {et~al.}(1995)\citenamefont
  {Stickel}, \citenamefont {Fischer},\ and\ \citenamefont
  {Richert}}]{stickel95}%
  \BibitemOpen
  \bibfield  {author} {\bibinfo {author} {\bibfnamefont {F.}~\bibnamefont
  {Stickel}}, \bibinfo {author} {\bibfnamefont {E.~W.}\ \bibnamefont
  {Fischer}}, \ and\ \bibinfo {author} {\bibfnamefont {R.}~\bibnamefont
  {Richert}},\ }\href@noop {} {\bibfield  {journal} {\bibinfo  {journal} {J.
  Chem. Phys.}\ }\textbf {\bibinfo {volume} {102}},\ \bibinfo {pages} {6251}
  (\bibinfo {year} {1995})}\BibitemShut {NoStop}%
\bibitem [{\citenamefont {Frenkel}\ and\ \citenamefont
  {Ladd}(1984)}]{frenkel1984new}%
  \BibitemOpen
  \bibfield  {author} {\bibinfo {author} {\bibfnamefont {D.}~\bibnamefont
  {Frenkel}}\ and\ \bibinfo {author} {\bibfnamefont {A.~J.}\ \bibnamefont
  {Ladd}},\ }\href@noop {} {\bibfield  {journal} {\bibinfo  {journal} {J. Chem.
  Phys.}\ }\textbf {\bibinfo {volume} {81}},\ \bibinfo {pages} {3188} (\bibinfo
  {year} {1984})}\BibitemShut {NoStop}%
\bibitem [{\citenamefont {Frenkel}(2014)}]{frenkel2014colloidal}%
  \BibitemOpen
  \bibfield  {author} {\bibinfo {author} {\bibfnamefont {D.}~\bibnamefont
  {Frenkel}},\ }\href@noop {} {\bibfield  {journal} {\bibinfo  {journal} {Mol.
  Phys.}\ }\textbf {\bibinfo {volume} {112}},\ \bibinfo {pages} {2325}
  (\bibinfo {year} {2014})}\BibitemShut {NoStop}%
\bibitem [{\citenamefont {Ozawa}\ and\ \citenamefont
  {Berthier}(2017)}]{ozawa2017does}%
  \BibitemOpen
  \bibfield  {author} {\bibinfo {author} {\bibfnamefont {M.}~\bibnamefont
  {Ozawa}}\ and\ \bibinfo {author} {\bibfnamefont {L.}~\bibnamefont
  {Berthier}},\ }\href@noop {} {\bibfield  {journal} {\bibinfo  {journal} {J.
  Chem. Phys.}\ }\textbf {\bibinfo {volume} {146}},\ \bibinfo {pages} {014502}
  (\bibinfo {year} {2017})}\BibitemShut {NoStop}%
\bibitem [{\citenamefont {Stillinger}\ and\ \citenamefont
  {Weber}(1985)}]{stillinger1985inherent}%
  \BibitemOpen
  \bibfield  {author} {\bibinfo {author} {\bibfnamefont {F.~H.}\ \bibnamefont
  {Stillinger}}\ and\ \bibinfo {author} {\bibfnamefont {T.~A.}\ \bibnamefont
  {Weber}},\ }\href@noop {} {\bibfield  {journal} {\bibinfo  {journal} {J.
  Chem. Phys.}\ }\textbf {\bibinfo {volume} {83}},\ \bibinfo {pages} {4767}
  (\bibinfo {year} {1985})}\BibitemShut {NoStop}%
\bibitem [{\citenamefont {Biroli}\ and\ \citenamefont
  {Monasson}(2000)}]{monasson-biroli}%
  \BibitemOpen
  \bibfield  {author} {\bibinfo {author} {\bibfnamefont {G.}~\bibnamefont
  {Biroli}}\ and\ \bibinfo {author} {\bibfnamefont {R.}~\bibnamefont
  {Monasson}},\ }\href@noop {} {\bibfield  {journal} {\bibinfo  {journal}
  {Europhys. Lett.}\ }\textbf {\bibinfo {volume} {50}},\ \bibinfo {pages} {155}
  (\bibinfo {year} {2000})}\BibitemShut {NoStop}%
\bibitem [{\citenamefont {Franz}\ and\ \citenamefont {Parisi}(1997)}]{FP97}%
  \BibitemOpen
  \bibfield  {author} {\bibinfo {author} {\bibfnamefont {S.}~\bibnamefont
  {Franz}}\ and\ \bibinfo {author} {\bibfnamefont {G.}~\bibnamefont {Parisi}},\
  }\href@noop {} {\bibfield  {journal} {\bibinfo  {journal} {Phys. Rev. Lett.}\
  }\textbf {\bibinfo {volume} {79}},\ \bibinfo {pages} {2486} (\bibinfo {year}
  {1997})}\BibitemShut {NoStop}%
\bibitem [{\citenamefont {Adam}\ and\ \citenamefont
  {Gibbs}(1965)}]{adam1965temperature}%
  \BibitemOpen
  \bibfield  {author} {\bibinfo {author} {\bibfnamefont {G.}~\bibnamefont
  {Adam}}\ and\ \bibinfo {author} {\bibfnamefont {J.~H.}\ \bibnamefont
  {Gibbs}},\ }\href@noop {} {\bibfield  {journal} {\bibinfo  {journal} {J.
  Chem. Phys.}\ }\textbf {\bibinfo {volume} {43}},\ \bibinfo {pages} {139}
  (\bibinfo {year} {1965})}\BibitemShut {NoStop}%
\bibitem [{\citenamefont {Bouchaud}\ and\ \citenamefont {Biroli}(2004)}]{BB04}%
  \BibitemOpen
  \bibfield  {author} {\bibinfo {author} {\bibfnamefont {J.-P.}\ \bibnamefont
  {Bouchaud}}\ and\ \bibinfo {author} {\bibfnamefont {G.}~\bibnamefont
  {Biroli}},\ }\href@noop {} {\bibfield  {journal} {\bibinfo  {journal} {J.
  Chem. Phys.}\ }\textbf {\bibinfo {volume} {121}},\ \bibinfo {pages} {7347}
  (\bibinfo {year} {2004})}\BibitemShut {NoStop}%
\bibitem [{\citenamefont {Biroli}\ \emph {et~al.}(2008)\citenamefont {Biroli},
  \citenamefont {Bouchaud}, \citenamefont {Cavagna}, \citenamefont {Grigera},\
  and\ \citenamefont {Verrocchio}}]{BBCGV08}%
  \BibitemOpen
  \bibfield  {author} {\bibinfo {author} {\bibfnamefont {G.}~\bibnamefont
  {Biroli}}, \bibinfo {author} {\bibfnamefont {J.-P.}\ \bibnamefont
  {Bouchaud}}, \bibinfo {author} {\bibfnamefont {A.}~\bibnamefont {Cavagna}},
  \bibinfo {author} {\bibfnamefont {T.~S.}\ \bibnamefont {Grigera}}, \ and\
  \bibinfo {author} {\bibfnamefont {P.}~\bibnamefont {Verrocchio}},\
  }\href@noop {} {\bibfield  {journal} {\bibinfo  {journal} {Nat. Phys.}\
  }\textbf {\bibinfo {volume} {4}},\ \bibinfo {pages} {771} (\bibinfo {year}
  {2008})}\BibitemShut {NoStop}%
\bibitem [{\citenamefont {Lubchenko}(2015)}]{reviewlub}%
  \BibitemOpen
  \bibfield  {author} {\bibinfo {author} {\bibfnamefont {V.}~\bibnamefont
  {Lubchenko}},\ }\href@noop {} {\bibfield  {journal} {\bibinfo  {journal}
  {Adv. Phys.}\ }\textbf {\bibinfo {volume} {64}},\ \bibinfo {pages} {283}
  (\bibinfo {year} {2015})}\BibitemShut {NoStop}%
\bibitem [{\citenamefont {Albert}\ \emph {et~al.}(2016)\citenamefont {Albert},
  \citenamefont {Bauer}, \citenamefont {Michl}, \citenamefont {Biroli},
  \citenamefont {Bouchaud}, \citenamefont {Loidl}, \citenamefont
  {Lunkenheimer}, \citenamefont {Tourbot}, \citenamefont {Wiertel-Gasquet},\
  and\ \citenamefont {Ladieu}}]{ABMBBLLTWL16}%
  \BibitemOpen
  \bibfield  {author} {\bibinfo {author} {\bibfnamefont {S.}~\bibnamefont
  {Albert}}, \bibinfo {author} {\bibfnamefont {T.}~\bibnamefont {Bauer}},
  \bibinfo {author} {\bibfnamefont {M.}~\bibnamefont {Michl}}, \bibinfo
  {author} {\bibfnamefont {G.}~\bibnamefont {Biroli}}, \bibinfo {author}
  {\bibfnamefont {J.-P.}\ \bibnamefont {Bouchaud}}, \bibinfo {author}
  {\bibfnamefont {A.}~\bibnamefont {Loidl}}, \bibinfo {author} {\bibfnamefont
  {P.}~\bibnamefont {Lunkenheimer}}, \bibinfo {author} {\bibfnamefont
  {R.}~\bibnamefont {Tourbot}}, \bibinfo {author} {\bibfnamefont
  {C.}~\bibnamefont {Wiertel-Gasquet}}, \ and\ \bibinfo {author} {\bibfnamefont
  {F.}~\bibnamefont {Ladieu}},\ }\href@noop {} {\bibfield  {journal} {\bibinfo
  {journal} {Science}\ }\textbf {\bibinfo {volume} {352}},\ \bibinfo {pages}
  {1308} (\bibinfo {year} {2016})}\BibitemShut {NoStop}%
\bibitem [{\citenamefont {Rabochiy}\ \emph
  {et~al.}(2013{\natexlab{b}})\citenamefont {Rabochiy}, \citenamefont
  {Wolynes},\ and\ \citenamefont {Lubchenko}}]{rabo2013}%
  \BibitemOpen
  \bibfield  {author} {\bibinfo {author} {\bibfnamefont {P.}~\bibnamefont
  {Rabochiy}}, \bibinfo {author} {\bibfnamefont {P.~G.}\ \bibnamefont
  {Wolynes}}, \ and\ \bibinfo {author} {\bibfnamefont {V.}~\bibnamefont
  {Lubchenko}},\ }\href@noop {} {\bibfield  {journal} {\bibinfo  {journal} {J.
  Phys. Chem. B}\ }\textbf {\bibinfo {volume} {117}},\ \bibinfo {pages} {15204}
  (\bibinfo {year} {2013}{\natexlab{b}})}\BibitemShut {NoStop}%
\bibitem [{\citenamefont {Berthier}\ \emph {et~al.}(2005)\citenamefont
  {Berthier}, \citenamefont {Biroli}, \citenamefont {Bouchaud}, \citenamefont
  {Cipelletti}, \citenamefont {Masri}, \citenamefont {L'Hote}, \citenamefont
  {Ladieu},\ and\ \citenamefont {Pierno}}]{science05}%
  \BibitemOpen
  \bibfield  {author} {\bibinfo {author} {\bibfnamefont {L.}~\bibnamefont
  {Berthier}}, \bibinfo {author} {\bibfnamefont {G.}~\bibnamefont {Biroli}},
  \bibinfo {author} {\bibfnamefont {J.-P.}\ \bibnamefont {Bouchaud}}, \bibinfo
  {author} {\bibfnamefont {L.}~\bibnamefont {Cipelletti}}, \bibinfo {author}
  {\bibfnamefont {D.~E.}\ \bibnamefont {Masri}}, \bibinfo {author}
  {\bibfnamefont {D.}~\bibnamefont {L'Hote}}, \bibinfo {author} {\bibfnamefont
  {F.}~\bibnamefont {Ladieu}}, \ and\ \bibinfo {author} {\bibfnamefont
  {M.}~\bibnamefont {Pierno}},\ }\href@noop {} {\bibfield  {journal} {\bibinfo
  {journal} {Science}\ }\textbf {\bibinfo {volume} {310}},\ \bibinfo {pages}
  {1797} (\bibinfo {year} {2005})}\BibitemShut {NoStop}%
\bibitem [{\citenamefont {Charbonneau}\ and\ \citenamefont
  {Tarjus}(2013)}]{sho}%
  \BibitemOpen
  \bibfield  {author} {\bibinfo {author} {\bibfnamefont {P.}~\bibnamefont
  {Charbonneau}}\ and\ \bibinfo {author} {\bibfnamefont {G.}~\bibnamefont
  {Tarjus}},\ }\href@noop {} {\bibfield  {journal} {\bibinfo  {journal} {Phys.
  Rev. E}\ }\textbf {\bibinfo {volume} {87}},\ \bibinfo {pages} {042305}
  (\bibinfo {year} {2013})}\BibitemShut {NoStop}%
\bibitem [{\citenamefont {Allen}\ and\ \citenamefont
  {Tildesley}(1989)}]{allen1989computer}%
  \BibitemOpen
  \bibfield  {author} {\bibinfo {author} {\bibfnamefont {M.~P.}\ \bibnamefont
  {Allen}}\ and\ \bibinfo {author} {\bibfnamefont {D.~J.}\ \bibnamefont
  {Tildesley}},\ }\href@noop {} {\emph {\bibinfo {title} {Computer simulation
  of liquids}}}\ (\bibinfo  {publisher} {Oxford University Press},\ \bibinfo
  {year} {1989})\BibitemShut {NoStop}%
\bibitem [{\citenamefont {Santos}\ \emph {et~al.}(2005)\citenamefont {Santos},
  \citenamefont {Yuste},\ and\ \citenamefont {de~Haro}}]{Santos05}%
  \BibitemOpen
  \bibfield  {author} {\bibinfo {author} {\bibfnamefont {A.}~\bibnamefont
  {Santos}}, \bibinfo {author} {\bibfnamefont {S.~B.}\ \bibnamefont {Yuste}}, \
  and\ \bibinfo {author} {\bibfnamefont {M.~L.}\ \bibnamefont {de~Haro}},\
  }\href@noop {} {\bibfield  {journal} {\bibinfo  {journal} {J. Chem. Phys.}\
  }\textbf {\bibinfo {volume} {123}},\ \bibinfo {pages} {234512} (\bibinfo
  {year} {2005})}\BibitemShut {NoStop}%
\bibitem [{\citenamefont {Gazzillo}\ and\ \citenamefont
  {Pastore}(1989)}]{gazzillo1989equation}%
  \BibitemOpen
  \bibfield  {author} {\bibinfo {author} {\bibfnamefont {D.}~\bibnamefont
  {Gazzillo}}\ and\ \bibinfo {author} {\bibfnamefont {G.}~\bibnamefont
  {Pastore}},\ }\href@noop {} {\bibfield  {journal} {\bibinfo  {journal} {Chem.
  Phys. Lett.}\ }\textbf {\bibinfo {volume} {159}},\ \bibinfo {pages} {388}
  (\bibinfo {year} {1989})}\BibitemShut {NoStop}%
\bibitem [{\citenamefont {Sindzingre}\ \emph {et~al.}(1989)\citenamefont
  {Sindzingre}, \citenamefont {Massobrio}, \citenamefont {Ciccotti},\ and\
  \citenamefont {Frenkel}}]{sindzingre1989calculation}%
  \BibitemOpen
  \bibfield  {author} {\bibinfo {author} {\bibfnamefont {P.}~\bibnamefont
  {Sindzingre}}, \bibinfo {author} {\bibfnamefont {C.}~\bibnamefont
  {Massobrio}}, \bibinfo {author} {\bibfnamefont {G.}~\bibnamefont {Ciccotti}},
  \ and\ \bibinfo {author} {\bibfnamefont {D.}~\bibnamefont {Frenkel}},\
  }\href@noop {} {\bibfield  {journal} {\bibinfo  {journal} {Chem. Phys.}\
  }\textbf {\bibinfo {volume} {129}},\ \bibinfo {pages} {213} (\bibinfo {year}
  {1989})}\BibitemShut {NoStop}%
\bibitem [{\citenamefont {Santen}\ and\ \citenamefont
  {Krauth}(2001)}]{santen2001liquid}%
  \BibitemOpen
  \bibfield  {author} {\bibinfo {author} {\bibfnamefont {L.}~\bibnamefont
  {Santen}}\ and\ \bibinfo {author} {\bibfnamefont {W.}~\bibnamefont
  {Krauth}},\ }\href@noop {} {\bibfield  {journal} {\bibinfo  {journal} {arXiv
  preprint cond-mat/0107459}\ } (\bibinfo {year} {2001})}\BibitemShut {NoStop}%
\bibitem [{\citenamefont {Pronk}\ and\ \citenamefont
  {Frenkel}(2004)}]{pronk2004melting}%
  \BibitemOpen
  \bibfield  {author} {\bibinfo {author} {\bibfnamefont {S.}~\bibnamefont
  {Pronk}}\ and\ \bibinfo {author} {\bibfnamefont {D.}~\bibnamefont
  {Frenkel}},\ }\href@noop {} {\bibfield  {journal} {\bibinfo  {journal} {Phys.
  Rev. E}\ }\textbf {\bibinfo {volume} {69}},\ \bibinfo {pages} {066123}
  (\bibinfo {year} {2004})}\BibitemShut {NoStop}%
\bibitem [{\citenamefont {Frenkel}\ and\ \citenamefont {Smit}(2001)}]{FS01}%
  \BibitemOpen
  \bibfield  {author} {\bibinfo {author} {\bibfnamefont {D.}~\bibnamefont
  {Frenkel}}\ and\ \bibinfo {author} {\bibfnamefont {B.}~\bibnamefont {Smit}},\
  }\href@noop {} {\emph {\bibinfo {title} {Understanding Molecular
  Simulation}}}\ (\bibinfo  {publisher} {Academic Press, New York, ed. 2.},\
  \bibinfo {year} {2001})\BibitemShut {NoStop}%
\bibitem [{\citenamefont {Brumer}\ and\ \citenamefont
  {Reichman}(2004)}]{brumer2004numerical}%
  \BibitemOpen
  \bibfield  {author} {\bibinfo {author} {\bibfnamefont {Y.}~\bibnamefont
  {Brumer}}\ and\ \bibinfo {author} {\bibfnamefont {D.~R.}\ \bibnamefont
  {Reichman}},\ }\href@noop {} {\bibfield  {journal} {\bibinfo  {journal} {J.
  Phys. Chem. B}\ }\textbf {\bibinfo {volume} {108}},\ \bibinfo {pages} {6832}
  (\bibinfo {year} {2004})}\BibitemShut {NoStop}%
\bibitem [{\citenamefont {Guti{\'e}rrez}\ \emph {et~al.}(2015)\citenamefont
  {Guti{\'e}rrez}, \citenamefont {Karmakar}, \citenamefont {Pollack},\ and\
  \citenamefont {Procaccia}}]{gutierrez2015static}%
  \BibitemOpen
  \bibfield  {author} {\bibinfo {author} {\bibfnamefont {R.}~\bibnamefont
  {Guti{\'e}rrez}}, \bibinfo {author} {\bibfnamefont {S.}~\bibnamefont
  {Karmakar}}, \bibinfo {author} {\bibfnamefont {Y.~G.}\ \bibnamefont
  {Pollack}}, \ and\ \bibinfo {author} {\bibfnamefont {I.}~\bibnamefont
  {Procaccia}},\ }\href@noop {} {\bibfield  {journal} {\bibinfo  {journal}
  {Europhys. Lett.}\ }\textbf {\bibinfo {volume} {111}},\ \bibinfo {pages}
  {56009} (\bibinfo {year} {2015})}\BibitemShut {NoStop}%
\bibitem [{\citenamefont {Sastry}\ \emph {et~al.}(1998)\citenamefont {Sastry},
  \citenamefont {Debenedetti},\ and\ \citenamefont
  {Stillinger}}]{sastry_signatures_1998}%
  \BibitemOpen
  \bibfield  {author} {\bibinfo {author} {\bibfnamefont {S.}~\bibnamefont
  {Sastry}}, \bibinfo {author} {\bibfnamefont {P.~G.}\ \bibnamefont
  {Debenedetti}}, \ and\ \bibinfo {author} {\bibfnamefont {F.~H.}\ \bibnamefont
  {Stillinger}},\ }\href@noop {} {\bibfield  {journal} {\bibinfo  {journal}
  {Nature}\ }\textbf {\bibinfo {volume} {393}},\ \bibinfo {pages} {554}
  (\bibinfo {year} {1998})}\BibitemShut {NoStop}%
\bibitem [{\citenamefont {Charbonneau}\ \emph
  {et~al.}(2014{\natexlab{b}})\citenamefont {Charbonneau}, \citenamefont {Jin},
  \citenamefont {Parisi},\ and\ \citenamefont
  {Zamponi}}]{charbonneau2014hopping}%
  \BibitemOpen
  \bibfield  {author} {\bibinfo {author} {\bibfnamefont {P.}~\bibnamefont
  {Charbonneau}}, \bibinfo {author} {\bibfnamefont {Y.}~\bibnamefont {Jin}},
  \bibinfo {author} {\bibfnamefont {G.}~\bibnamefont {Parisi}}, \ and\ \bibinfo
  {author} {\bibfnamefont {F.}~\bibnamefont {Zamponi}},\ }\href@noop {}
  {\bibfield  {journal} {\bibinfo  {journal} {Proc. Natl. Acad. Sci. U.S.A.}\
  }\textbf {\bibinfo {volume} {111}},\ \bibinfo {pages} {15025} (\bibinfo
  {year} {2014}{\natexlab{b}})}\BibitemShut {NoStop}%
\bibitem [{\citenamefont {Boubl{\'\i}k}(1970)}]{boublik1970hard}%
  \BibitemOpen
  \bibfield  {author} {\bibinfo {author} {\bibfnamefont {T.}~\bibnamefont
  {Boubl{\'\i}k}},\ }\href@noop {} {\bibfield  {journal} {\bibinfo  {journal}
  {J. Chem. Phys.}\ }\textbf {\bibinfo {volume} {53}},\ \bibinfo {pages} {471}
  (\bibinfo {year} {1970})}\BibitemShut {NoStop}%
\bibitem [{\citenamefont {Mansoori}\ \emph {et~al.}(1971)\citenamefont
  {Mansoori}, \citenamefont {Carnahan}, \citenamefont {Starling},\ and\
  \citenamefont {Leland~Jr}}]{mansoori1971equilibrium}%
  \BibitemOpen
  \bibfield  {author} {\bibinfo {author} {\bibfnamefont {G.~A.}\ \bibnamefont
  {Mansoori}}, \bibinfo {author} {\bibfnamefont {N.~F.}\ \bibnamefont
  {Carnahan}}, \bibinfo {author} {\bibfnamefont {K.~E.}\ \bibnamefont
  {Starling}}, \ and\ \bibinfo {author} {\bibfnamefont {T.~W.}\ \bibnamefont
  {Leland~Jr}},\ }\href@noop {} {\bibfield  {journal} {\bibinfo  {journal} {J.
  Chem. Phys.}\ }\textbf {\bibinfo {volume} {54}},\ \bibinfo {pages} {1523}
  (\bibinfo {year} {1971})}\BibitemShut {NoStop}%
\bibitem [{\citenamefont {Berthier}\ \emph
  {et~al.}(2016{\natexlab{c}})\citenamefont {Berthier}, \citenamefont
  {Charbonneau}, \citenamefont {Jin}, \citenamefont {Parisi}, \citenamefont
  {Seoane},\ and\ \citenamefont {Zamponi}}]{berthier2016growing}%
  \BibitemOpen
  \bibfield  {author} {\bibinfo {author} {\bibfnamefont {L.}~\bibnamefont
  {Berthier}}, \bibinfo {author} {\bibfnamefont {P.}~\bibnamefont
  {Charbonneau}}, \bibinfo {author} {\bibfnamefont {Y.}~\bibnamefont {Jin}},
  \bibinfo {author} {\bibfnamefont {G.}~\bibnamefont {Parisi}}, \bibinfo
  {author} {\bibfnamefont {B.}~\bibnamefont {Seoane}}, \ and\ \bibinfo {author}
  {\bibfnamefont {F.}~\bibnamefont {Zamponi}},\ }\href@noop {} {\bibfield
  {journal} {\bibinfo  {journal} {Proc. Natl. Acad. Sci. U.S.A.}\ }\textbf
  {\bibinfo {volume} {113}},\ \bibinfo {pages} {8397} (\bibinfo {year}
  {2016}{\natexlab{c}})}\BibitemShut {NoStop}%
\bibitem [{\citenamefont {Vasisht}\ \emph {et~al.}(2011)\citenamefont
  {Vasisht}, \citenamefont {Saw},\ and\ \citenamefont {Sastry}}]{vashist}%
  \BibitemOpen
  \bibfield  {author} {\bibinfo {author} {\bibfnamefont {V.~V.}\ \bibnamefont
  {Vasisht}}, \bibinfo {author} {\bibfnamefont {S.}~\bibnamefont {Saw}}, \ and\
  \bibinfo {author} {\bibfnamefont {S.}~\bibnamefont {Sastry}},\ }\href@noop {}
  {\bibfield  {journal} {\bibinfo  {journal} {Nat. Phys.}\ }\textbf {\bibinfo
  {volume} {7}},\ \bibinfo {pages} {549} (\bibinfo {year} {2011})}\BibitemShut
  {NoStop}%
\bibitem [{\citenamefont {Shiba}\ \emph {et~al.}(2012)\citenamefont {Shiba},
  \citenamefont {Kawasaki},\ and\ \citenamefont {Onuki}}]{kawasaki}%
  \BibitemOpen
  \bibfield  {author} {\bibinfo {author} {\bibfnamefont {H.}~\bibnamefont
  {Shiba}}, \bibinfo {author} {\bibfnamefont {T.}~\bibnamefont {Kawasaki}}, \
  and\ \bibinfo {author} {\bibfnamefont {A.}~\bibnamefont {Onuki}},\
  }\href@noop {} {\bibfield  {journal} {\bibinfo  {journal} {Phys. Rev. E}\
  }\textbf {\bibinfo {volume} {86}},\ \bibinfo {pages} {041504} (\bibinfo
  {year} {2012})}\BibitemShut {NoStop}%
\bibitem [{\citenamefont {Rycroft}(2009)}]{voro++}%
  \BibitemOpen
  \bibfield  {author} {\bibinfo {author} {\bibfnamefont {C.}~\bibnamefont
  {Rycroft}},\ }\href@noop {} {\  (\bibinfo {year} {2009})}\BibitemShut
  {NoStop}%
\bibitem [{\citenamefont {Ninarello}\ \emph
  {et~al.}(2017{\natexlab{b}})\citenamefont {Ninarello}, \citenamefont
  {Berthier},\ and\ \citenamefont {Coslovich}}]{ninarello}%
  \BibitemOpen
  \bibfield  {author} {\bibinfo {author} {\bibfnamefont {A.}~\bibnamefont
  {Ninarello}}, \bibinfo {author} {\bibfnamefont {L.}~\bibnamefont {Berthier}},
  \ and\ \bibinfo {author} {\bibfnamefont {D.}~\bibnamefont {Coslovich}},\
  }\href@noop {} {\bibfield  {journal} {\bibinfo  {journal} {Phys. Rev. X}\
  }\textbf {\bibinfo {volume} {7}},\ \bibinfo {pages} {021039} (\bibinfo {year}
  {2017}{\natexlab{b}})}\BibitemShut {NoStop}%
\bibitem [{\citenamefont {Elmatad}\ \emph {et~al.}(2009)\citenamefont
  {Elmatad}, \citenamefont {Chandler},\ and\ \citenamefont
  {Garrahan}}]{elmatad2009corresponding}%
  \BibitemOpen
  \bibfield  {author} {\bibinfo {author} {\bibfnamefont {Y.~S.}\ \bibnamefont
  {Elmatad}}, \bibinfo {author} {\bibfnamefont {D.}~\bibnamefont {Chandler}}, \
  and\ \bibinfo {author} {\bibfnamefont {J.~P.}\ \bibnamefont {Garrahan}},\
  }\href@noop {} {\bibfield  {journal} {\bibinfo  {journal} {J. Phys. Chem. B}\
  }\textbf {\bibinfo {volume} {113}},\ \bibinfo {pages} {5563} (\bibinfo {year}
  {2009})}\BibitemShut {NoStop}%
\bibitem [{\citenamefont {Isobe}\ \emph {et~al.}(2016)\citenamefont {Isobe},
  \citenamefont {Keys}, \citenamefont {Chandler},\ and\ \citenamefont
  {Garrahan}}]{isobe2016applicability}%
  \BibitemOpen
  \bibfield  {author} {\bibinfo {author} {\bibfnamefont {M.}~\bibnamefont
  {Isobe}}, \bibinfo {author} {\bibfnamefont {A.~S.}\ \bibnamefont {Keys}},
  \bibinfo {author} {\bibfnamefont {D.}~\bibnamefont {Chandler}}, \ and\
  \bibinfo {author} {\bibfnamefont {J.~P.}\ \bibnamefont {Garrahan}},\
  }\href@noop {} {\bibfield  {journal} {\bibinfo  {journal} {Phys. Rev. Lett.}\
  }\textbf {\bibinfo {volume} {117}},\ \bibinfo {pages} {145701} (\bibinfo
  {year} {2016})}\BibitemShut {NoStop}%
\bibitem [{\citenamefont {Richert}\ and\ \citenamefont
  {Angell}(1998)}]{richert_dynamics_1998}%
  \BibitemOpen
  \bibfield  {author} {\bibinfo {author} {\bibfnamefont {R.}~\bibnamefont
  {Richert}}\ and\ \bibinfo {author} {\bibfnamefont {C.}~\bibnamefont
  {Angell}},\ }\href@noop {} {\bibfield  {journal} {\bibinfo  {journal} {J.
  Chem. Phys.}\ }\textbf {\bibinfo {volume} {108}},\ \bibinfo {pages} {9016}
  (\bibinfo {year} {1998})}\BibitemShut {NoStop}%
\bibitem [{\citenamefont {Sastry}(2001)}]{Sastry01}%
  \BibitemOpen
  \bibfield  {author} {\bibinfo {author} {\bibfnamefont {S.}~\bibnamefont
  {Sastry}},\ }\href@noop {} {\bibfield  {journal} {\bibinfo  {journal}
  {Nature}\ }\textbf {\bibinfo {volume} {409}},\ \bibinfo {pages} {164}
  (\bibinfo {year} {2001})}\BibitemShut {NoStop}%
\bibitem [{\citenamefont {Salacuse}\ and\ \citenamefont
  {Stell}(1982)}]{salacuse1982polydisperse}%
  \BibitemOpen
  \bibfield  {author} {\bibinfo {author} {\bibfnamefont {J.}~\bibnamefont
  {Salacuse}}\ and\ \bibinfo {author} {\bibfnamefont {G.}~\bibnamefont
  {Stell}},\ }\href@noop {} {\bibfield  {journal} {\bibinfo  {journal} {J.
  Chem. Phys.}\ }\textbf {\bibinfo {volume} {77}},\ \bibinfo {pages} {3714}
  (\bibinfo {year} {1982})}\BibitemShut {NoStop}%
\bibitem [{\citenamefont {Sollich}(2001)}]{sollich2001predicting}%
  \BibitemOpen
  \bibfield  {author} {\bibinfo {author} {\bibfnamefont {P.}~\bibnamefont
  {Sollich}},\ }\href@noop {} {\bibfield  {journal} {\bibinfo  {journal} {J.
  Phys. Cond. Matter}\ }\textbf {\bibinfo {volume} {14}},\ \bibinfo {pages}
  {R79} (\bibinfo {year} {2001})}\BibitemShut {NoStop}%
\bibitem [{\citenamefont {Xu}\ \emph {et~al.}(2005)\citenamefont {Xu},
  \citenamefont {Blawzdziewicz},\ and\ \citenamefont {O'Hern}}]{xu2005random}%
  \BibitemOpen
  \bibfield  {author} {\bibinfo {author} {\bibfnamefont {N.}~\bibnamefont
  {Xu}}, \bibinfo {author} {\bibfnamefont {J.}~\bibnamefont {Blawzdziewicz}}, \
  and\ \bibinfo {author} {\bibfnamefont {C.~S.}\ \bibnamefont {O'Hern}},\
  }\href@noop {} {\bibfield  {journal} {\bibinfo  {journal} {Phys. Rev. E}\
  }\textbf {\bibinfo {volume} {71}},\ \bibinfo {pages} {061306} (\bibinfo
  {year} {2005})}\BibitemShut {NoStop}%
\bibitem [{\citenamefont {Desmond}\ and\ \citenamefont
  {Weeks}(2009)}]{desmond2009random}%
  \BibitemOpen
  \bibfield  {author} {\bibinfo {author} {\bibfnamefont {K.~W.}\ \bibnamefont
  {Desmond}}\ and\ \bibinfo {author} {\bibfnamefont {E.~R.}\ \bibnamefont
  {Weeks}},\ }\href@noop {} {\bibfield  {journal} {\bibinfo  {journal} {Phys.
  Rev. E}\ }\textbf {\bibinfo {volume} {80}},\ \bibinfo {pages} {051305}
  (\bibinfo {year} {2009})}\BibitemShut {NoStop}%
\bibitem [{\citenamefont {Berthier}(2013)}]{Berthier13}%
  \BibitemOpen
  \bibfield  {author} {\bibinfo {author} {\bibfnamefont {L.}~\bibnamefont
  {Berthier}},\ }\href@noop {} {\bibfield  {journal} {\bibinfo  {journal}
  {Phys. Rev. E}\ }\textbf {\bibinfo {volume} {88}},\ \bibinfo {pages} {022313}
  (\bibinfo {year} {2013})}\BibitemShut {NoStop}%
\bibitem [{\citenamefont {Berthier}\ and\ \citenamefont {Jack}(2015)}]{BJ15}%
  \BibitemOpen
  \bibfield  {author} {\bibinfo {author} {\bibfnamefont {L.}~\bibnamefont
  {Berthier}}\ and\ \bibinfo {author} {\bibfnamefont {R.~L.}\ \bibnamefont
  {Jack}},\ }\href@noop {} {\bibfield  {journal} {\bibinfo  {journal} {Phys.
  Rev. Lett.}\ }\textbf {\bibinfo {volume} {114}},\ \bibinfo {pages} {205701}
  (\bibinfo {year} {2015})}\BibitemShut {NoStop}%
\bibitem [{\citenamefont {Hukushima}\ and\ \citenamefont
  {Nemoto}(1996)}]{Hukushima96}%
  \BibitemOpen
  \bibfield  {author} {\bibinfo {author} {\bibfnamefont {K.}~\bibnamefont
  {Hukushima}}\ and\ \bibinfo {author} {\bibfnamefont {K.}~\bibnamefont
  {Nemoto}},\ }\href@noop {} {\bibfield  {journal} {\bibinfo  {journal} {J.
  Phys. Soc. Jpn}\ }\textbf {\bibinfo {volume} {65}},\ \bibinfo {pages} {1604}
  (\bibinfo {year} {1996})}\BibitemShut {NoStop}%
\bibitem [{\citenamefont {Coslovich}\ \emph {et~al.}(2017)\citenamefont
  {Coslovich}, \citenamefont {Berthier},\ and\ \citenamefont
  {Ozawa}}]{ozawa2017exploring}%
  \BibitemOpen
  \bibfield  {author} {\bibinfo {author} {\bibfnamefont {D.}~\bibnamefont
  {Coslovich}}, \bibinfo {author} {\bibfnamefont {L.}~\bibnamefont {Berthier}},
  \ and\ \bibinfo {author} {\bibfnamefont {M.}~\bibnamefont {Ozawa}},\
  }\href@noop {} {\bibfield  {journal} {\bibinfo  {journal} {SciPost Physics}\
  }\textbf {\bibinfo {volume} {3}},\ \bibinfo {pages} {027} (\bibinfo {year}
  {2017})}\BibitemShut {NoStop}%
\bibitem [{\citenamefont {Fukunishi}\ \emph {et~al.}(2002)\citenamefont
  {Fukunishi}, \citenamefont {Watanabe},\ and\ \citenamefont {Takada}}]{FWT02}%
  \BibitemOpen
  \bibfield  {author} {\bibinfo {author} {\bibfnamefont {H.}~\bibnamefont
  {Fukunishi}}, \bibinfo {author} {\bibfnamefont {O.}~\bibnamefont {Watanabe}},
  \ and\ \bibinfo {author} {\bibfnamefont {S.}~\bibnamefont {Takada}},\
  }\href@noop {} {\bibfield  {journal} {\bibinfo  {journal} {J. Chem. Phys.}\
  }\textbf {\bibinfo {volume} {116}},\ \bibinfo {pages} {9058} (\bibinfo {year}
  {2002})}\BibitemShut {NoStop}%
\bibitem [{\citenamefont {Katzgraber}\ \emph {et~al.}(2006)\citenamefont
  {Katzgraber}, \citenamefont {Trebst}, \citenamefont {Huse},\ and\
  \citenamefont {Troyer}}]{KTHT06}%
  \BibitemOpen
  \bibfield  {author} {\bibinfo {author} {\bibfnamefont {H.~G.}\ \bibnamefont
  {Katzgraber}}, \bibinfo {author} {\bibfnamefont {S.}~\bibnamefont {Trebst}},
  \bibinfo {author} {\bibfnamefont {D.~A.}\ \bibnamefont {Huse}}, \ and\
  \bibinfo {author} {\bibfnamefont {M.}~\bibnamefont {Troyer}},\ }\href@noop {}
  {\bibfield  {journal} {\bibinfo  {journal} {J. Stat. Mech.}\ }\textbf
  {\bibinfo {volume} {2006}},\ \bibinfo {pages} {P03018} (\bibinfo {year}
  {2006})}\BibitemShut {NoStop}%
\bibitem [{\citenamefont {Vogel}\ and\ \citenamefont {Perez}(2015)}]{VP15}%
  \BibitemOpen
  \bibfield  {author} {\bibinfo {author} {\bibfnamefont {T.}~\bibnamefont
  {Vogel}}\ and\ \bibinfo {author} {\bibfnamefont {D.}~\bibnamefont {Perez}},\
  }\href@noop {} {\bibfield  {journal} {\bibinfo  {journal} {Phys. Rev. Lett.}\
  }\textbf {\bibinfo {volume} {115}},\ \bibinfo {pages} {190602} (\bibinfo
  {year} {2015})}\BibitemShut {NoStop}%
\bibitem [{\citenamefont {Cavagna}\ \emph {et~al.}(2012)\citenamefont
  {Cavagna}, \citenamefont {Grigera},\ and\ \citenamefont
  {Verrocchio}}]{BICtest12}%
  \BibitemOpen
  \bibfield  {author} {\bibinfo {author} {\bibfnamefont {A.}~\bibnamefont
  {Cavagna}}, \bibinfo {author} {\bibfnamefont {T.~S.}\ \bibnamefont
  {Grigera}}, \ and\ \bibinfo {author} {\bibfnamefont {P.}~\bibnamefont
  {Verrocchio}},\ }\href@noop {} {\bibfield  {journal} {\bibinfo  {journal} {J.
  Chem. Phys.}\ }\textbf {\bibinfo {volume} {136}},\ \bibinfo {pages} {204502}
  (\bibinfo {year} {2012})}\BibitemShut {NoStop}%
\bibitem [{\citenamefont {Coluzzi}\ \emph {et~al.}(2000)\citenamefont
  {Coluzzi}, \citenamefont {Parisi},\ and\ \citenamefont
  {Verrocchio}}]{coluzzi2000lennard}%
  \BibitemOpen
  \bibfield  {author} {\bibinfo {author} {\bibfnamefont {B.}~\bibnamefont
  {Coluzzi}}, \bibinfo {author} {\bibfnamefont {G.}~\bibnamefont {Parisi}}, \
  and\ \bibinfo {author} {\bibfnamefont {P.}~\bibnamefont {Verrocchio}},\
  }\href@noop {} {\bibfield  {journal} {\bibinfo  {journal} {J. Chem. Phys.}\
  }\textbf {\bibinfo {volume} {112}},\ \bibinfo {pages} {2933} (\bibinfo {year}
  {2000})}\BibitemShut {NoStop}%
\bibitem [{\citenamefont {Ozawa}\ \emph {et~al.}(2015)\citenamefont {Ozawa},
  \citenamefont {Kob}, \citenamefont {Ikeda},\ and\ \citenamefont
  {Miyazaki}}]{ozawa2015equilibrium}%
  \BibitemOpen
  \bibfield  {author} {\bibinfo {author} {\bibfnamefont {M.}~\bibnamefont
  {Ozawa}}, \bibinfo {author} {\bibfnamefont {W.}~\bibnamefont {Kob}}, \bibinfo
  {author} {\bibfnamefont {A.}~\bibnamefont {Ikeda}}, \ and\ \bibinfo {author}
  {\bibfnamefont {K.}~\bibnamefont {Miyazaki}},\ }\href@noop {} {\bibfield
  {journal} {\bibinfo  {journal} {Proc. Natl. Acad. Sci. U.S.A.}\ }\textbf
  {\bibinfo {volume} {112}},\ \bibinfo {pages} {6914} (\bibinfo {year}
  {2015})}\BibitemShut {NoStop}%
\bibitem [{\citenamefont {Rosenfeld}\ and\ \citenamefont
  {Tarazona}(1998)}]{rosenfeld1998density}%
  \BibitemOpen
  \bibfield  {author} {\bibinfo {author} {\bibfnamefont {Y.}~\bibnamefont
  {Rosenfeld}}\ and\ \bibinfo {author} {\bibfnamefont {P.}~\bibnamefont
  {Tarazona}},\ }\href@noop {} {\bibfield  {journal} {\bibinfo  {journal} {Mol.
  Phys.}\ }\textbf {\bibinfo {volume} {95}},\ \bibinfo {pages} {141} (\bibinfo
  {year} {1998})}\BibitemShut {NoStop}%
\bibitem [{\citenamefont {Sciortino}(2005)}]{sciortino2005potential}%
  \BibitemOpen
  \bibfield  {author} {\bibinfo {author} {\bibfnamefont {F.}~\bibnamefont
  {Sciortino}},\ }\href@noop {} {\bibfield  {journal} {\bibinfo  {journal} {J.
  Stat. Mech.}\ }\textbf {\bibinfo {volume} {2005}},\ \bibinfo {pages} {P05015}
  (\bibinfo {year} {2005})}\BibitemShut {NoStop}%
\bibitem [{\citenamefont {Flenner}\ and\ \citenamefont
  {Szamel}(2006)}]{flenner2006hybrid}%
  \BibitemOpen
  \bibfield  {author} {\bibinfo {author} {\bibfnamefont {E.}~\bibnamefont
  {Flenner}}\ and\ \bibinfo {author} {\bibfnamefont {G.}~\bibnamefont
  {Szamel}},\ }\href@noop {} {\bibfield  {journal} {\bibinfo  {journal} {Phys.
  Rev. E}\ }\textbf {\bibinfo {volume} {73}},\ \bibinfo {pages} {061505}
  (\bibinfo {year} {2006})}\BibitemShut {NoStop}%
\end{thebibliography}%

\end{document}